\documentclass[]{article}

\usepackage{multirow, amsmath, amsthm, graphicx, amssymb, amsfonts, latexsym, enumerate, amscd, xcolor, float, caption}
\usepackage{listings, hyperref, manyfoot}
\usepackage{geometry}
\usepackage{amsmath}
\usepackage{amsfonts}
\usepackage{amssymb}
\usepackage{graphicx}
\usepackage{url} 
\usepackage{amsthm}
\usepackage{academicons}
\usepackage{breakcites} 
\usepackage{color}
\usepackage{epsfig}
\usepackage{epstopdf}
\usepackage{etoolbox}
\usepackage{enumitem}
\usepackage{float}
\usepackage{fancyhdr}
\usepackage{graphics}
\usepackage{latexsym,amsfonts}
\usepackage{longtable}
\usepackage{listings}
\usepackage{lscape} 
\usepackage{lipsum}
\usepackage{multirow}             
\usepackage{pdfpages}
\usepackage[figuresright]{rotating} 
\usepackage{sectsty}
\usepackage{setspace}
\usepackage{subfigure}
\usepackage{textcomp}
\usepackage{wrapfig} 
\usepackage{wasysym} 
\usepackage{xcolor}
\usepackage{amsmath,bm}
\usepackage{dsfont}
\usepackage{alltt}
\RequirePackage[numbers]{natbib}

\DeclareMathOperator*{\argmin}{arg\,min}
\DeclareMathOperator{\Trace}{Tr}

\begin{document}
\title{Scalable skewed Bayesian inference for latent Gaussian models}

\author{Shourya Dutta$^1$ (shourya.dutta@kaust.edu.sa),\\ and \\ Janet van Niekerk$^{1,2}$(janet.vanNiekerk@kaust.edu.sa)\\ and \\
H\aa vard Rue$^1$ (haavard.rue@kaust.edu.sa)\\ \\
$^1$ Statistics Program, CEMSE Division,
King Abdullah University of \\ Science and Technology,
Kingdom of Saudi Arabia\\
$^2$ Department of Statistics, University of Pretoria, South Africa}

\maketitle

\begin{abstract}
Approximate Bayesian inference for the class of latent Gaussian models can be achieved efficiently with integrated nested Laplace approximations (INLA). Based on recent reformulations in the INLA methodology, we propose a further extension that is necessary in some cases like heavy-tailed likelihoods or binary regression with imbalanced data. This extension formulates a skewed version of the Laplace method such that some marginals are skewed and some are kept Gaussian while the dependence is maintained with the Gaussian copula from the Laplace method. Our approach is formulated to be scalable in model and data size, using a variational inferential framework enveloped in INLA. We illustrate the necessity and performance using simulated cases, as well as a case study of a rare disease where class imbalance is naturally present.
\end{abstract}

\section{Introduction}
\label{sec:intro}
Latent Gaussian models (LGMs) contain various well-known statistical models such as generalized additive mixed models, spatial models like the Besag model for irregular lattice and the Matern model for point-indexed data, temporal smoothers like random walks or autoregressive processes, stochastic volatility models, survival and reliability analysis models, joint models, models for multivariate regression and many more. Bayesian inference of LGMs can be done accurately and efficiently using the integrated nested Laplace approximations (INLA) methodology \citep{rue2009approximate}. The development of the R-INLA package have enabled scientist to construct complex models and infer them using INLA, see \citet{konstantinoudis2022regional, van2021bayesian, msemburi2023estimates, bhatt2015effect} for some recent works using LGMs and INLA.\\
For the purpose of presenting the proposed methodology in the case of Bayesian inference for LGMs, we will assume that our statistical model is an LGM for data $\pmb y$ of size $n$ with likelihood function $L(.)$, covariates $\pmb x$, unobservable latent field $\pmb f$ of size $p$ with a Gaussian prior $p_N(.)$, and hyper/nuisance parameters $\pmb \theta$ of size $k$, from the likelihood or prior model, respectively, such that
\begin{eqnarray}
	y_i|\pmb f,\pmb\theta &\sim& L_i(y_i;\eta_i,\pmb\theta)\quad i = 1,2,...,n\text{   conditionally independent}\notag\\
	\pmb f |\pmb\theta &\sim& p_N(\pmb f;\pmb 0, \pmb Q_{\pmb\theta})\notag\\
	\pmb\theta &\sim& p(\pmb\theta)\label{eq:LGM} 
\end{eqnarray}
where $\pmb Q_{\pmb\theta}$ is the sparse prior precision matrix of $\pmb f$ and $p(\pmb\theta)$ is the prior density of $\pmb\theta$. The additive linear predictor that connects the data to the latent field is specified as $\pmb\eta = \pmb A\pmb f$, or $\eta_i = \pmb A_i \pmb f$, where the design matrix $\pmb A$ is constructed from the covariates $\pmb x$, and $\pmb A_i$ is the $i^{\text{th}}$ row of $\pmb A$. The parameters to be estimated are $\pmb f$ and $\pmb\theta$.\\
INLA approximates the marginal posteriors semi-parametrically through numerical integration with respect to $\pmb\theta$, using the (corrected) Laplace method for the conditional posterior of the latent field. The Gaussian approximation emanating from the Laplace method \citep{tierney1986accurate}, by matching the mode, and the curvature around the mode, is very efficient but can be inaccurate in some cases. Various other Gaussian approximation strategies exist, but few are as scalable as the Laplace method in both model complexity and data size. Moreover, placing the mean of the multivariate Gaussian target at the mode of the high-dimensional parameter space can be problematic due to the domain of attraction being away from the mode, also known as the curse of dimensionality. \\ \\
Recently, a hybrid method for a Gaussian approximation has been proposed where the mean from the Laplace method is corrected with an efficient low-rank Variational Bayes correction \citep{van2021correcting}, and this has subsequently been adopted by the INLA methodology \citep{van2023new}. This approach holds promise for algorithms using Laplace's method, either in the marginal sense or in a conditional sense (like the case for INLA) for models with nuisance/hyperparameters. The advantage of this work, is that the mean-field assumption is not necessary, and a (fully correlated) multivariate approximation is possible at a much lower cost than with sampling-based inference. In most cases the covariance matrix is approximated well using only the Laplace method. Corrections of the variance can be addressed with an extension of the mean-correcting methodology, although the implementation is not straightforward (this variance correction methodology using low-rank Variational Bayes is proposed in Section \ref{sec:varcorr}). \\ \\
However, the resultant conditional approximation is still Gaussian and thus symmetric. The question we pondered is how we can move away from multivariate Gaussian to a multivariate skew model that is scalable in practice (in data size and model complexity) with a computational cost similar to that of the Laplace method, and can be used in Bayesian inference for complex models and/or large data.\\ \\
In this work, we propose an extension of the INLA methodology where the multivariate conditional posterior of the latent field contains some skewed marginals and some symmetric marginals, while the dependence is retained through a Gaussian copula. The inference of this approach is performed in two steps - firstly, a mean- and variance corrected Gaussian approximation with fully connected covariance matrix (not mean-field) from the Laplace method is computed, whereafter some elements are skewed and this skewness is inferred using a Variational Bayes procedure. This approach scales well with model complexity and data size, and provides an efficient and sufficiently accurate high-dimensional skew approximation. The details are presented in Section \ref{sec:skewcorr}. 
\subsection{Other related works}
Our proposal allows for some elements to have a skew distribution while other are kept symmetric in the multivariate conditional posterior in the INLA methodology. The multivariate skew-normal distribution \citep{azzalini1999statistical} and its variates come to mind, but are not directly applicable since they do not allow for some of the skewness components to be exactly zero and the inference of these models are particularly cumbersome. The literature contains many other proposals of multivariate distributions where the marginals are skewed, although most (if not all) of these are infeasible in practice for complex models or large datasets (see the discussions of \citet{mondal2024multivariate} for EM algorithm-based inference and \citet{tan2024variational} for mean-field variational inference). If we consider simple models or moderate data, we have many multivariate target densities that we can consider although approximate methods are probably not even necessary in this case, as exact methods like Markov Chain Monte Carlo (MCMC) \citep{plummer2003jags} or Hamiltonion Monte Carlo (HMC) \citep{carpenter2017stan} can be implemented (some examples of EM or MCMC algorithms for multivariate skew models are \citet{mondal2024multivariate}, \citet{fruhwirth2010bayesian}, \citet{lachos2007likelihood} and \citet{liseo2013bayesian}). Parametric approximations have been proposed by \citet{zhou2024tractable} who proposed a matching-based scheme for a multivariate skew-normal approximation, and \citet{durante2024skewed} proposed a skew-modal approximation based on a third-order extension of the Laplace method, respectively. When the model complexity is increased by including various structured and unstructured random effects such as spatial, temporal, frailties or group effects, and/or the data size is large, then direct parametric methods and abstract choices of multivariate approximations can be infeasible and impractical. Moreover, the mean-field assumption of the latent posteriors is often assumed in Variational Inference algorithms, as in \citet{tan2024variational}. However, if we want to capture the posterior dependence of the parameter space, then the mean-field assumption can be too restrictive, and this misspecification of the relationships among variables could lead to erroneous conclusions. 
\subsection{Structure of the paper}
In Section \ref{sec:basics} we provide a necessary, albeit concise, overview and details surrounding the Laplace method for a Gaussian approximation, the concept of variational inference and the INLA methodology. A low-rank Variational Bayes correction to the mean and covariance of the Laplace method is presented in Section \ref{sec:meanvarcorr}. We present the methodology of our high-dimensional skewed approximation proposal for LGMs using INLA in Section \ref{sec:skewcorr} and illustrate the necessity and novelty of our method on simulated examples. An application to multiplex autoimmune assay data is presented in Section \ref{sec:app}, followed by a discussion in Section \ref{sec:disc}.

\section{Preliminaries}\label{sec:basics}
\subsection{Laplace's method}
In Bayesian inference, it's a common occurrence to encounter complex multivariate posterior probability distributions that are challenging to analyze directly. To address this issue, the Laplace method was proposed to construct an efficient multivariate Gaussian approximation, resulting in advantageous Gaussian marginal and conditional densities.

Consider the unknown density function of a vector $\pmb \phi$, $p(\pmb \phi)$. The Laplace method constructs the Gaussian approximation, $p_N(\pmb \phi;\pmb\mu, \pmb Q)$, to $p(\pmb \phi)$ based on first and second-order derivatives at the mode of $\log p(\pmb \phi)$, $\pmb\mu$. The unknown parameters of the approximation, the mean vector, $\pmb\mu$ and the precision matrix $\pmb Q$, are obtained from the solution of the following linear system

\begin{equation}
	-\pmb H \left(\log p(\pmb \phi)\right)|_{\pmb \phi = \pmb \mu_0}\ \ \pmb\mu_0 = \pmb b|_{\pmb \phi = \pmb \mu_0},
	\label{eq:lm}
\end{equation}

where $\pmb H \left(\log p(\pmb \phi)\right)|_{\pmb \phi = \pmb \mu_0}$ is the Hessian matrix of the log density $\log p(\pmb \phi)$ and $ \pmb b|_{\pmb \phi = \pmb \mu_0} = \nabla \log p(\pmb \phi)|_{\pmb \phi = \pmb \mu_0} -\pmb H \left(\log p(\pmb \phi)\right)|_{\pmb \phi = \pmb \mu_0}\pmb\mu_0$, evaluated at $\pmb \phi = \pmb \mu_0$. The system \eqref{eq:lm} is solved when $\nabla \log p(\pmb \phi)|_{{\pmb \phi = \pmb \mu_0}} = 0$, hence $\pmb \mu_0$ is the mode of $\log p(\pmb \phi)$ and subsequently the mode of $p(\pmb \phi)$. The proposed approximate distribution of $\pmb \phi$ from the Laplace method is then
\begin{equation}
	\pmb \phi \overset{.}{\sim} \text{N}\left(\pmb\mu = \pmb\mu_0 \ ,  \pmb Q = -\pmb H \left(\log p(\pmb \phi)\right)_{\pmb \phi = \pmb \mu_0}\right).
\end{equation}

\subsubsection{Laplace method for approximate Bayesian inference of LGMs} \label{sec2.1.2}

The Laplace method provides a general framework for a multivariate Gaussian approximation, and we can also apply this to approximate the posterior distribution in terms of Bayesian inference. Consider an LGM as defined in \eqref{eq:LGM}, the unknown parameters are the latent field $\pmb f$ and the hyperparameters $\pmb\theta$. 
The joint density of the data, latent field and hyperparameters is
\begin{eqnarray*}
	p(\pmb y, \pmb f, \pmb\theta) &=& p(\pmb y| \pmb f, \pmb\theta)\ p(\pmb f|\pmb\theta) p(\pmb\theta)\\
	&=& \prod_{i=1}^n L_i(y_i; \pmb f, \pmb\theta) p_N (\pmb f;\pmb 0, \pmb Q_{\pmb\theta} ) p(\pmb\theta).
\end{eqnarray*}
Hence the conditional posterior of the latent field is
\begin{eqnarray}
	p(\pmb f|\pmb y, \pmb\theta) &=& \frac{p(\pmb f, \pmb y, \pmb\theta)}{p(\pmb y, \pmb\theta)}
	\propto p(\pmb f, \pmb y, \pmb\theta)\notag\\
	&\propto& \exp\left(-\frac{1}{2}\pmb f^\top \pmb Q_{\pmb\theta}\pmb f -\frac{1}{2} \pmb f ^\top \pmb A^\top \pmb C(\pmb\theta,\pmb y)  \pmb A \pmb f +  \pmb b^\top(\pmb\theta,\pmb y)\pmb A \pmb f\right),\label{eq:jointbayes}
\end{eqnarray}
where $\pmb C(\pmb\theta,\pmb y)$ is a diagonal matrix with entries $-(\log L_i)^{''}((\pmb A \pmb f_0)_i)$, and $\pmb b(\pmb\theta,\pmb y)$ is a vector with entries $(\log L_i)^{'}((\pmb A \pmb f_0)_i)-(\log L_i)^{''}((\pmb A \pmb f_0)_i)\ (\pmb f_0)_i$, for $i=1,2,...,n$ from a second order Taylor series expansion of the log-likelihood around the point $\pmb f = \pmb f_0$.
Thus the Laplace method's linear system is
\begin{equation*}
	\left( \pmb A^\top \pmb C(\pmb\theta,\pmb y)  \pmb A + \pmb Q_{\pmb\theta} \right)|_{\pmb f = \pmb f_0 } \ \pmb{f}_0 =  \pmb b(\pmb\theta, \pmb y)|_{\pmb f = \pmb f_0},
\end{equation*}
such that the Gaussian approximation to $p(\pmb f|\pmb y, \pmb\theta)$, $p_N(\pmb f|\pmb y, \pmb\theta)$, at the mode $\pmb{\mu_f} = \pmb f_0$ is then defined as 
\begin{equation}
	\pmb f|\pmb y, \pmb\theta \overset{\cdot}{\sim} \text{N}\left(\pmb{\mu_f}, \pmb {Q_f}
	\right),
	\label{eq:lm_bayes}
\end{equation}
where $\pmb {Q_f }= \pmb A^\top \pmb C(\pmb\theta,\pmb y)  \pmb A + \pmb Q_{\pmb\theta}$ is the posterior precision matrix. From $p_N(\pmb f |\pmb\theta,\pmb y)$ we can derive the Gaussian approximation of the posterior of the linear predictors, $p_N(\eta_i|\pmb\theta, \pmb y)$, as follows:

\begin{equation} \label{eqn2.14}
	\eta_i \mid \pmb{\theta}, \pmb{y} \overset{\cdot}{\sim} \text{N} (\pmb{A}_i\pmb{\mu_f}, \tau_i (\pmb{A}, \pmb{Q_f}) ),
\end{equation}
where $\tau_i (\pmb{A}, \pmb{Q_f})$ is the precision for $\eta_i$ and is calculated as described in Section 3.2. of \citet{van2023new}.

\subsection{Variational Bayes}\label{sec:vb}
A Variational framework poses a question in an optimization-based view rather than a marginalization-based view \citep{blei2017variational}. Due to the computational cost and convergence challenges of various sampling-based frameworks, variational inference became a common framework in Bayesian inference. However, most literature on Variational Bayes focus on different ways of optimizing the objective function such as various gradient-based optimization algorithms. This is not our focus. Instead we use Variational Bayes as the optimization-based formulation of Bayes' rule as presented by \citet{zellner1988optimal}. From a decision-theoretic point of view,  \citet{zellner1988optimal} showed that Bayes' Rule is an efficient information processing rule since the postdata density that preserves all input information without hallucination, is the posterior density as derived from Bayes' rule.\\ \\
Consider data $\pmb{y}$ and unknown parameters $\pmb\phi$. 
As inputs we have $L(\pmb{\phi} \mid \pmb{y})$ which is the likelihood function for the parameters $\pmb{\phi}$ in the parameter space $\pmb{\Phi}$, and the antedata belief (prior belief not depending on the data) about $\pmb\phi$, $p_a (\pmb{\phi} \mid \mathcal{\pmb{I}})$. From these inputs, an information processing rule (IPR) provides as outputs a postdata belief about $\pmb\phi$, $p_b (\pmb{\phi} \mid \pmb{D} = \{\pmb{y}, \mathcal{\pmb{I}}\})$ and evidence $p(\pmb y)$. If the IPR is efficient, then the amount of output information is the same as the amount if input information in an entropy sense, thus the information is conserved and this IPR adheres to the information conservation principle (ICP).\\
\\
Consider the Shannon entropy with regards to the postdata belief, then
\begin{align}
	\Delta
	&=  (\textbf{postdata belief} + \textbf{evidence}) - (\textbf{likelihood} + \textbf{antedata belief}) \notag \\
	&= \int_{\pmb{\phi}} p_b (\pmb{\phi} \mid \pmb{D}) \log p_b (\pmb{\phi} \mid \pmb{D}) d \pmb{\phi} +  \int_{\pmb{\phi}} p_b (\pmb{\phi} \mid \pmb{D}) \log p (\pmb{y}) d \pmb{\phi} \notag \\
	& -\left(\int_{\pmb{\phi}} p_b (\pmb{\phi} \mid \pmb{D}) \log L(\pmb{\phi} \mid \pmb{y}) d \pmb{\phi} +\int_{\pmb{\phi}} p_b (\pmb{\phi} \mid \pmb{D}) \log p_a (\pmb{\phi} \mid \mathcal{I}) d \pmb{\phi} \right)\notag \\
	&= \int_{\pmb{\phi}} p_b (\pmb{\phi} \mid \pmb{D}) \log \left[ \dfrac{p_b (\pmb{\phi} \mid \pmb{D})}{p_a (\pmb{\phi} \mid \mathcal{I})} \times \dfrac{p (\pmb{y} )}{L(\pmb{\phi} \mid \pmb{y})} \right] d\pmb{\phi} \notag \\
	&= \text{KLD} \left( p_b(\pmb{\phi} \mid \pmb{D}) \left\vert \left\vert \dfrac{p_a (\pmb{\phi} ) L(\pmb{y} \mid \pmb{\phi})}{p (\pmb{y} )} \right. \right. \right) \label{eqn2.18}
\end{align}
Under ICP, $\Delta = 0$, which implies a postdata belief as follows,
\begin{equation*}
	p_b(\pmb{\phi} \mid \pmb{D}) = \dfrac{p_a (\pmb{\phi} \mid \mathcal{I}) L(\pmb{y} \mid \pmb{\phi})}{p (\pmb{y})}, 
\end{equation*}
which is exactly the posterior density of $\pmb\phi$ from Bayes' rule. This implies that the posterior density is the postdata belief that minimizes $\Delta$, and is exact for $\Delta = 0$.  Thus, the posterior density $p_b(\pmb{\phi} \mid \pmb{D})$ can be approximated using a family $P$ to minimize $\Delta$, i.e. 
\begin{equation*}
	\tilde{p}_b (\pmb{\phi} \mid \pmb{D}) = \argmin_{p_b(\pmb{\phi} \mid \pmb{D}) \in P} \left \{ E_{p_b(\pmb{\phi} \mid \pmb{D})} [- \log L(\pmb{y} \mid \pmb{\phi})] + \text{KLD} (p_b(\pmb{\phi} \mid \pmb{D}) \vert \vert p_a(\pmb{\phi} \mid \mathcal{I}))\right \} ,
\end{equation*}
which defines a variational form of Bayes' rule. 

\subsection{INLA}\label{sec:INLA}
The INLA methodology was introduced by \citet{rue2009approximate} as an approximate method for posterior inference of LGMs. The resulting posteriors are often almost exact, at a fraction of the computational cost of sampling-based approaches. Based on recent developments \citep{van2023new}, the modern INLA methodology can be summarized for the model described in \eqref{eq:LGM} as follows:

\begin{enumerate}
	\item Stage 1: The joint posterior of the hyperparameters is approximated using the full joint density and a Gaussian approximation for the conditional posterior of the latent field as described in Section \ref{sec2.1.2}.
	\begin{equation*}
		\tilde{p}(\pmb \theta|\pmb y ) \propto \left. \frac{p(\pmb y,\pmb f, \pmb \theta)}{p_N(\pmb f| \pmb y, \pmb\theta)} \right|_{\pmb f = \pmb \mu_f} 
	\end{equation*}
	
	\item Stage 2: Based on the joint posterior approximation in Stage 1, $\tilde{p}(\pmb\theta|\pmb y)$ is explored and integration points and corresponding integration weights are calculated. Subsequently, the marginal posteriors of the hyperparameters are approximated using numerical integration.
	\begin{equation*} \tilde{p}(\theta_j|\pmb y) = \int \tilde{p}(\pmb\theta|\pmb y)d\pmb\theta_{-j} 
	\end{equation*}
	
	\item Stage 3: From Stage 1, we extract the univariate approximate conditional posteriors of the latent field from the (corrected) Gaussian approximation and numerically integrate over $\pmb\theta$, using the integration points and weights from Stage 2, to obtain the marginal posteriors of the latent field. 
	\begin{equation} 
		\tilde{p}(f_l|\pmb y) = \int p_N(f_l|\pmb y,\pmb\theta) \tilde{p}(\pmb\theta|\pmb y)d\pmb\theta
		\label{eq:inla}
	\end{equation}
\end{enumerate}
The univariate conditional posterior of each latent element in Stage 3, is approximated as a Gaussian approximation, trivially extracted from the corrected joint Gaussian approximation (see Section \ref{sec:meancorr} for a brief description) in the denominator of Stage 1 in \eqref{eq:inla}. The uncorrected Gaussian approximation from Laplace's method is not accurate enough for non-Gaussian likelihoods to use this approach (see \citet{rue2009approximate} for details), but with the new corrected Gaussian approximation, as described in \citet{van2021correcting}, we can surpass the nested step of the classic INLA methodology while maintaining the same level of accuracy.

\section{Low-rank correction of the mean and variance of the Laplace method}\label{sec:meanvarcorr}

The Laplace method as described in Section \ref{sec2.1.2} provides a way to obtain an efficient high-dimensional posterior density function, although it provides crude estimates in particular cases. Instead of using the variational framework as another way to find an approximate posterior distribution, we use a variational framework to \textit{correct for inaccuracies} in the approximate method from the Laplace method. Due to the linear system constructed by the Laplace method, we can infer these corrections using a low-rank variational Bayes framework. This idea was proposed by \citet{van2021correcting} and applied to correction of the posterior mean of the latent field. The computational cost of this approach is similar to that of the Laplace method and thus very efficient. \\
We present a short review of this low-rank mean correction of the posterior mean and then we extend this approach to corrections of the posterior marginal variances.\\ \\
From \eqref{eq:lm_bayes}, we obtained $\pmb{f} \mid \pmb{\theta}, \pmb{y} \sim \text{N}(\pmb{\mu_f}, \pmb{Q_f})$ from the Laplace method and this is our departure point.

\subsection{Mean correction (VB-M)}\label{sec:meancorr}

Our objective is to enhance the accuracy of the posterior mean from the Laplace method in \eqref{eq:lm_bayes}.
Thus we want to find the approximation $\pmb{f} \mid \pmb{\theta}, \pmb{y} \overset{.}{\sim} \text{N}(\pmb{\mu_f}^{corr}, \pmb{Q_f})$, where $\pmb{\mu_f}^{corr} = \pmb{\mu_f} + \pmb\psi$.\\

\noindent A Variational framework can be invoked to explicitly solve for $\pmb\psi$. While this is theoretically sound, it becomes computationally burdensome due to the optimization's dimensionality, which aligns with the latent variable $\pmb{f}$ and can range up to millions. Recognizing this challenge, an alternative strategy was adopted in \citet{van2021correcting}. Instead of a full-rank correction, the authors opted for a low-rank implicit solution of the correction, significantly reducing the computational cost without compromising on accuracy.\\

\noindent From \eqref{eq:lm_bayes}, the posterior mean of the latent field is related to the first and second order derivatives of the true posterior as follows:

\begin{equation} \label{eqn3.2}
	\pmb{Q_f}\pmb{\mu_f} = \pmb{b_f}. 
\end{equation} 

\noindent Instead of an explicit full-rank correction to $\pmb{\mu_f}$, a low-rank additive correction $\pmb{\delta}$ is proposed for $\pmb{b_f}$ defined as 

\begin{equation}  \label{eqn3.3}
	\pmb{\delta} = 
	\begin{cases}
		\delta_i, \ i \in \mathcal{C} \\
		0, \ \text{otherwise},
	\end{cases}
\end{equation}

\noindent where $\mathcal{C}$ is the index set of components of $\pmb{b_f}$ to be corrected, such that the linear system of the Laplace method can be reformulated as follows
\begin{equation} \label{eqn3.4}
	\pmb{Q_f} \pmb{\mu_f}^{corr} = \pmb{b_f} + \pmb{\delta}.
\end{equation}

\noindent The corrected posterior mean is then calculated by solving the linear system in \eqref{eqn3.4}.
Note that the estimated correction to $\pmb {b_f}$ is propagated to every element in the posterior mean vector, hence the \textit{low-rank} variational Bayes framework.
For further details, see \citet{van2021correcting}.\\

\noindent Naturally, this approach can be extended to correct elements in the variance-covariance matrix calculated from the Laplace method, and this is presented in the next section.

\subsection{Variance correction (VB-M+V)}\label{sec:varcorr}
Consider the approximate posterior after the mean correction, i.e.
\begin{equation*}
	\pmb{f} \mid \pmb{\theta}, \pmb{y} \overset{.}{\sim} \text{N}(\pmb{\mu_f}^{corr}, \pmb{Q_f}).
\end{equation*}

\noindent Similarly to Section \ref{sec:meancorr}, we formulate a low-rank implicit marginal variance correction of the latent field. Define a vector $ \pmb{\delta} \in \mathbb{R}^{p}$ ($p$ is the dimension of the latent field) where some of the elements are set to be structurally zero (similar to \eqref{eqn3.3}). Then the linear system of the Laplace method can be reformulated as 

\begin{equation*} 
	\left(\pmb{Q_f} + \text{diag}(\pmb\delta) \right)\pmb{\mu_f}^{corr} = \pmb{b_f}^{corr},
\end{equation*}
such that $\pmb{Q_f}^{corr} = \pmb{Q_f} + \text{diag}(\pmb\delta)$, implying the approximate posterior density 
\begin{equation} \label{eqn4.1}
	\pmb{f} \mid \pmb{\theta}, \pmb{y} \overset{.}{\sim} \text{N} \left( \pmb{\mu_f}^{corr}, \pmb{Q_f}^{corr} \right).
\end{equation}

\noindent We employ the variational framework of Section \ref{sec:vb} to estimate $\pmb{\delta}$ as follows,

\begin{eqnarray}
	\widehat{\pmb{\delta}}  &=& \argmin_{\pmb{\delta}}\left\{  E_{\pmb{\eta}} [- \log L(\pmb{y} | \pmb{f}, \pmb{\theta}) ] \right. \notag \\ 
	&&+ \left. \text{KLD}(p_N \left(\pmb{\mu_f}^{corr}, \pmb{Q_f}+ \text{diag}(\pmb{\delta}) \right) \mid \mid p_N (\pmb{0}, \pmb{Q_\theta}) ) \right\} \label{eqn4.3} \notag\\
	& =& \argmin_{\pmb{\delta}} \  \{  E_{\pmb{f}} [- \log L(\pmb{y} | \pmb{f}, \pmb{\theta}) ] + \dfrac{1}{2} \left\{ \Trace \left( \pmb{Q_\theta} \left( \pmb{Q_f} + \text{diag}(\pmb{\delta})\right)^{-1} \right)   \right. \notag \\
	&& \left. +   \log \left( \det \left( \pmb{Q_f}+ \text{diag}(\pmb{\delta}) \right)  \right) \right\} \} \label{eqn4.4}
\end{eqnarray}
\noindent Although the optimization of \eqref{eqn4.4} appears straightforward, practical implementation encounters significant hurdles, particularly when dealing with a latent field of very large dimension, i.e. millions. The challenge arises from the necessity of accessing the inverse of $\pmb{Q_f} + \text{diag}(\pmb{\delta})$ or evaluating the derivatives of both the trace and log-determinant terms. Overcoming this obstacle requires the utilization of partial inverse techniques for the precision matrix. Furthermore, the application of partial inverse methods significantly enhances computational efficiency, thereby accelerating the optimization process.\\ 

Routine calculation of the approximate posterior densities of the linear predictors using \eqref{eqn2.14} and the corrected mean and precision matrix, allows efficient predictions and subsequent model validation techniques.

\subsection{Illustrative examples} 
The applicability of the variational Bayes marginal variance correction extends to scenarios where a Gaussian approximation of the posterior distribution is utilized (or in general, for any Gaussian approximation based on the Laplace method). 

\subsubsection{Low-count Poisson process}
Consider the illustrative example presented in \citet{ferkingstad2015improving}. Here, we simulate observations $y_i \sim \text{Poisson}(\exp(\eta_i))$ for $i = 1,2,\ldots,n$, where $\eta_i$ relates to a parameter $\beta$ as follows,

\begin{equation} \label{eqn4.8}
	\eta_i = \beta + u_i,
\end{equation}

\noindent where $u_i \sim \text{N}(0, 1/\sigma^2) $, iid. We select $n = 300$ and $\sigma^2 = 1$ for data generation. We then proceed to compare the marginal posterior distribution of $\beta$, obtained by the variational Bayes marginal mean and variance correction, against the true posterior distribution derived from an extensive MCMC run.

\begin{figure}
	\centering
	\includegraphics[width=1\textwidth]{ 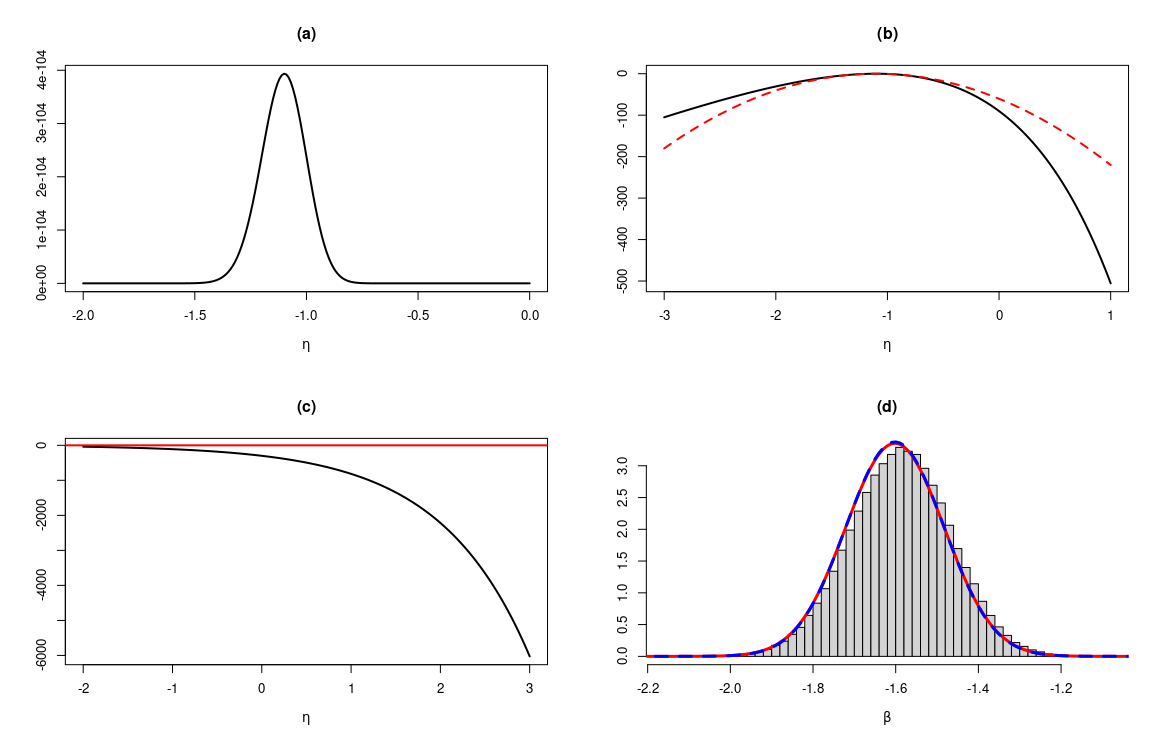}
	\caption{(a) The likelihood plotted against the linear predictor $\eta$. (b) The black curve represents the log-likelihood of the model specified in \eqref{eqn4.8} with respect to the linear predictor $\eta$. The red dashed curve depicts the Gaussian approximation at the mode of the Poisson likelihood. (c) Plot illustrating the second derivative of the log-likelihood with respect to $\eta$. (d) Plot of the posterior distribution of $\beta$ in \eqref{eqn4.8}. The histogram denotes the result from a long MCMC run, while the solid red curve represents the outcome from VB mean correction, and the dashed blue curve portrays the result from VB mean and marginal variance correction.}
	\label{image4.1}
\end{figure}

\noindent Conducting the MCMC procedure involves employing JAGS with four parallel chains, each generating 50,000 samples with a burn-in period of 10,000 samples. Figure \ref{image4.1} provides a comprehensive summary of the comparison results, showcasing the effectiveness and implications of the variational Bayes marginal variance correction in practical Bayesian inference scenarios.

\noindent It is evident from the Figure \ref{image4.1} that the variational Bayes marginal variance correction does not significantly enhance the results beyond mean correction alone. This observation prompts the question: why does this occur? Upon examining the second derivative of the log-likelihood with respect to the linear predictor, we observe that it asymptotically approaches zero and remains consistently negative for all values of $\eta$. The negativity of the second derivative indicates that the likelihood is consistently concave. Referring to \eqref{eq:jointbayes}, we can notice that the prior distribution is already Gaussian, implying that the non-Gaussianity arises from the likelihood component. Considering the second derivative of the log-likelihood for a Gaussian distribution with respect to the linear predictor, we would obtain a constant negative line, akin to the behavior depicted in Figure \ref{image4.1}(c) of the Poisson case, however, the second derivative of the log-likelihood approaches zero asymptotically, indicating that the left tail of the likelihood is heavy and does not align with Gaussian distribution properties. Conversely, the right tail of the log-likelihood matches well and is even thinner than the Gaussian approximation, albeit being nudged towards Gaussianity by the Gaussian prior. Consequently, the variational Bayes variance correction yields only marginal improvement for the Poisson case.\\ \\
So if the variance correction is not necessary in this case, in which cases is it necessary? In the next few sections we present more heavy tailed likelihood models where the necessity of the variance correction framework is clear. 

\subsubsection{Student's t-likelihood}

\begin{figure}
	\centering
	\includegraphics[width=1\textwidth]{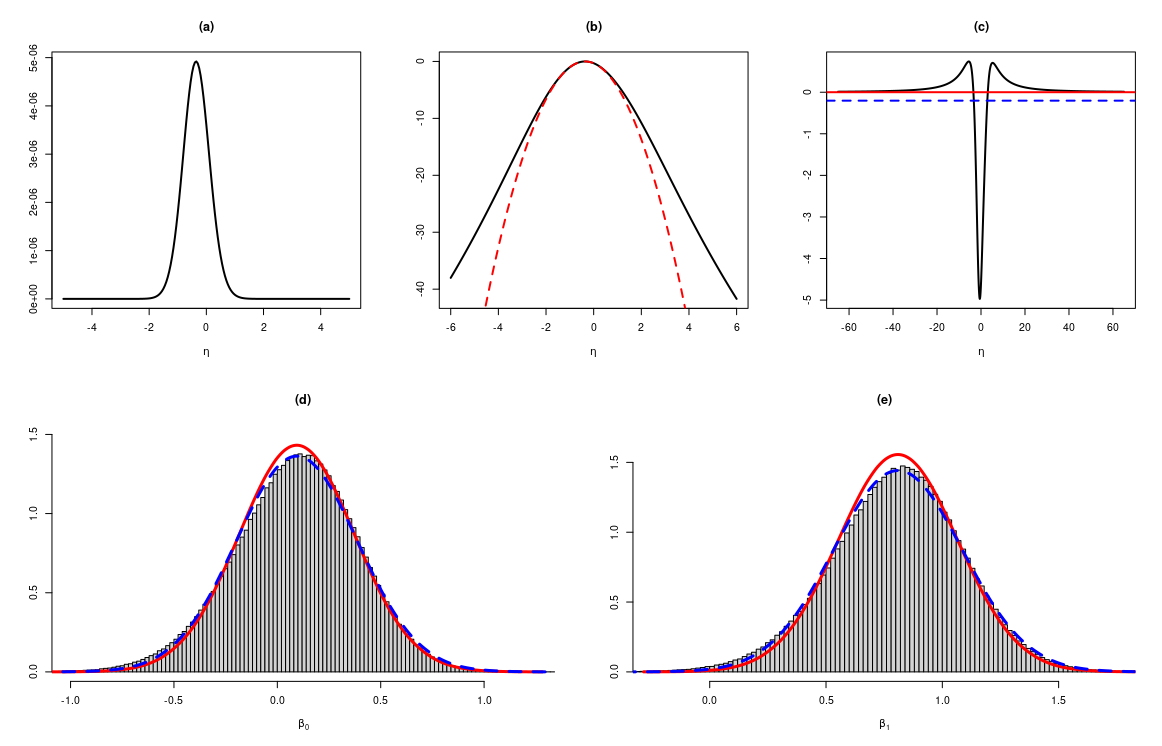}
	\caption{(a) The likelihood plotted against the linear predictor $\eta$. (b)The black curve represents the log-likelihood plot of the model outlined in \eqref{eqn4.10} in relation to the linear predictor $\eta$. The red dashed curve depicts the Gaussian approximation centered at the mode of the Student's t-likelihood. (c) Plot showing the second derivative of the log-likelihood with respect to $\eta$, along with the blue dashed curve representing the second derivative of the linear predictor of the Gaussian approximation. (d) and (e) Display of the marginal posterior distribution of $\beta_0$ and $\beta_1$ in \eqref{eqn4.10}. The histogram reflects the result obtained from long MCMC run, while the solid red curve illustrates the outcome from VB-M. The dashed blue curve portrays the result from VB-M+V. }
	\label{image4.2}
\end{figure}

We generate a dataset consisting of $n = 10$ independent and identically distributed (iid) samples, denoted by $y_i$, drawn from a standard Student's t-distribution with $4$ degrees of freedom. Each $y_i$ is generated according to:

\begin{equation} \label{eqn4.10}
	\sqrt{\tau}(y_i - \eta_i) \sim T_4,
\end{equation}

\noindent where $\eta_i = \beta_0 + \beta_1 x_i$. Here, $\tau$ is the precision parameter. Setting $\beta_0 = 0$ and $\beta_1 = 1$ for data generation, we sample $x_i$ from a Gaussian distribution. We then compare the results from Variational Bayes mean correction (VB-M) and Variational Bayes mean and marginal variance correction (VB-M+V) to the true marginal posterior distribution, determined via MCMC sampling using JAGS. The MCMC setup involves running $12$ parallel chains, each with $100,000$ samples and a burn-in of $10,000$. The summary of these results is illustrated in Figure \ref{image4.2}.

\noindent In Figure \ref{image4.2}(d) and (e), it is evident that the marginal variance correction accurately captures the true variance. However, in Figure \ref{image4.2}(b), we observe that the Gaussian approximation at the mode of the likelihood fails to accurately capture the tail of the distribution, as the t-distribution exhibits fat tails. This observation is confirmed by Figure \ref{image4.2}(c), where we observe positive values of the second derivative of the log-likelihood for certain values of the linear predictor $\eta$, indicating non-convexity in the likelihood function in contrast to the thoroughly concave nature of the log-likelihood for a Gaussian distribution.  The second derivative of the log-likelihood for the Gaussian approximation is depicted by the blue dashed line. Given the Gaussian prior, the likelihood is coerced towards a more Gaussian form, and following marginal variance correction, it effectively captures the variance of the data.

\noindent Furthermore, Table \ref{table4.1} provides numerical values for the standard deviation and the computational time required for VB-M, VB-M+V, and MCMC methods.

\begin{table}[h!]
	\centering
	\begin{tabular}{||c | c c c ||} 
		\hline
		& VB-M & VB-M+V & MCMC \\ [0.5ex] 
		\hline\hline
		$\beta_0$ & 0.2786 & 0.2927 &  0.2977 \\ 
		\hline
		$\beta_1$ & 0.2564 & 0.2767 & 0.2825   \\
		\hline
		Time (s) & 0.23 & 0.24 & 6.45 \\
		\hline 
	\end{tabular}
	\vspace{0.25cm}
	\caption{Posterior standard deviation of the Student-t likelihood from VB-M, VB-M+V, MCMC}
	\label{table4.1}
\end{table}

\subsubsection{Generalised Pareto Likelihood}

We generate $n = 10$ iid samples $y_i$, from the generalised Pareto distribution defined by:

\begin{equation} \label{eqn4.11}
	F(y_i; \sigma, \xi) = 1 - \left( 1 + \xi \dfrac{y_i}{\sigma} \right)^{-1/\xi}, \ y_i > 0
\end{equation}

\noindent where $\xi$ is the tail parameter and $\sigma$ is the scale parameter. The $\alpha$-th quantile of the generalised Pareto distribution is governed by the linear predictor $\eta$, such that $P(y_i \leq \alpha)  = \alpha$ and $q_{\alpha} = \exp(\eta_i) $. The scale parameter $\sigma_i$ is a function of $(q_{\alpha}, \xi)$, given by:

\begin{equation} \label{eqn4.12}
	\sigma_i = \dfrac{\xi \exp(\eta_i)}{(1 - \alpha)^{-\xi} - 1} \ , \ \text{where} \ \eta_i = \beta_0 + \beta_1 x_i
\end{equation}

\begin{figure}[h!]
	\centering
	\includegraphics[width=1\textwidth]{ 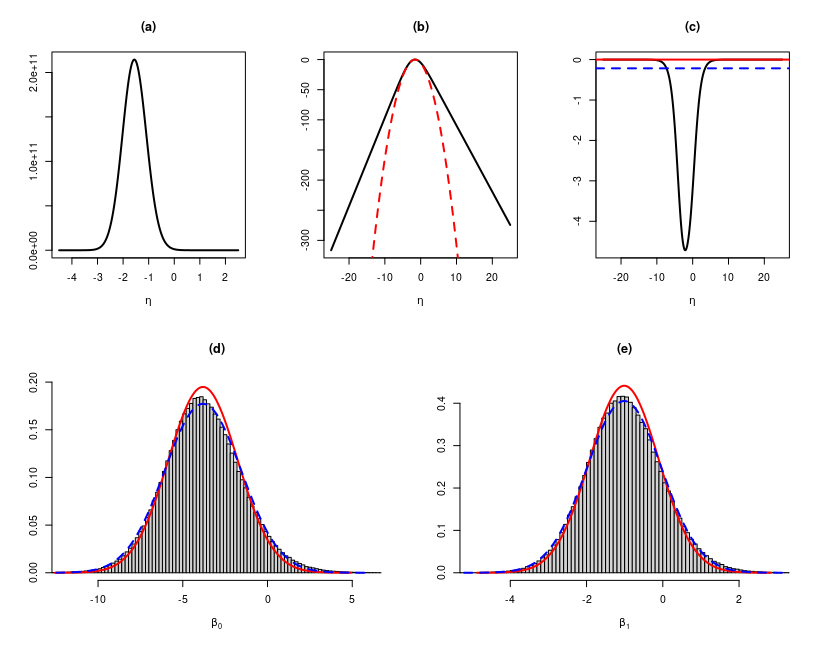}
	\caption{(a) The  likelihood is plotted against the linear predictor $\eta$. (b) The black curve represents the log-likelihood of the model specified in \eqref{eqn4.11} with respect to the linear predictor $\eta$. The red dashed curve depicts the Gaussian approximation at the mode of the generalised pareto likelihood. (c) The plot illustrates the second derivative of the log-likelihood with respect to the linear predictor $\eta$, with blue dashed curve representing the second derivative of the log-likelihood of the linear predictor of the Gaussian approximation. (d) and (e) The marginal posterior distribution of $\beta_0$ and $\beta_1$ in \ref{eqn4.12} is presented. The histogram represents the result from the long MCMC run, while the solid red curve depicts the outcome from the VB-M, and the dashed blue curve portrays the result from VB-M+V.}
	\label{image4.3}
\end{figure}

\noindent We set $\beta_0 = 1$ and $\beta_1 = 1$ to generate the data. To evaluate the performance of variational Bayes mean correction(VB-M) against the true posterior, we conduct a long MCMC run using RCpp \citep{Rcpp}, comprising $10$ million samples with a burn-in period of $10,000$ samples. The summary of the results is presented in Figure \ref{image4.3}, and the numerical values of the standard deviation and computation times for VB-M, VB-M+V and MCMC are provided in Table \ref{table4.2}.

\begin{table}[h!]
	\centering
	\begin{tabular}{||c | c c c ||} 
		\hline
		& VB-M & VB-M+V & MCMC \\ [0.5ex] 
		\hline\hline
		$\beta_0$ & 2.0470 & 2.2516 &  2.2821 \\ 
		\hline
		$\beta_1$ & 0.9040 & 0.9893 & 0.9965   \\
		\hline
		Time (s) & 0.21 & 0.22 & 28.34 \\
		\hline 
	\end{tabular}
	\vspace{0.25cm}
	\caption{Marginal posterior standard deviation of the generalised Pareto likelihood from VB-M, VB-M+V, MCMC}
	\label{table4.2}
\end{table}

\noindent From Figure \ref{image4.3}(d) and (e), the improvements brought about by VB-M+V are clearly evident. Particularly noteworthy is the plot in Figure \ref{image4.3}(c), where the second derivative of the log-likelihood with respect to the linear predictor $\eta$ exhibits two asymptotes approaching zero. This implies that neither of the tails of the distribution matches with that of a Gaussian distribution, contrasting with the case of the Poisson distribution where the right tail aligned well. Thus, in this scenario, the need for the marginal variance correction becomes apparent.

\subsubsection{Binomial Likelihood: Sensitivity Specificity Analysis}

\noindent We generate a dataset of size $n = 50$ from a binomial model defined in \eqref{eqn4.15}, as follows:
\begin{align} 
	& y_i \sim \text{Bernoulli}(f(\eta)) \notag \\
	& f(\eta) = \pi_0 \pi(\eta) + (1 - \pi_1) (1 - \pi(\eta))  \notag \\
	& \pi(\eta) = \dfrac{\exp(\eta)}{1 + \exp(\eta)} \label{eqn4.15}
\end{align}

\noindent In \eqref{eqn4.15}, $\pi_0 \in [0, 1]$ represents the sensitivity and $\pi_1 \in [0, 1]$ represents the specificity of a discrete test, with both values assumed to be fixed. We further assume $\eta$ follows a prior distribution of $\text{N}(0, 10^{-6}) $, and we set $\pi_0 = 0.8$ and $\pi_1 = 0.985$ to generate the data.

\begin{figure}[h!]
	\centering
	\includegraphics[width=1\textwidth]{ 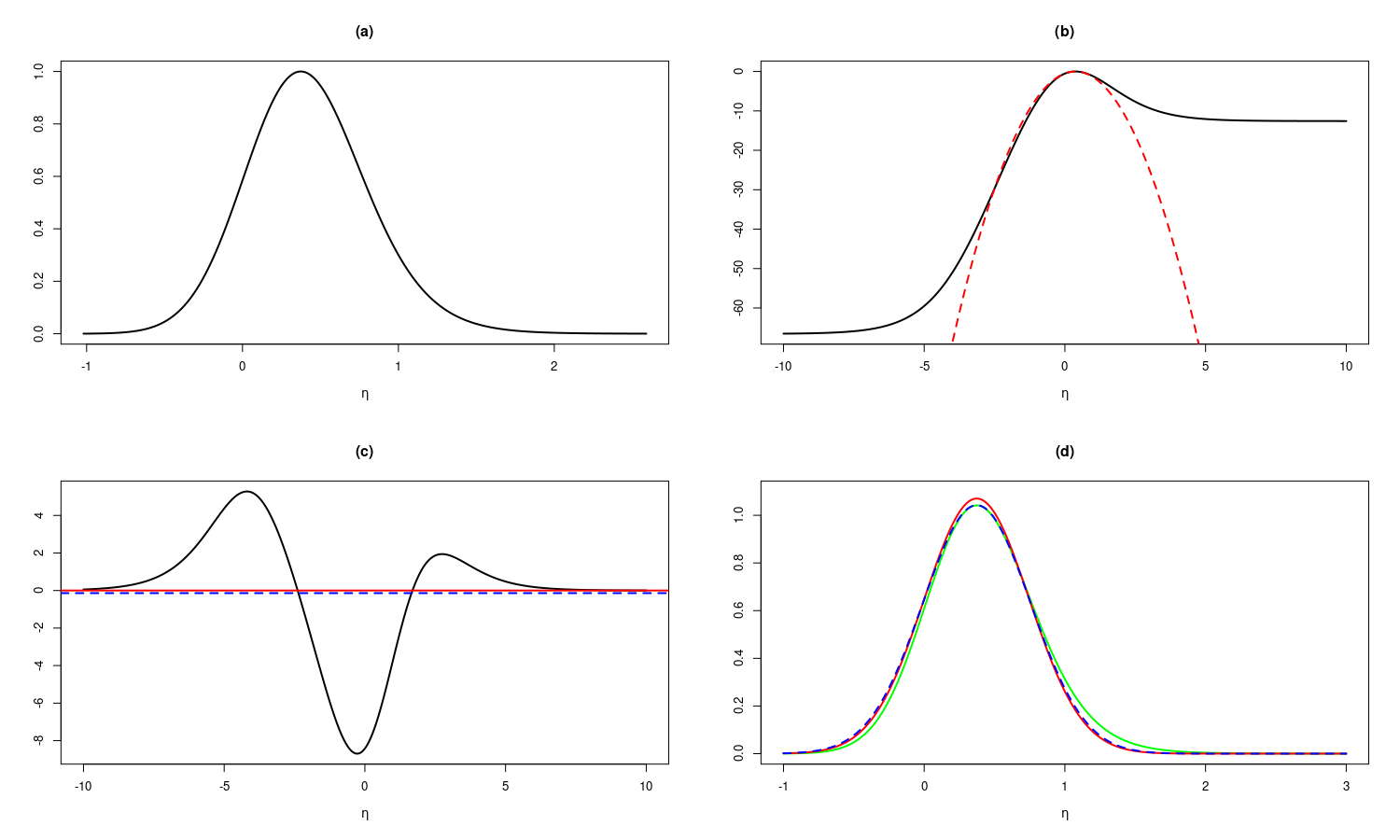}
	\caption{(a) The plot illustrates the likelihood with respect to the linear predictor $\eta$. (b) The black curve represents the log-likelihood of the model specified in \eqref{eqn4.15} with respect to the linear predictor $\eta$. The red dashed curve depicts the Gaussian approximation at the mode of the binomial likelihood. (c) This plot displays the second derivative of the log-likelihood with respect to $\eta$. The blue dashed curve represents the second derivative of the Gaussian approximation with respect to the linear predictor. (d) The posterior distribution of $\eta$ in \eqref{eqn4.15} is presented. The solid green curve denotes the true posterior, while the solid red curve represents the outcome from VB-M, and the dashed blue curve portrays the result from VB-M+V. } 
	\label{image4.4}
\end{figure}

\noindent The resulting posterior standard deviation of $\eta$ is summarized in Table \ref{table4.3} and illustrated in Figure \ref{image4.4}(d).

\begin{table}[h!]
	\centering
	\begin{tabular}{||c | c c c ||} 
		\hline
		& VB-M & VB-M+V & TRUE \\ [0.5ex] 
		\hline\hline
		$\eta$ & 0.1389 & 0.1465 &  0.1604 \\ 
		\hline
	\end{tabular}
	\vspace{0.25cm}
	\caption{Posterior standard deviation of the sensitivity and specificity example from VB-M, VB-M+V and the true value}
	\label{table4.3}
\end{table}

\noindent This example is particularly intriguing due to the characteristic of the log-likelihood and its second derivative, as depicted in Figure \ref{image4.4}(b) and (c). The second derivative exhibits three changes in curvature, reminiscent of the behavior observed in the t-likelihood.  Additionally, it becomes positive, indicating a curvature of the log-likelihood that significantly deviates from that of the Gaussian approximation, especially evident in the tail (as shown in Figure \ref{image4.4}(b)). Thus, variance correction becomes essential to rectify the underestimated variance.

\section{Skewing (selected) Gaussian marginals with Variational Bayes}\label{sec:skewcorr}
In the previous sections a hybrid method combining the Laplace method with Variational inference was proposed as a way to calculate a multivariate Gaussian approximation with unstructured corrected covariance matrix and a corrected mean vector, based on a computationally efficient low-rank optimization. We have shown that this method performs similar to Markov Chain Monte Carlo methods in terms of accuracy, but at the much lower cost of the Laplace method, for LGMs. This method provides a way to perform approximate Bayesian inference for complex models and/or huge data, with sufficient accuracy while maintaining a minimal cost. This cost is further lowered with specialized techniques for sparse precision matrices, as is the case for most latent Gaussian models in general.\\ \\
\noindent It is intriguing to attempt moving beyond the symmetry of the Gaussian, while maintaining an acceptable computational cost, so that the method admits sufficient flexibility, and can scale with model complexity and/or data size. Various non-symmetric multivariate distributions exist, and could be considered as a variational family but due to the necessary abandonment of the usual mean-field assumption for complex models, these multivariate proposals would incur a huge cost. This partly motivates the consideration and preference, even still today, of the multivariate Gaussian in various inferential frameworks such as machine learning algorithms, spatial process modeling and high-dimensional Bayesian inference. \\ \\
In this section we propose a new multivariate distribution that fits the objective of skewing marginal densities for some elements of a multivariate Gaussian vector. Then, we present a variational framework where the skewness of the densities are computed as "errors" from the Laplace method. This method is shown to be efficient and necessary in certain scenario's, especially in the case of unbalanced data. 

\subsection{Skewed multivariate proposal}\label{sec:sgc}
Since we already possess the corrected (marginal) mean and (marginal) variance of the posterior latent field, the primary goal is to retain these corrections while surpassing the inherent symmetry of the multivariate Gaussian distribution.

\noindent To achieve this goal, we propose a joint density that retains the corrected mean and variance of the random vector, but where some of the elements now have skew-normal marginals instead of Gaussian marginals. This density is thus constructed based on Gaussian and skew-normal marginals where the dependence is modeled with the Gaussian copula obtained from the corrected Gaussian approximation in Section \ref{sec:meanvarcorr}. We name this density the \textbf{S}kewed \textbf{G}aussian with Gaussian \textbf{C}opula abbreviated as \textbf{SGC}.\\ \\
Consider the corrected Gaussian approximation of the (conditional) posterior density of $\pmb f, 
\pmb{f} \mid \pmb\theta, \pmb y \overset{\cdot}{\sim} \text{N} \left( \pmb {\mu_f}, \pmb {Q_f}\right)$, 
where $\pmb{\mu_f}$ and $\pmb {Q_f}$ still depend on $\pmb\theta$ and $\pmb y$, and have been corrected using the low-rank variational Bayes frameworks from Section \ref{sec:meanvarcorr}. \\
\noindent Now, we want to change some of the Gaussian marginals to skew-normal, and we form a new random vector $ \tilde{\pmb f}$, where some $\tilde{f_i}$'s are exactly $f_i$, but then for those that are skewed, $\tilde{f_i}=g_i(f_i)$ where $g_i(.)$ is an element-wise mapping from the Gaussian to skew-normal density. The proposed density for $\tilde{ \pmb{f}}$ (assuming skew-normal marginals for all elements, without loss of generality) is 
derived as follows,  
\begin{equation} \label{eqn5.7}
	p_{\text{SGC}} (\tilde{\pmb{f}};\pmb\mu, \pmb Q, \pmb s) = p_N \left( { \pmb g^{-1} (\tilde{\pmb f})} ; \pmb\mu_f, \pmb Q_f\right) \left| \dfrac{\partial \pmb{f}}{\partial \tilde{\pmb{f}}} \right| =  p_N \left( { \pmb g^{-1} (\tilde{\pmb f})} ;\pmb\mu_f, \pmb Q_f\right)  \mathcal{J}(\pmb{f}, \tilde{\pmb{f}}) ,
\end{equation}
where 
\begin{equation*}
	\tilde{\pmb f} = \pmb g(\pmb f) = [g_1(f_1), g_2(f_2), ..., g_p(f_p)],
\end{equation*}
hence 
\begin{equation*}
	\pmb g^{-1}(\tilde{\pmb f}) = [g_1^{-1}(\tilde{f_1}), g_2^{-1}(\tilde{f_2}), ..., g_p^{-1}(\tilde{f_p})],
\end{equation*}
such that $\pmb f$ and $\tilde{\pmb f}$ have the same mean vector and precision matrix. 
Note that the Jacobian is 
\begin{align} \label{eqn5.8}
	\mathcal{J} \left(\pmb{f} , \pmb{\tilde{f}} \right)  
	= \left|
	\begin{bmatrix}
		\dfrac{\partial f_1}{\partial \tilde{f_1}} & 0 & \cdots & 0 \\
		0 & \dfrac{\partial f_2}{\partial \tilde{f}_2} & 0 & \vdots \\
		\vdots & 0 & \ddots & 0 \\
		0 & \cdots & 0 & \dfrac{\partial f_p}{\partial \tilde{f}_p}
	\end{bmatrix}
	\right| =  \prod_{i = 1}^{p} \left| \dfrac{\partial f_i}{\partial \tilde{f_i}} \right|
\end{align}

\begin{figure}[h!]
	
	\begin{tabular}{|c|c|c|c|c|c|c|c|}
		\hline
		&	$s_2 = -0.8$ & $s_2 = -0.5$ &  $s_2 = -0.2$ &  $s_2 = 0$ &  $s_2 = 0.2$ &  $s_2 = 0.5$ &  $s_2 = 0.8$ \\ \hline \hline
		\rotatebox{90}{$s_1 = -0.8$} & \includegraphics[width = 2cm]{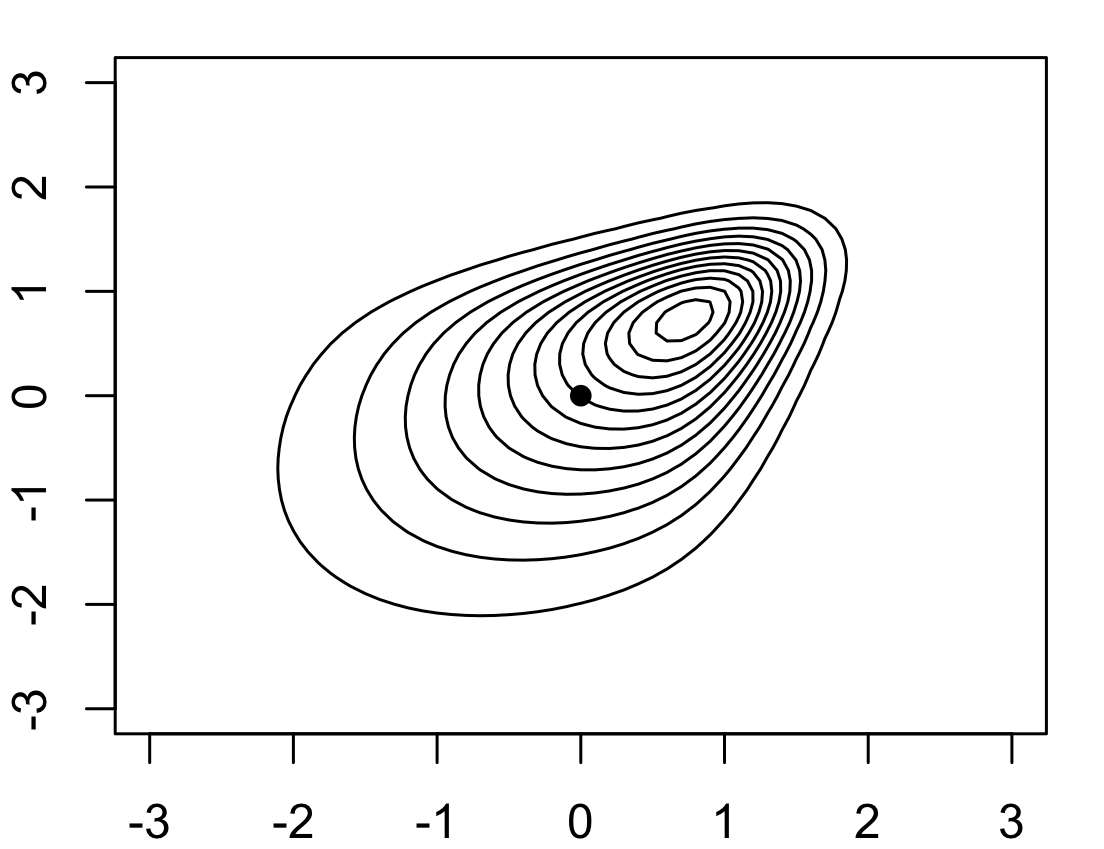} &
		\includegraphics[width = 2cm]{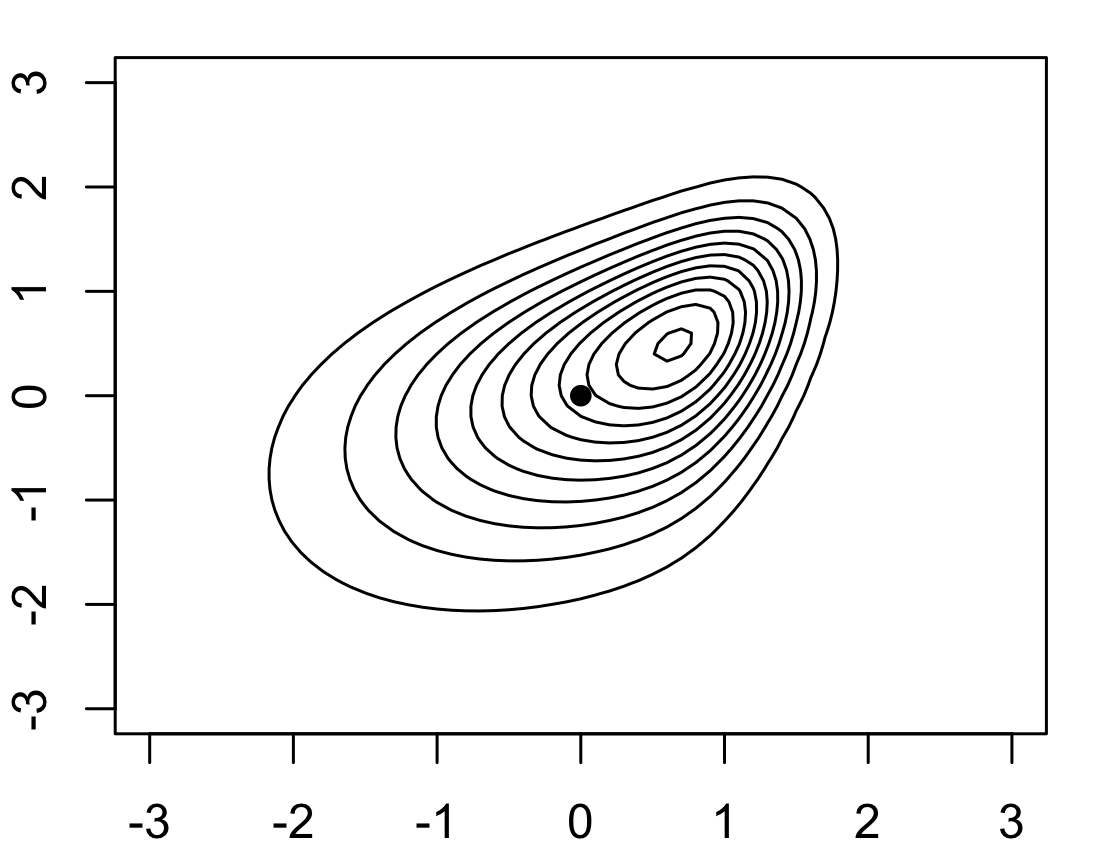} &
		\includegraphics[width = 2cm]{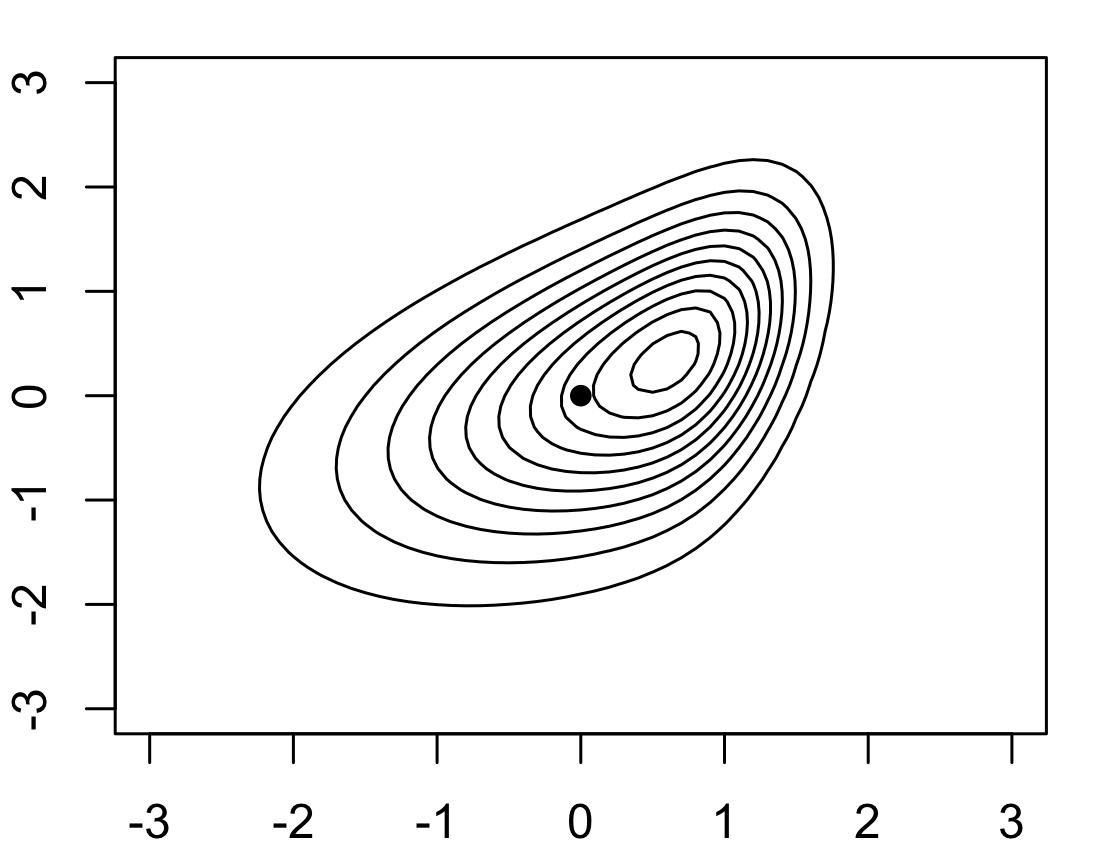} &
		\includegraphics[width = 2cm]{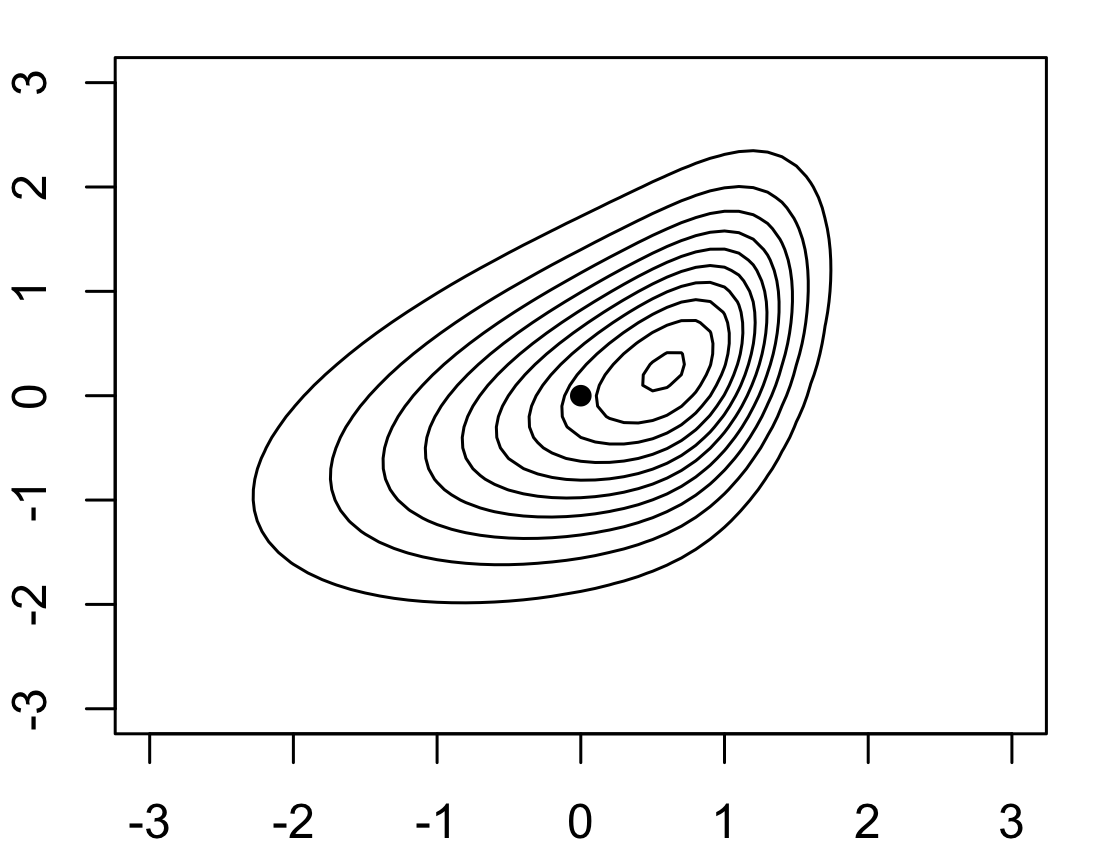} &
		\includegraphics[width = 2cm]{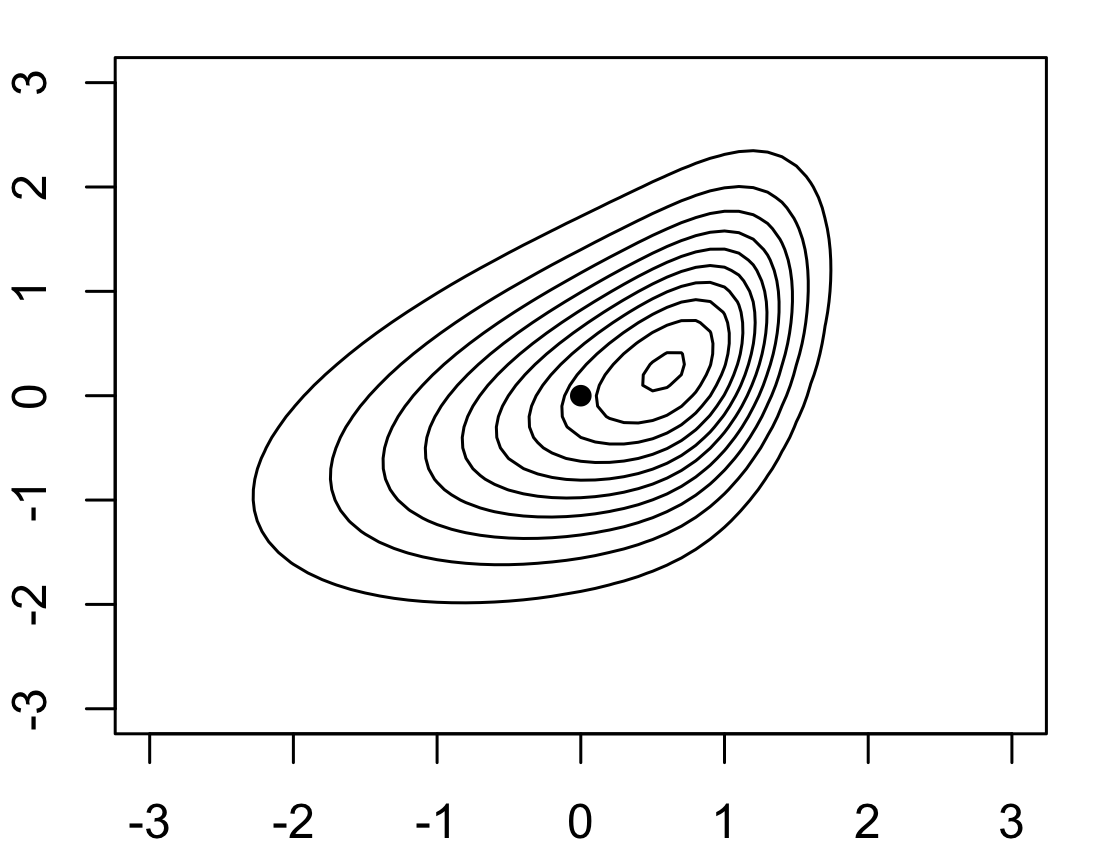} &
		\includegraphics[width = 2cm]{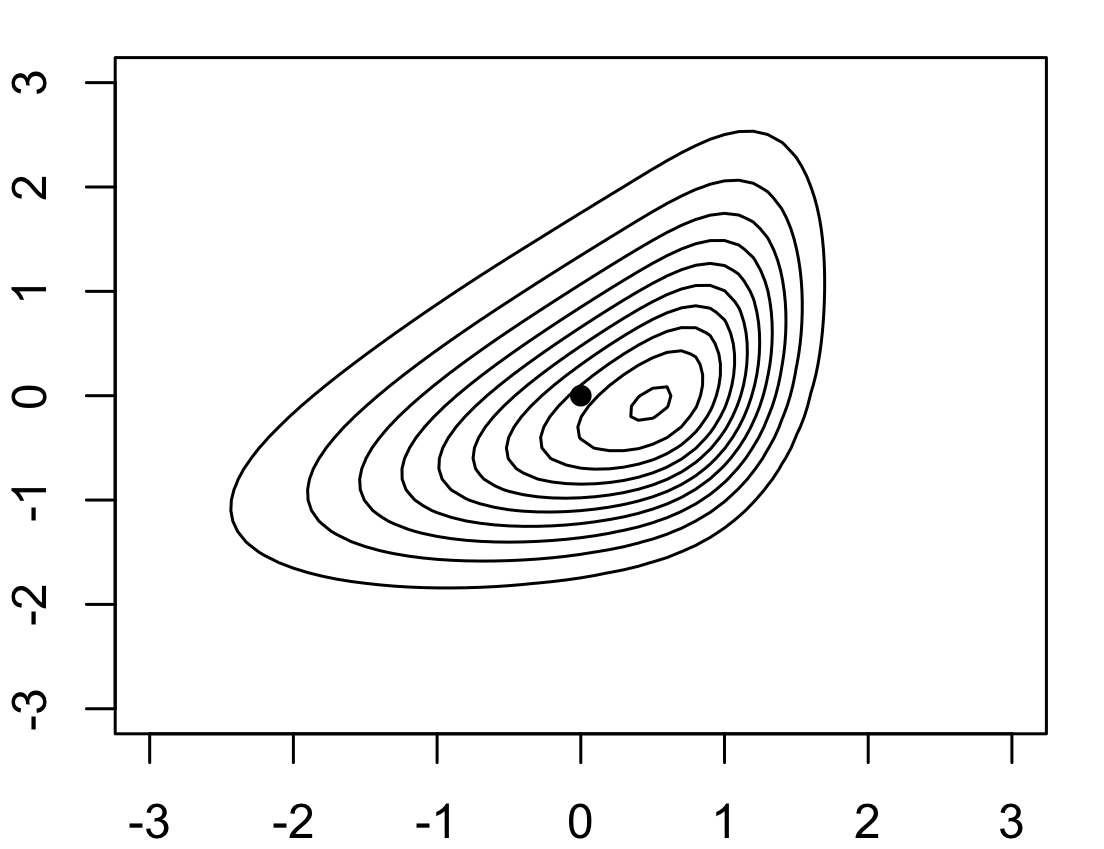} &
		\includegraphics[width = 2cm]{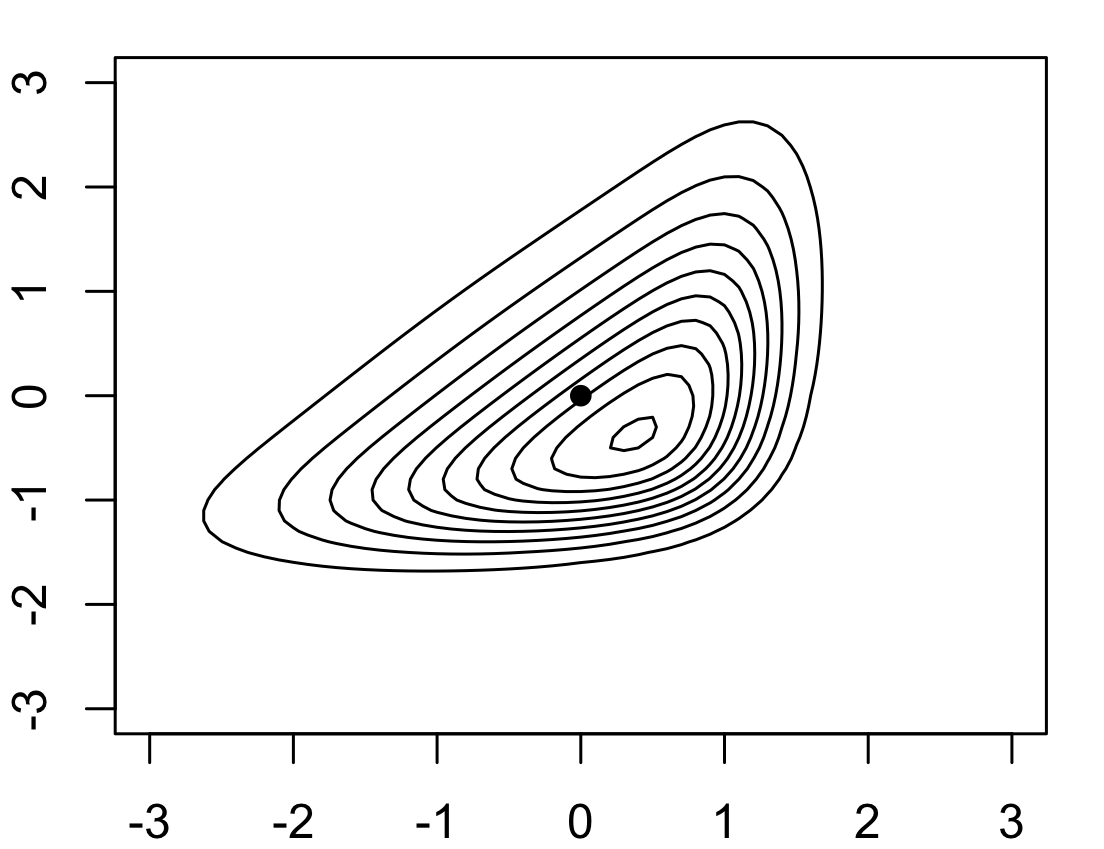}\\ \hline
		\rotatebox{90}{$s_1 = -0.5$} &\includegraphics[width = 2cm]{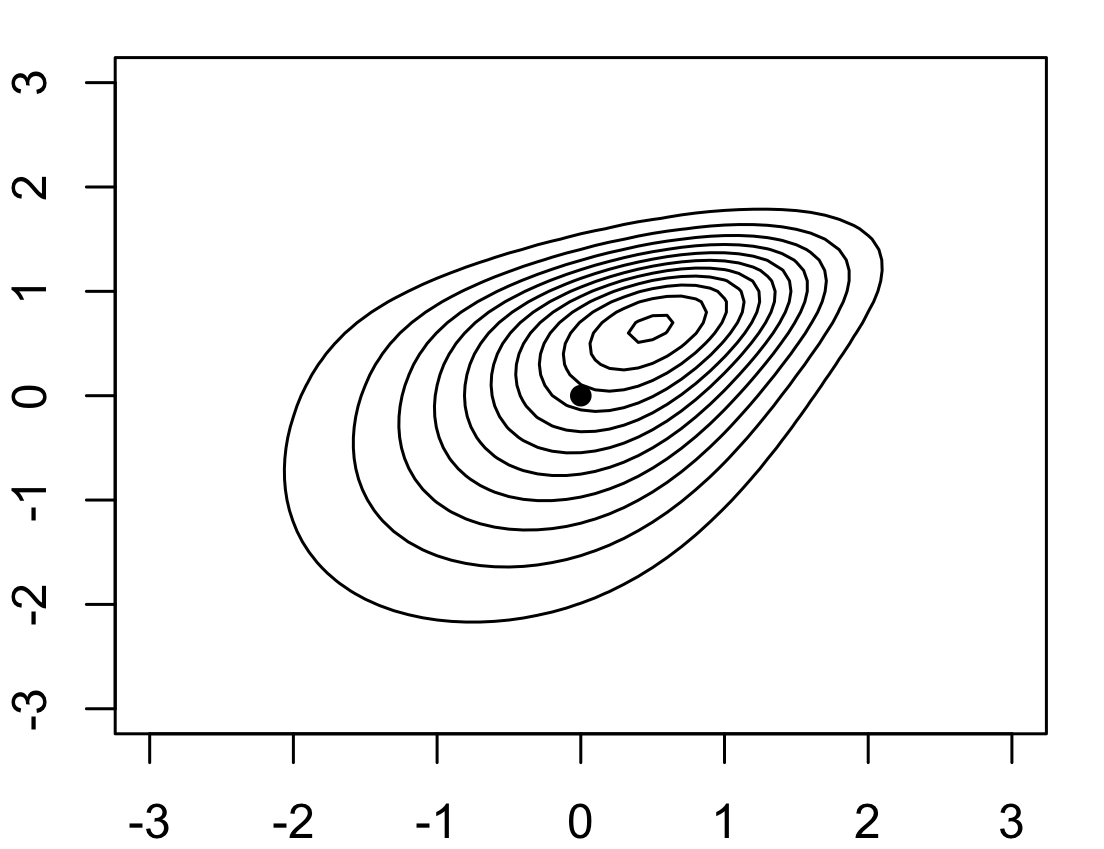} &
		\includegraphics[width = 2cm]{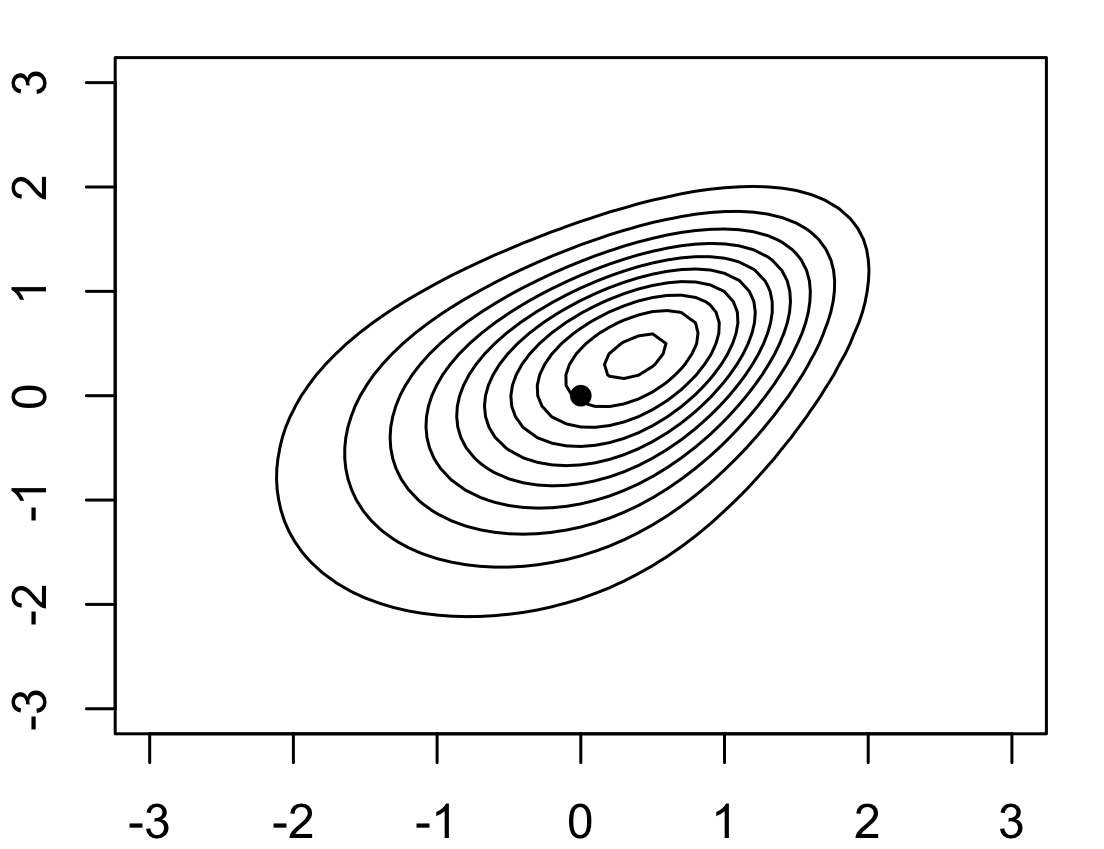} &
		\includegraphics[width = 2cm]{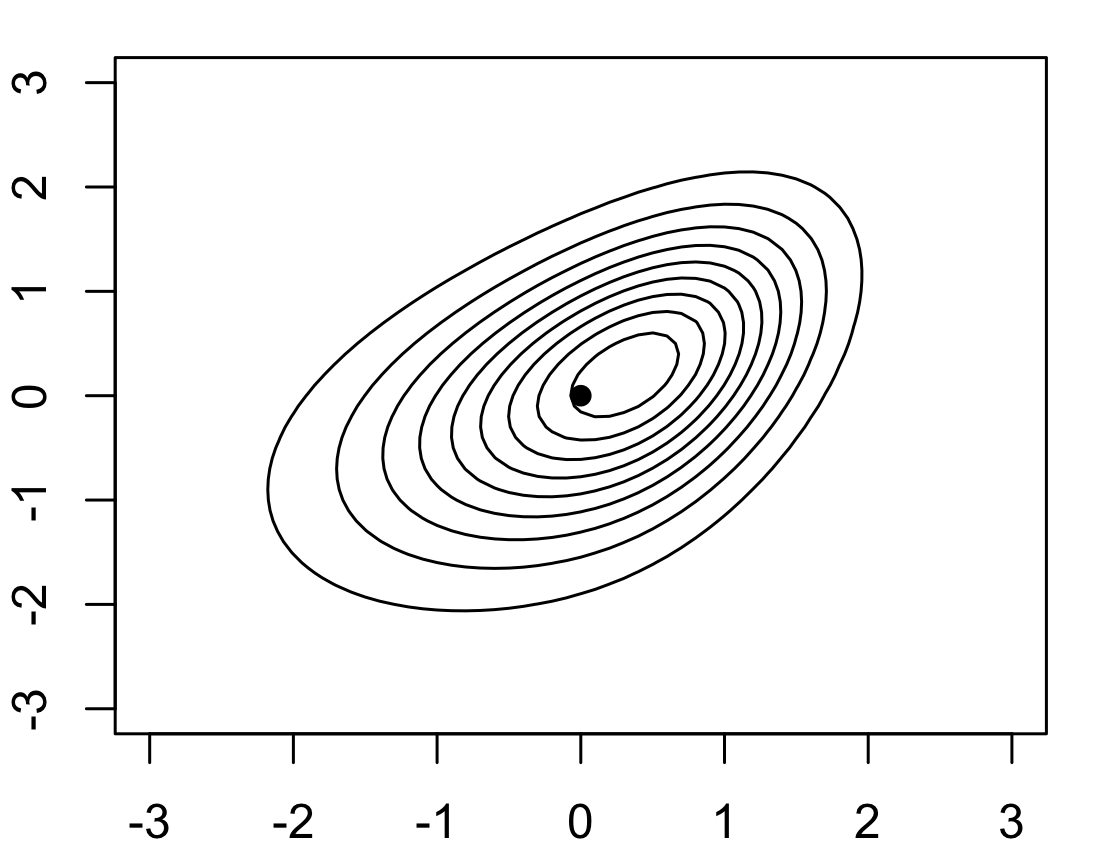} &
		\includegraphics[width = 2cm]{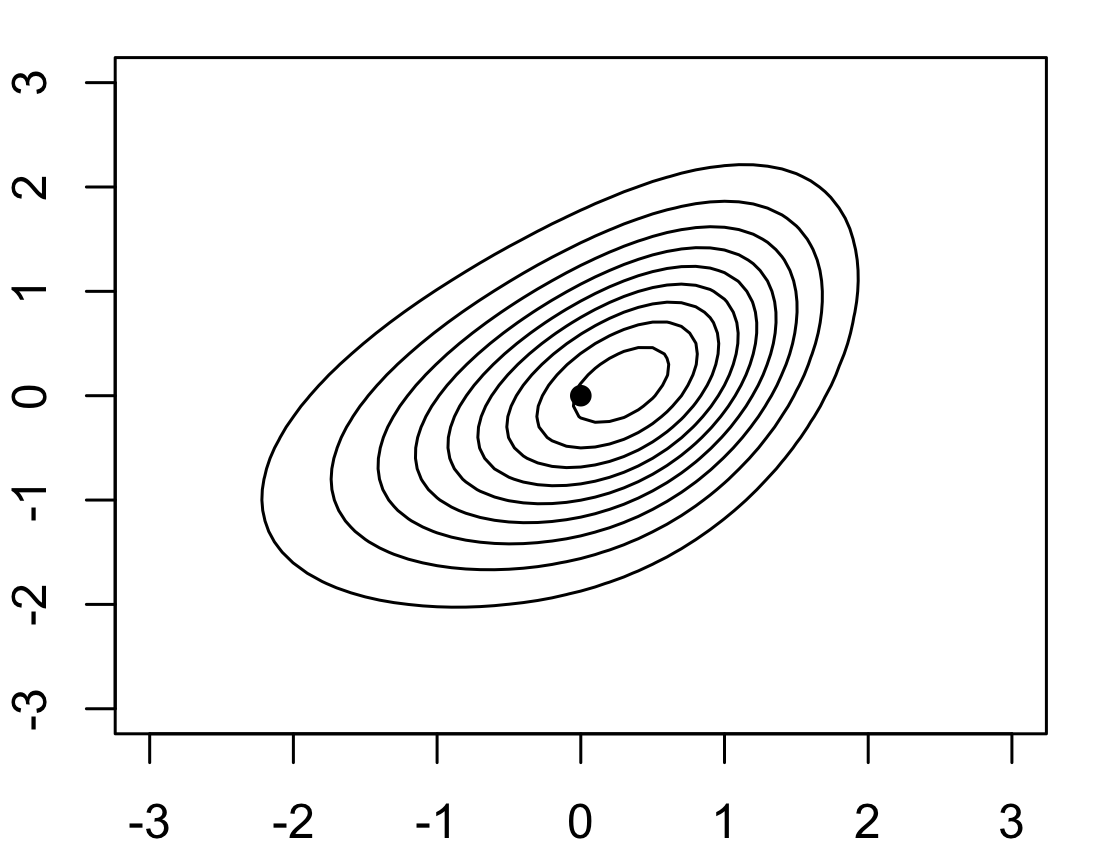} &
		\includegraphics[width = 2cm]{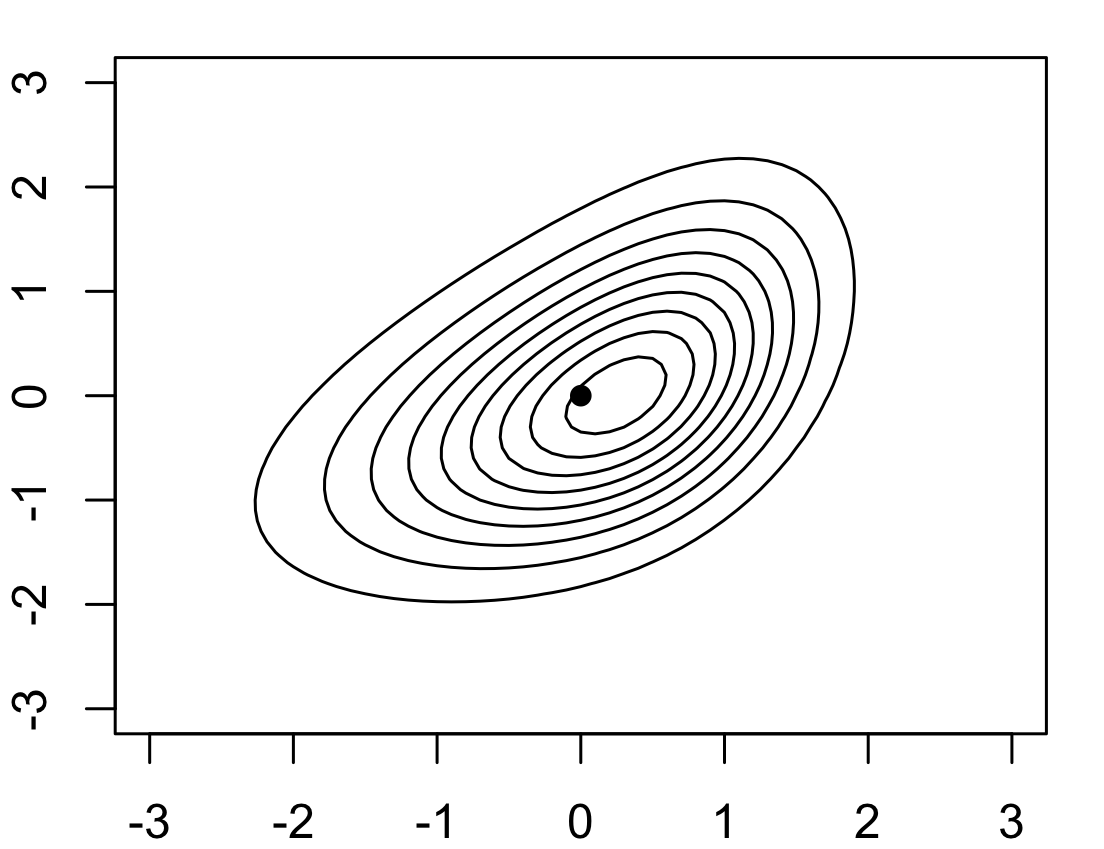} &
		\includegraphics[width = 2cm]{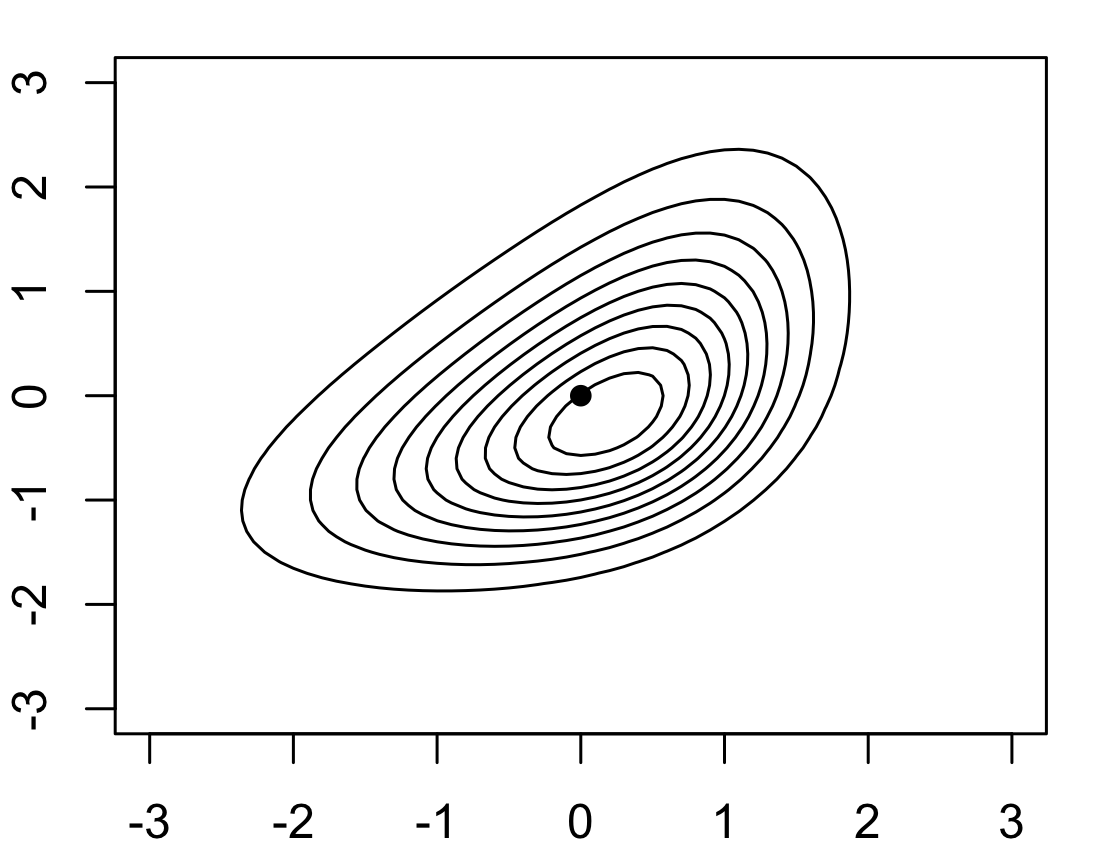} &
		\includegraphics[width = 2cm]{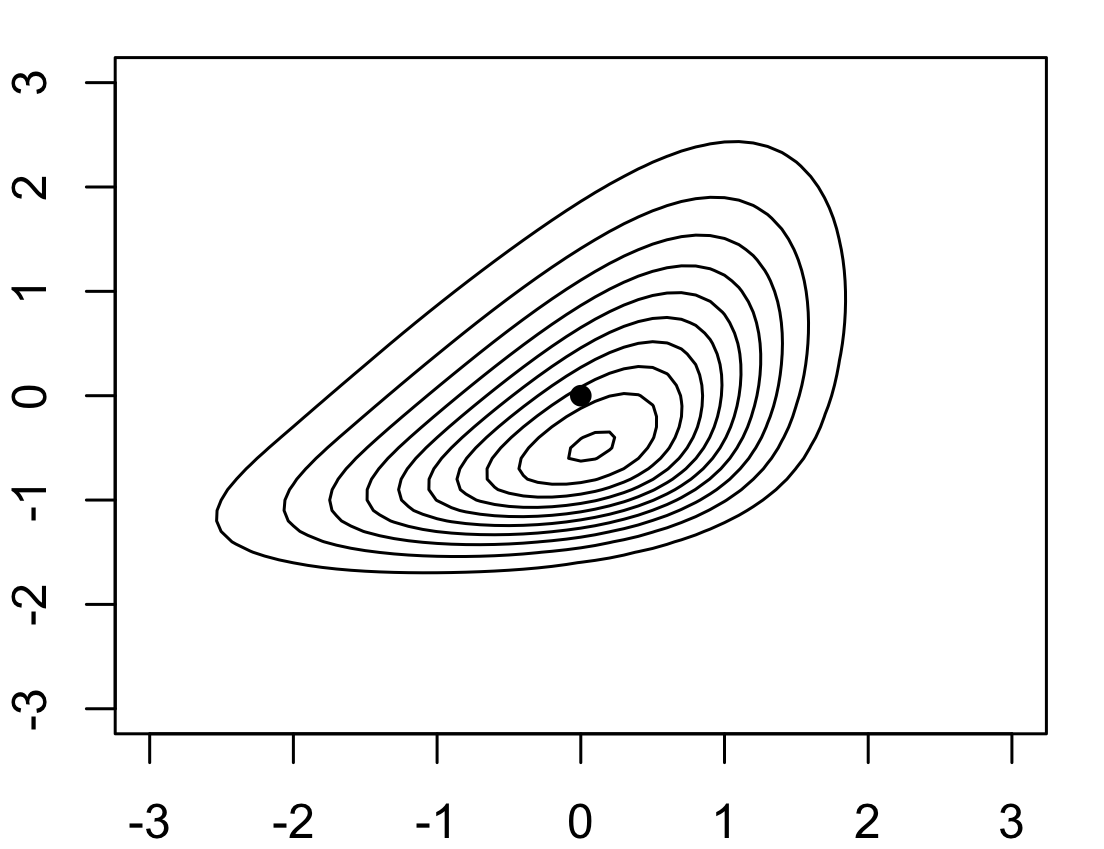}\\ \hline
		\rotatebox{90}{$s_1 = -0.2$} &\includegraphics[width = 2cm]{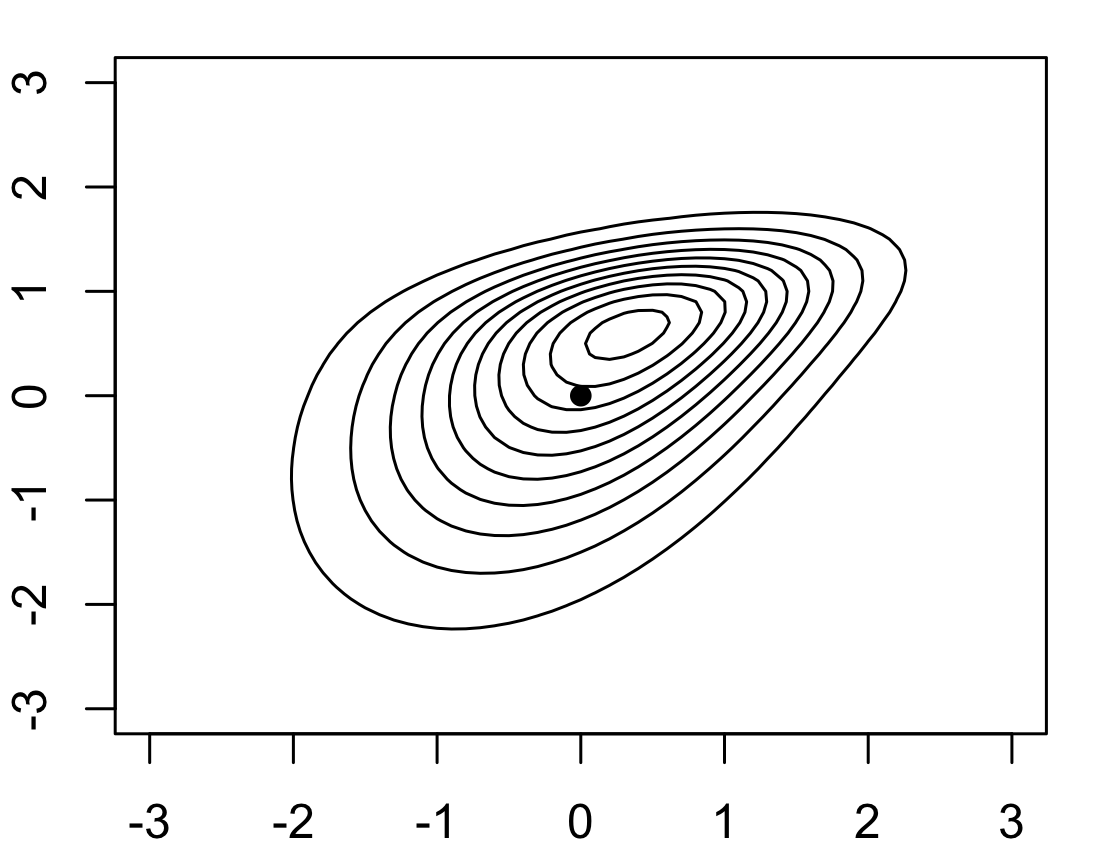} &
		\includegraphics[width = 2cm]{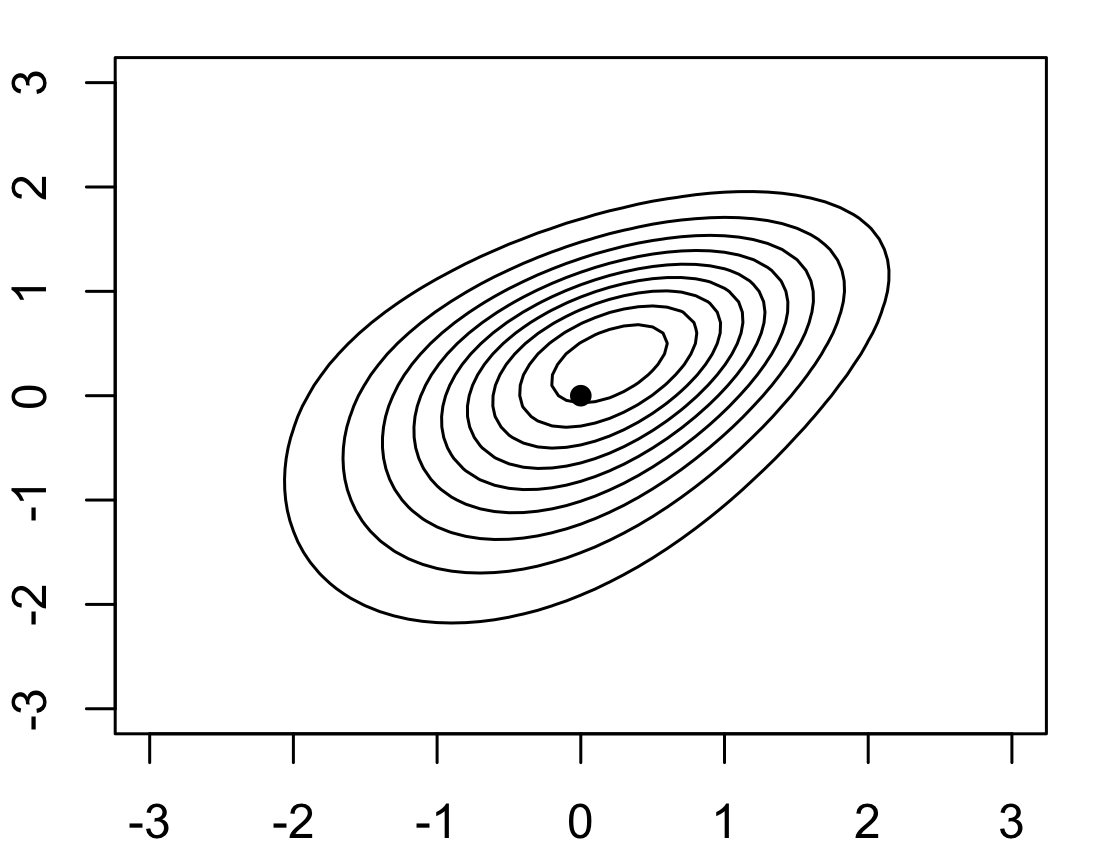} &
		\includegraphics[width = 2cm]{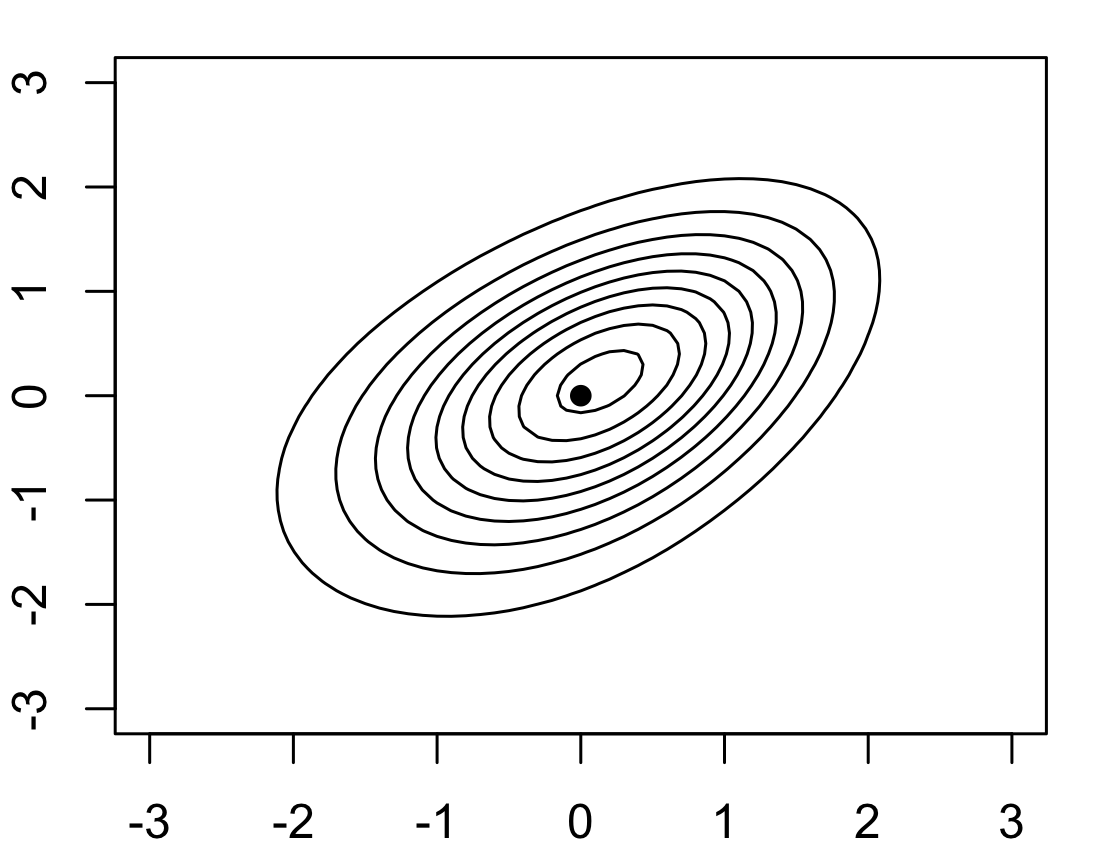} &
		\includegraphics[width = 2cm]{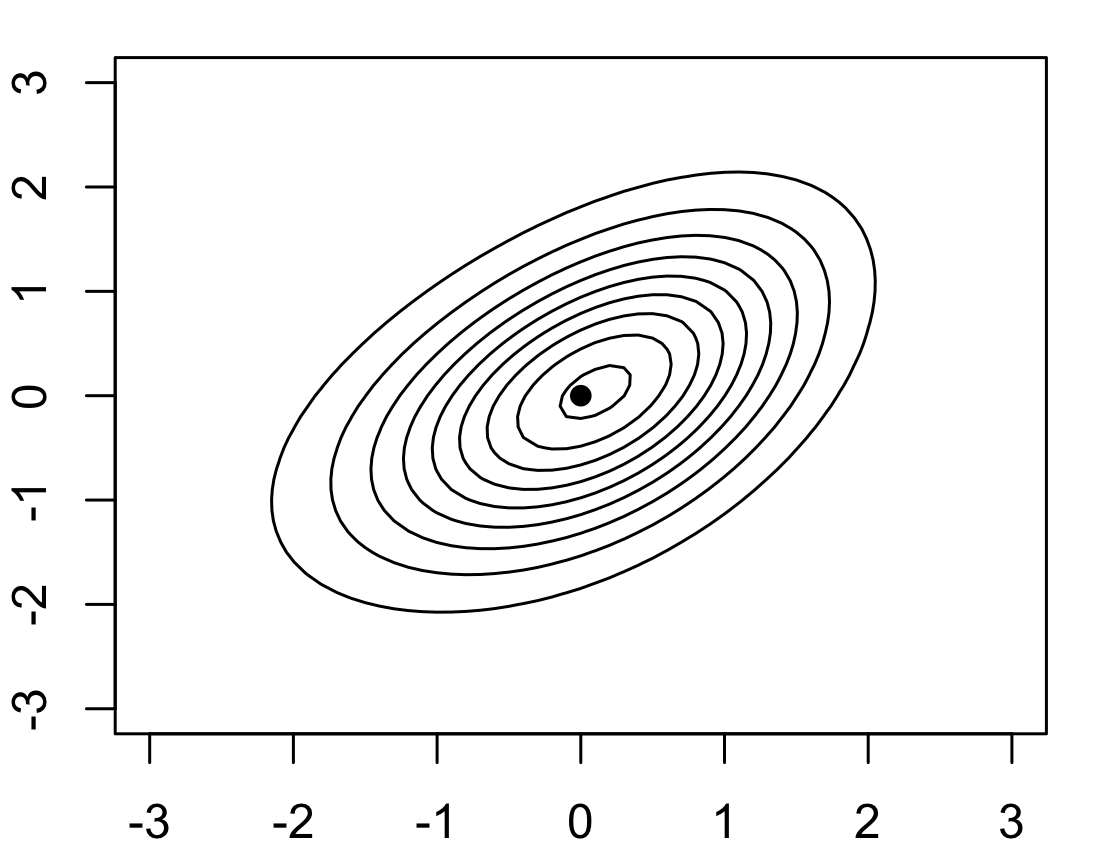} &
		\includegraphics[width = 2cm]{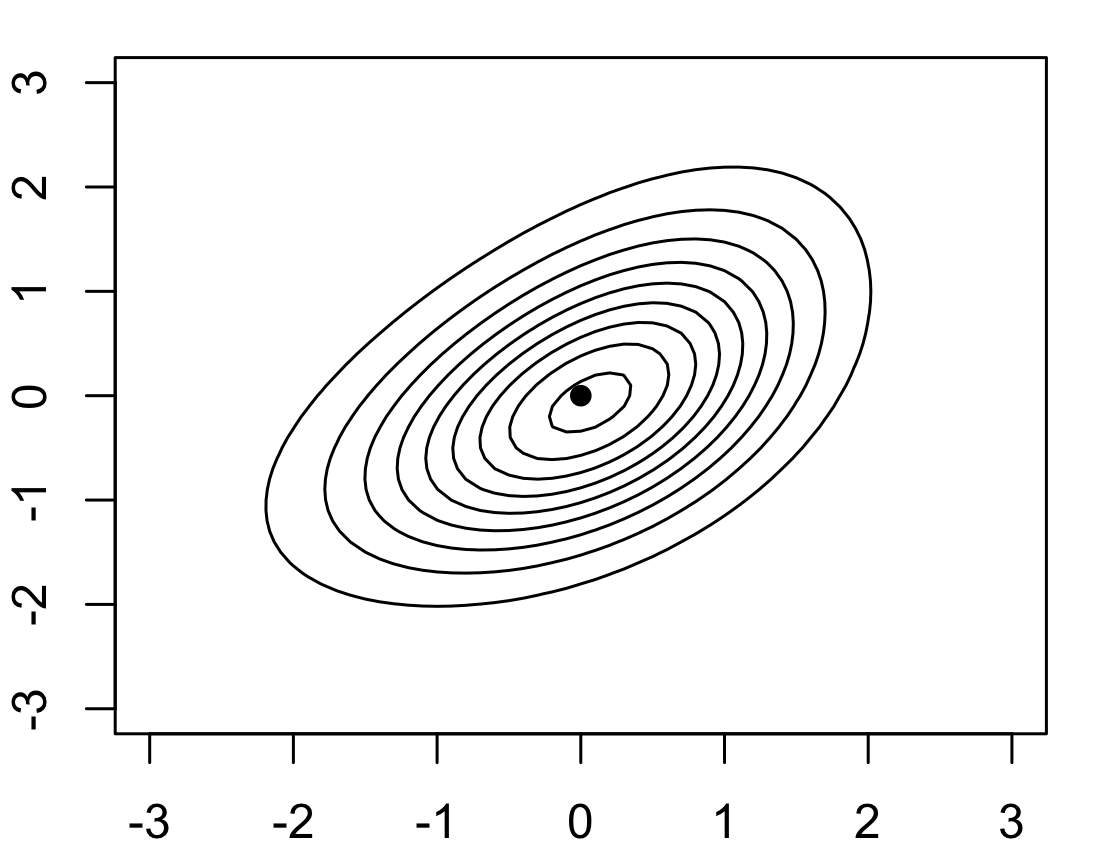} &
		\includegraphics[width = 2cm]{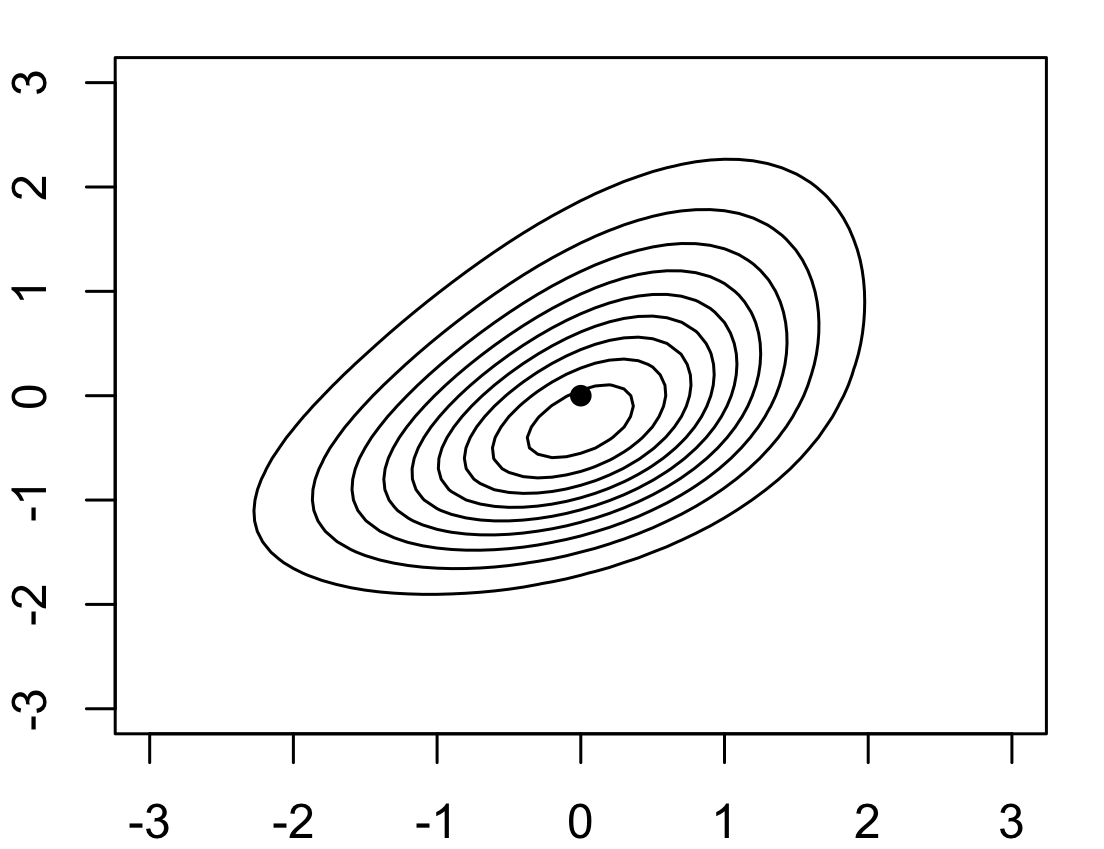} &
		\includegraphics[width = 2cm]{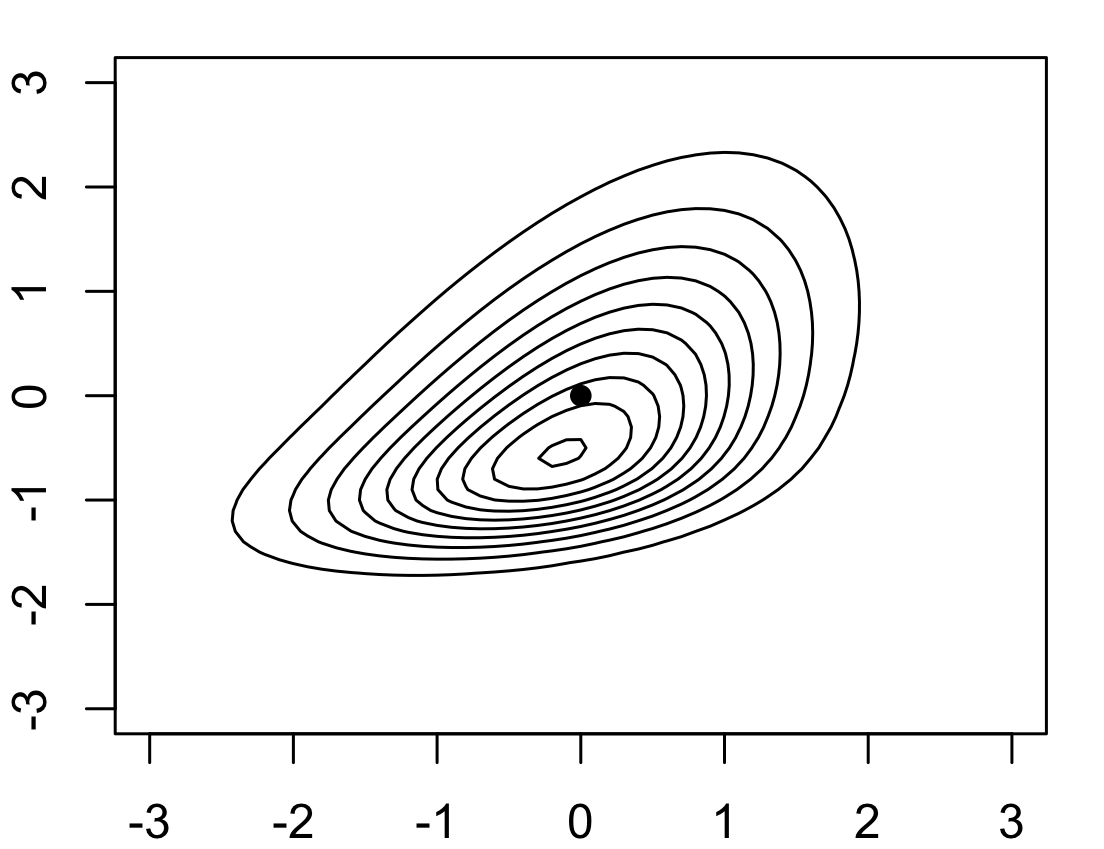}\\ \hline
		\rotatebox{90}{$s_1 = 0$} &\includegraphics[width = 2cm]{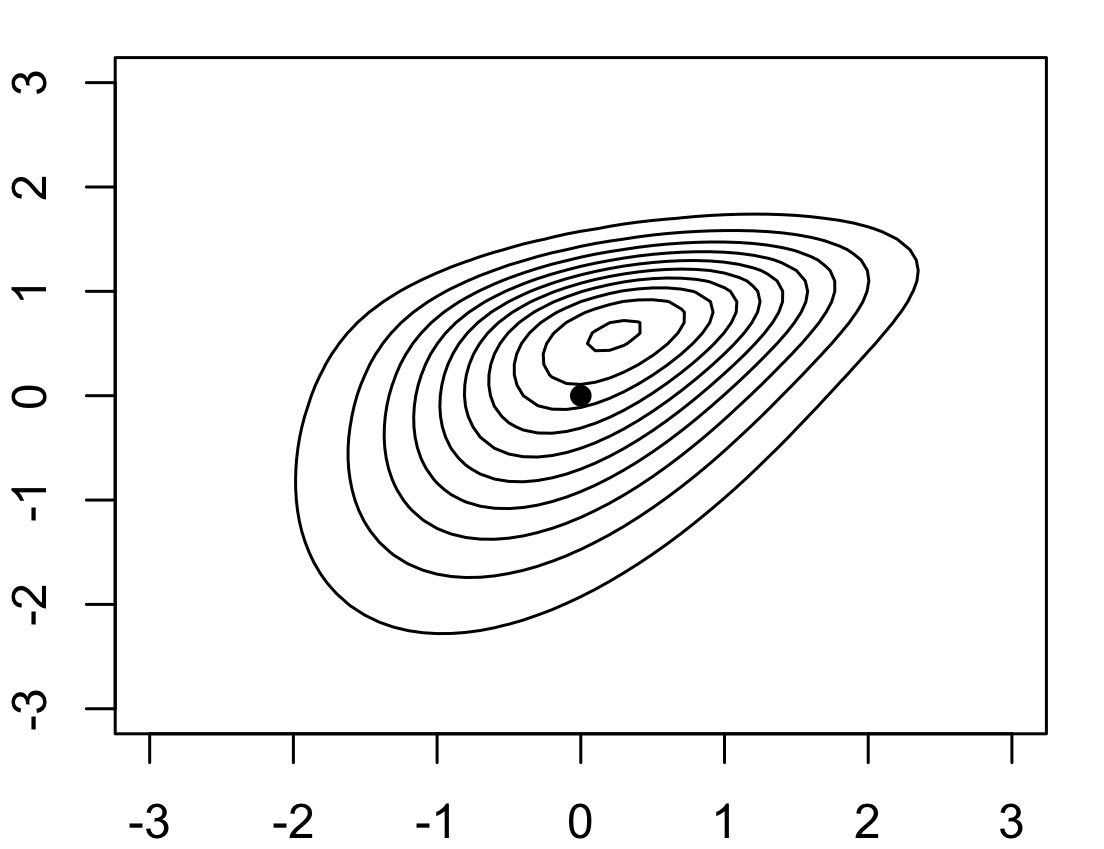} &
		\includegraphics[width = 2cm]{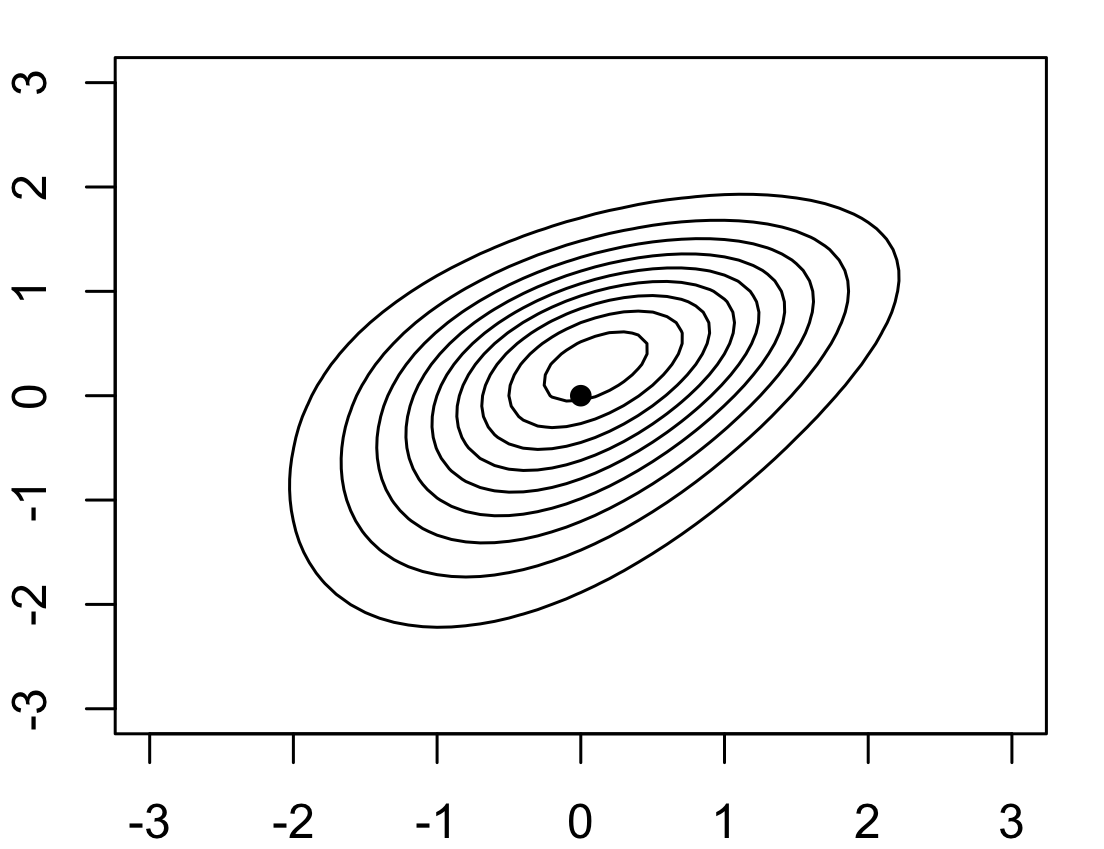} &
		\includegraphics[width = 2cm]{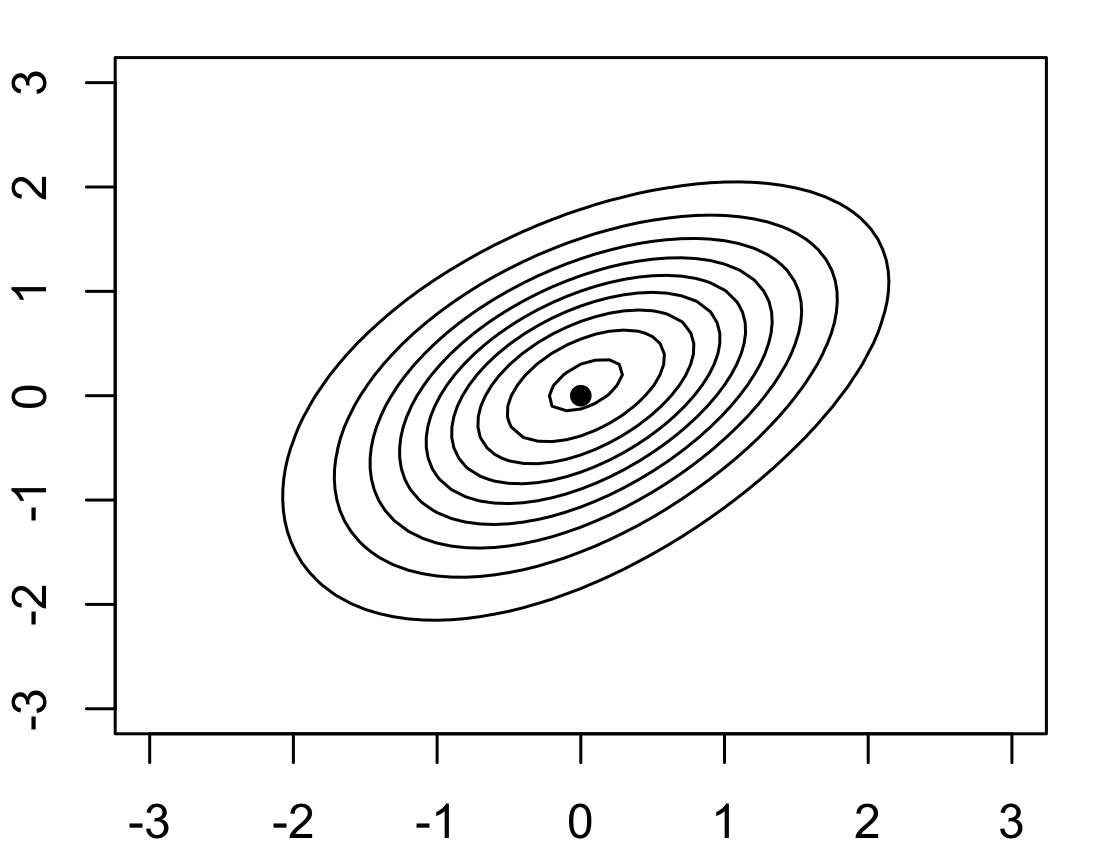} &
		\includegraphics[width = 2cm]{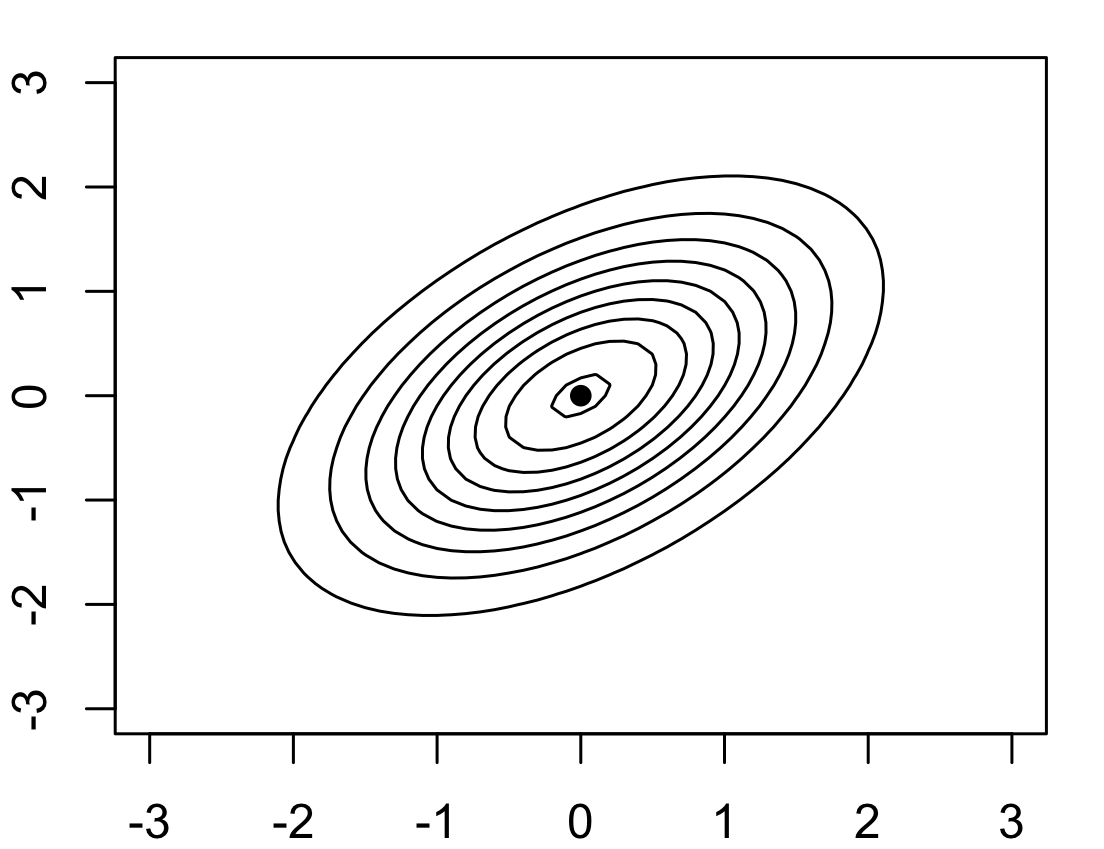} &
		\includegraphics[width = 2cm]{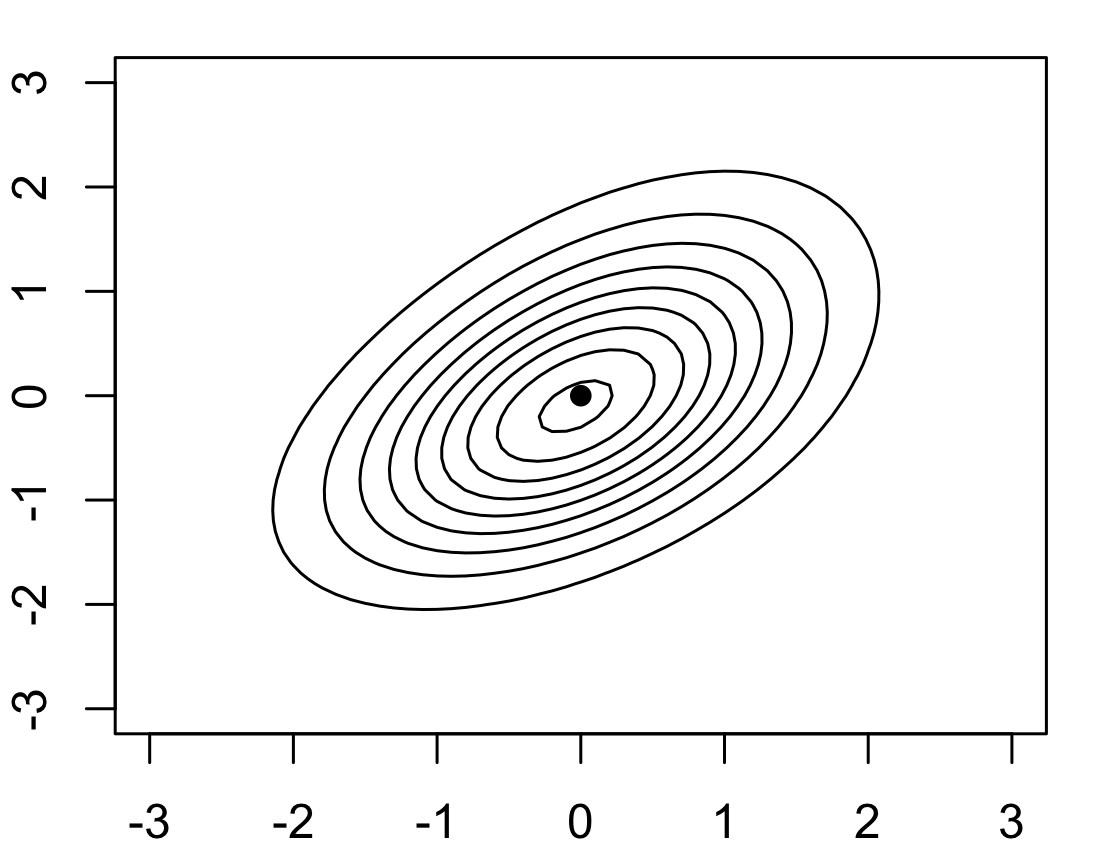} &
		\includegraphics[width = 2cm]{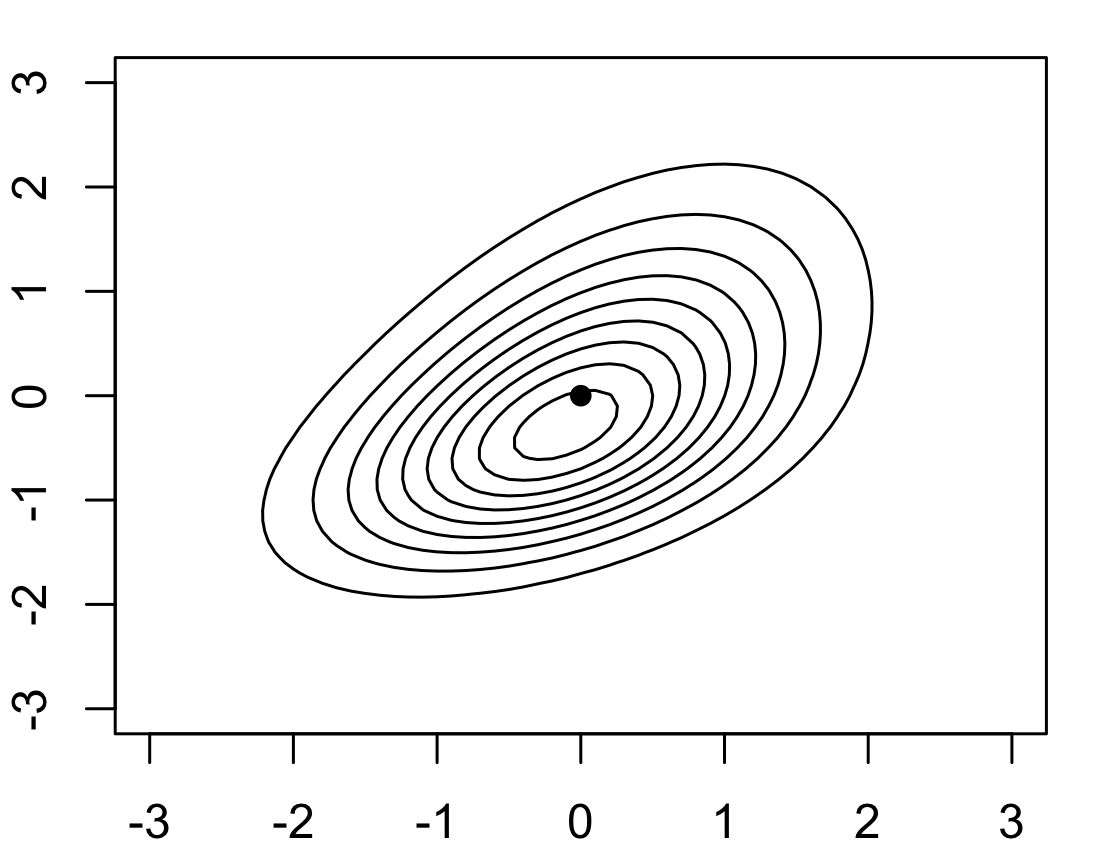} &
		\includegraphics[width = 2cm]{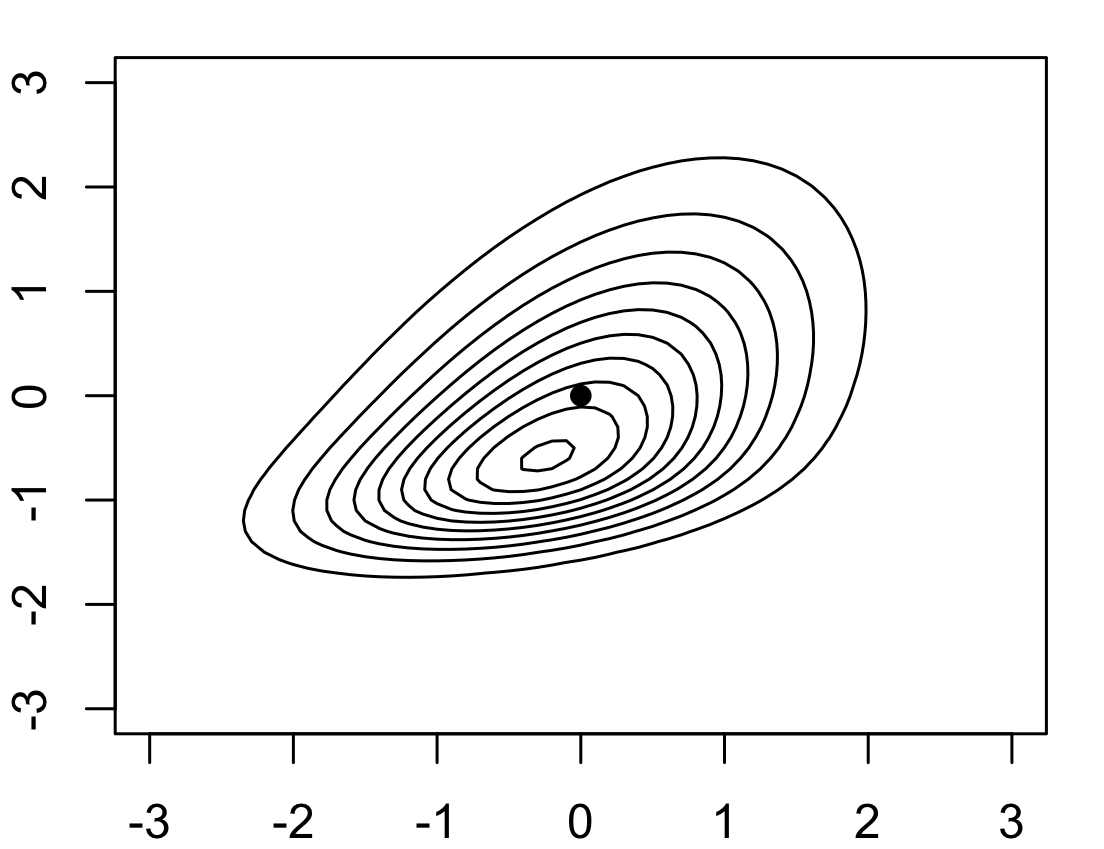}\\ \hline
		\rotatebox{90}{$s_1 = 0.2$} &\includegraphics[width = 2cm]{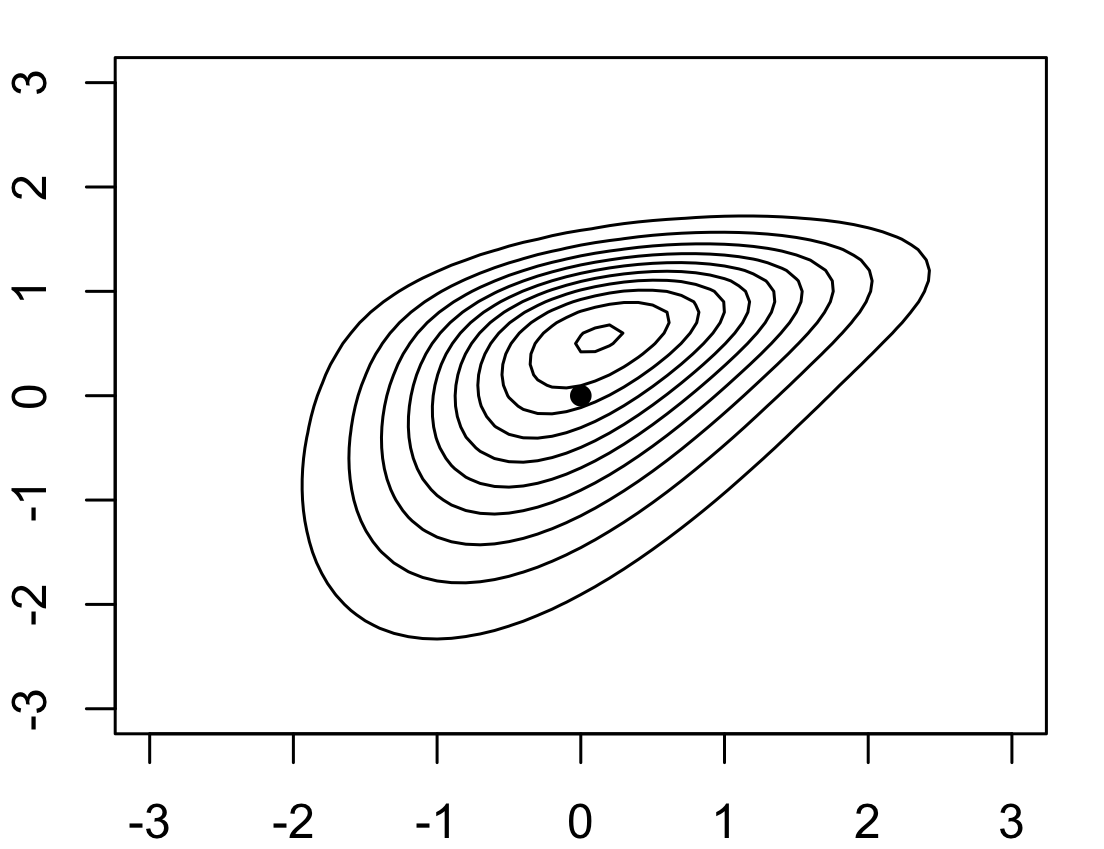} &
		\includegraphics[width = 2cm]{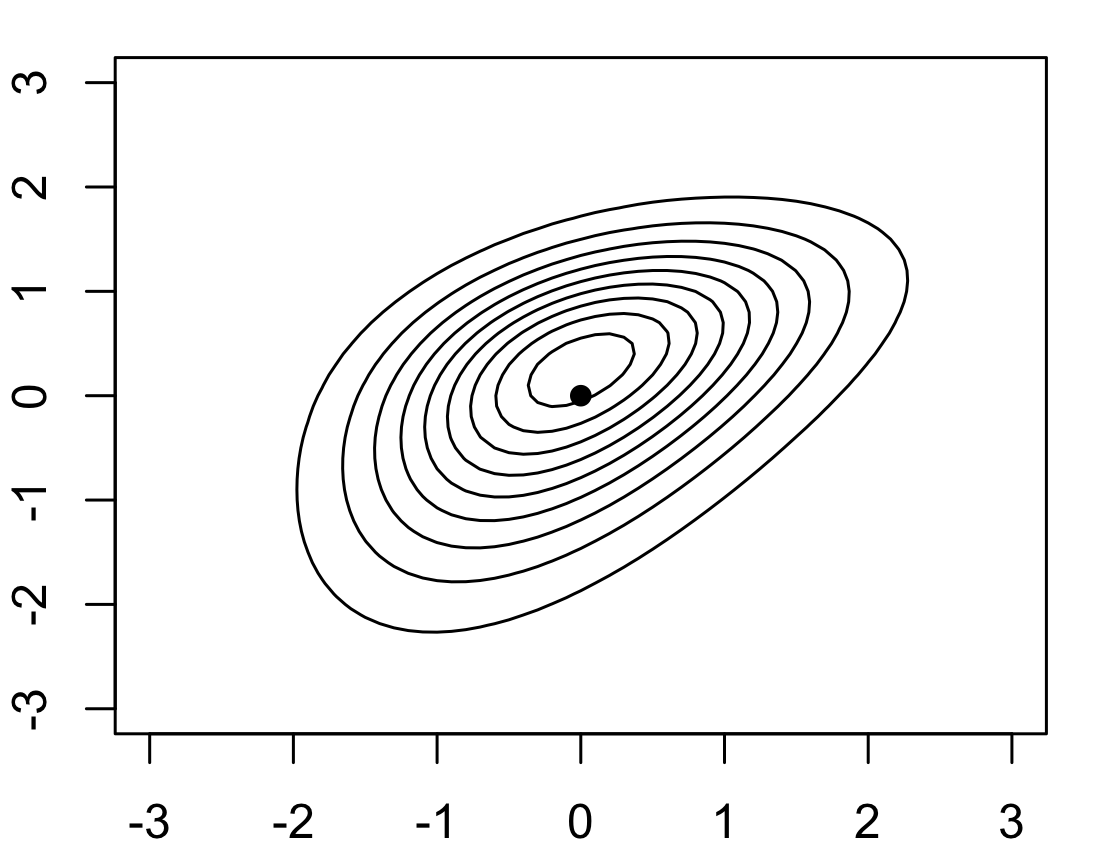} &
		\includegraphics[width = 2cm]{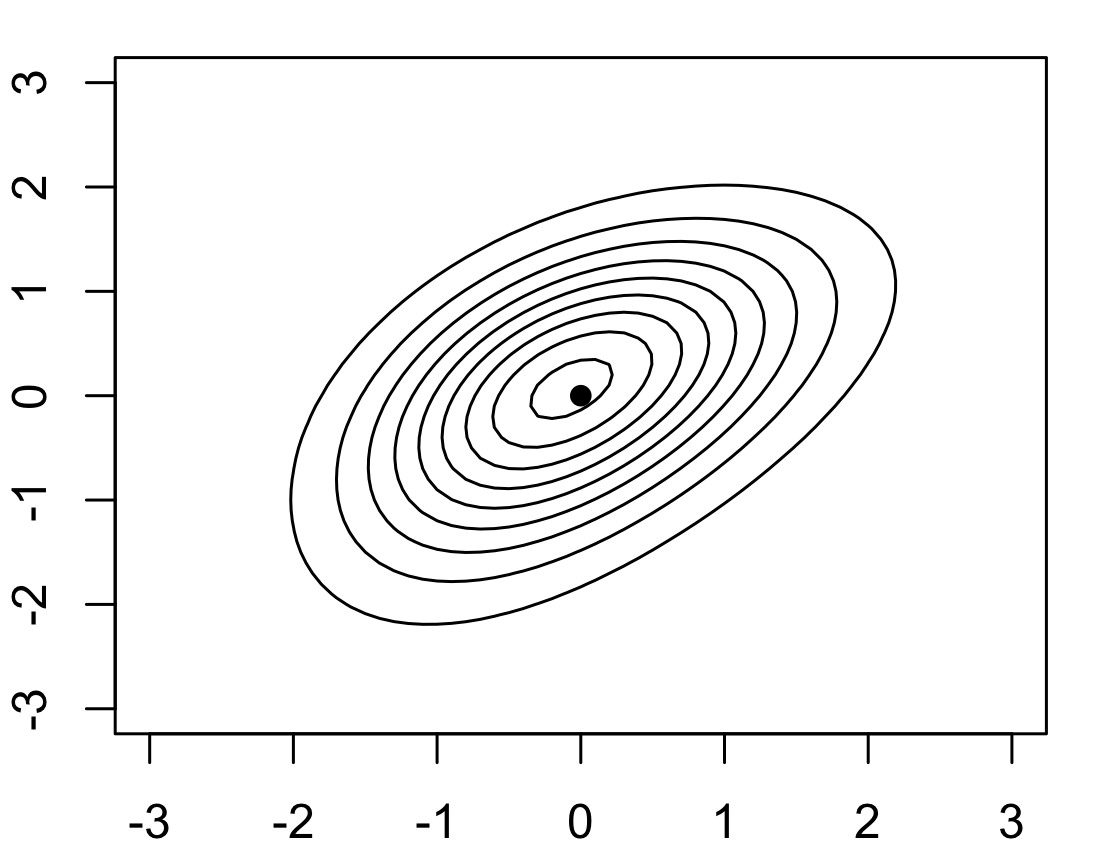} &
		\includegraphics[width = 2cm]{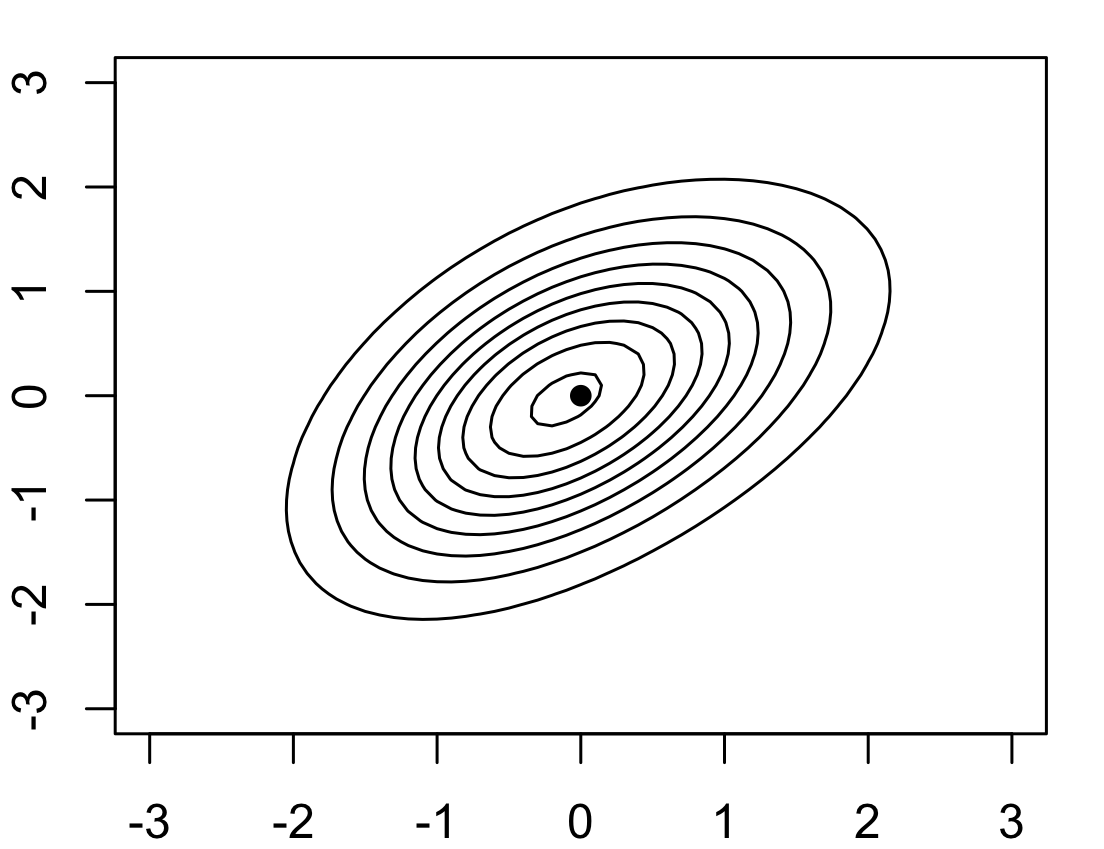} &
		\includegraphics[width = 2cm]{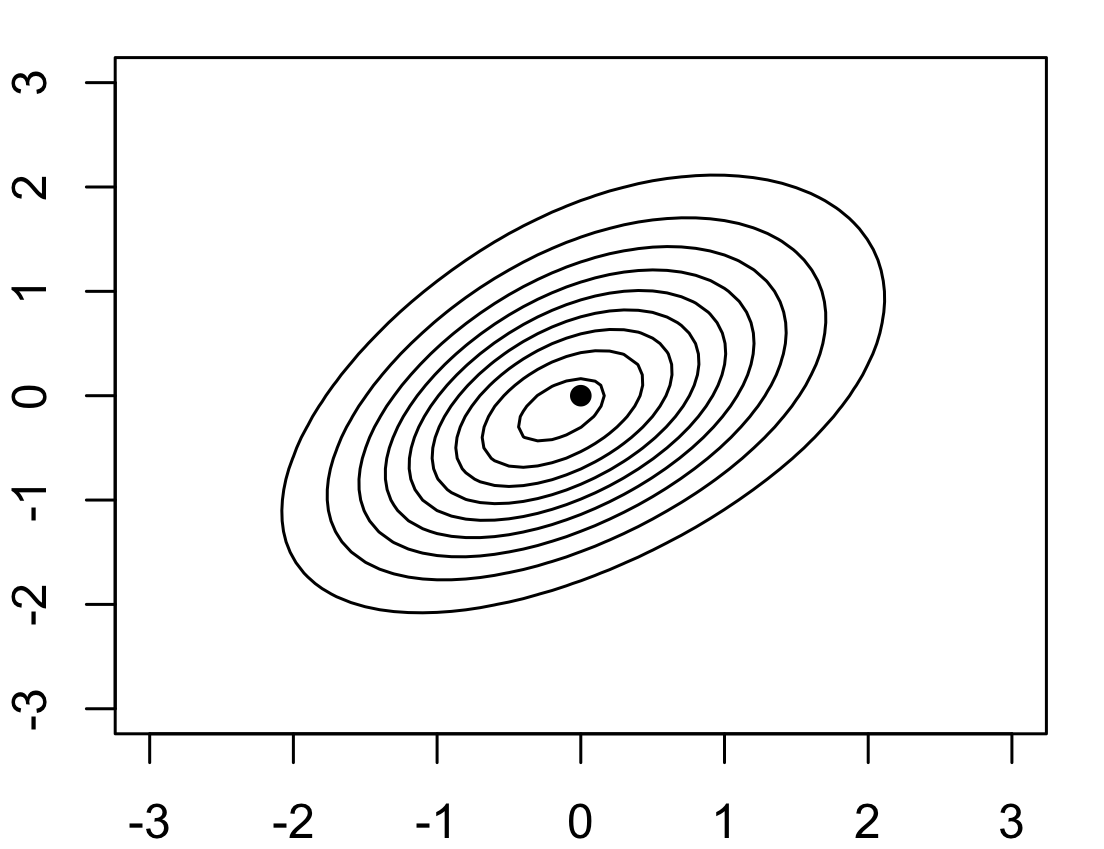} &
		\includegraphics[width = 2cm]{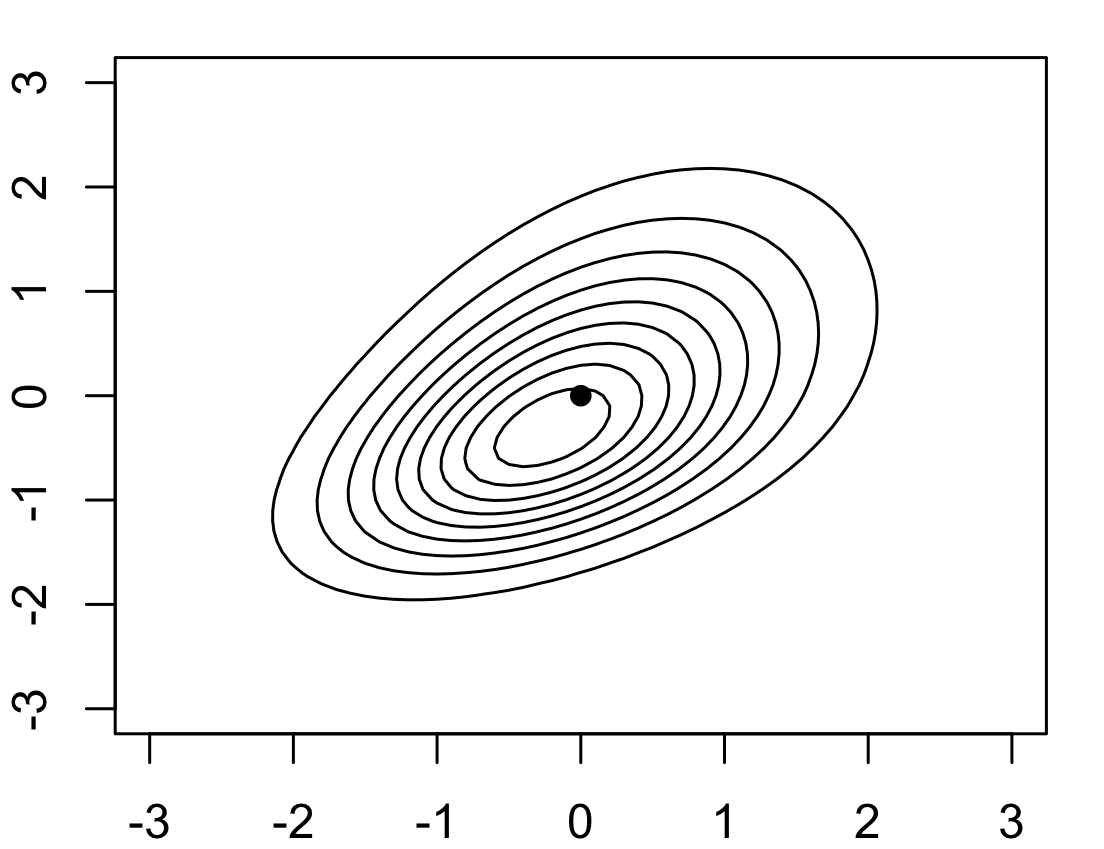} &
		\includegraphics[width = 2cm]{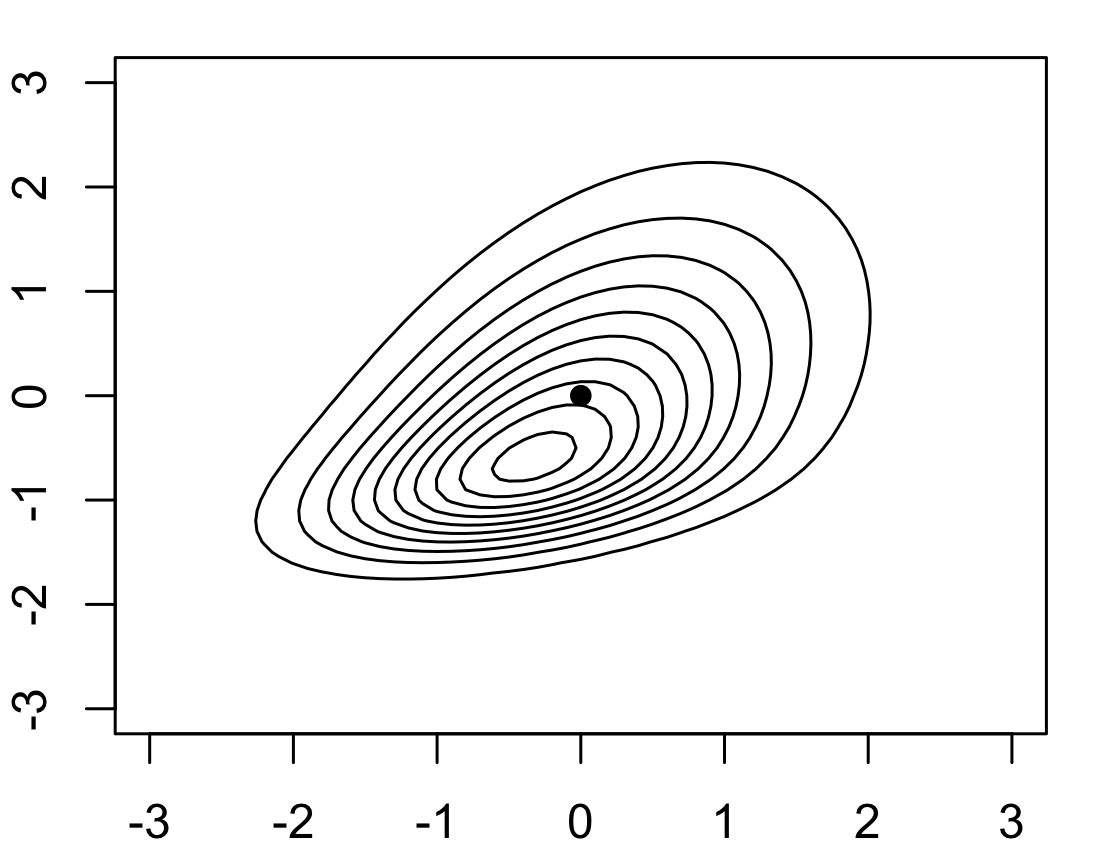}\\ \hline
		\rotatebox{90}{$s_1 = 0.5$} &\includegraphics[width = 2cm]{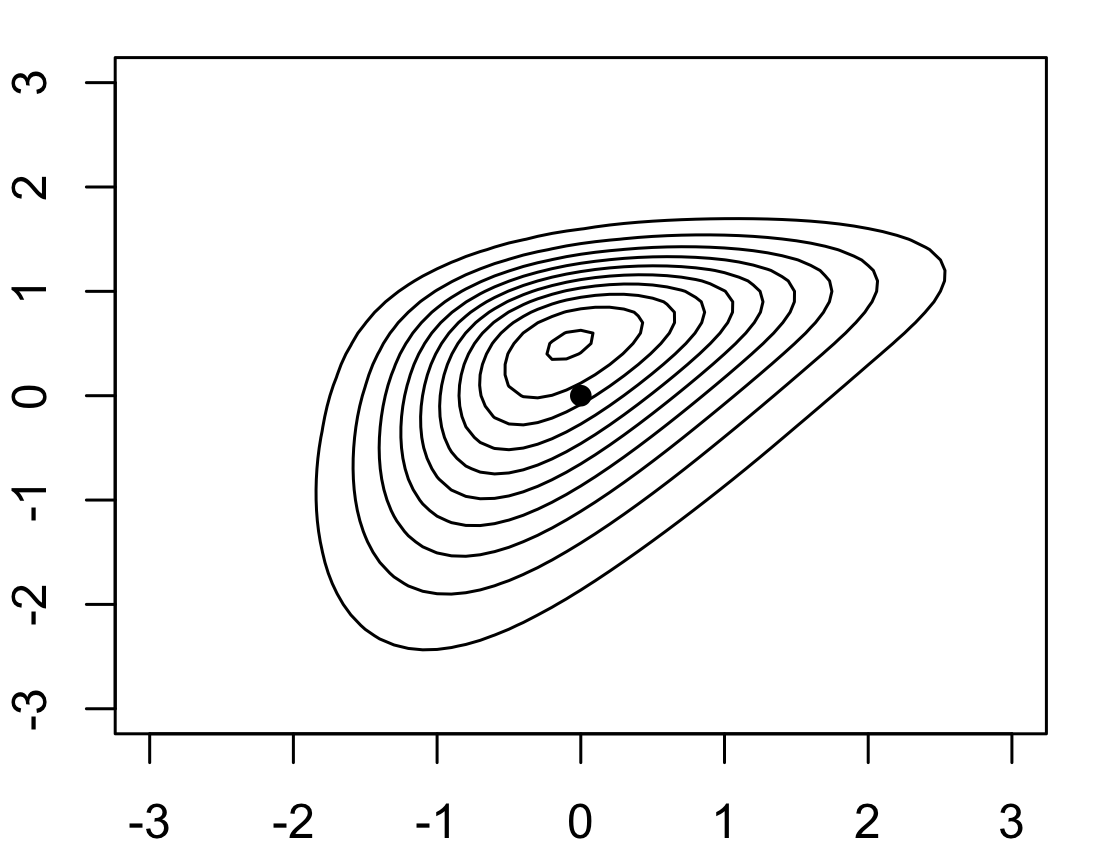} &
		\includegraphics[width = 2cm]{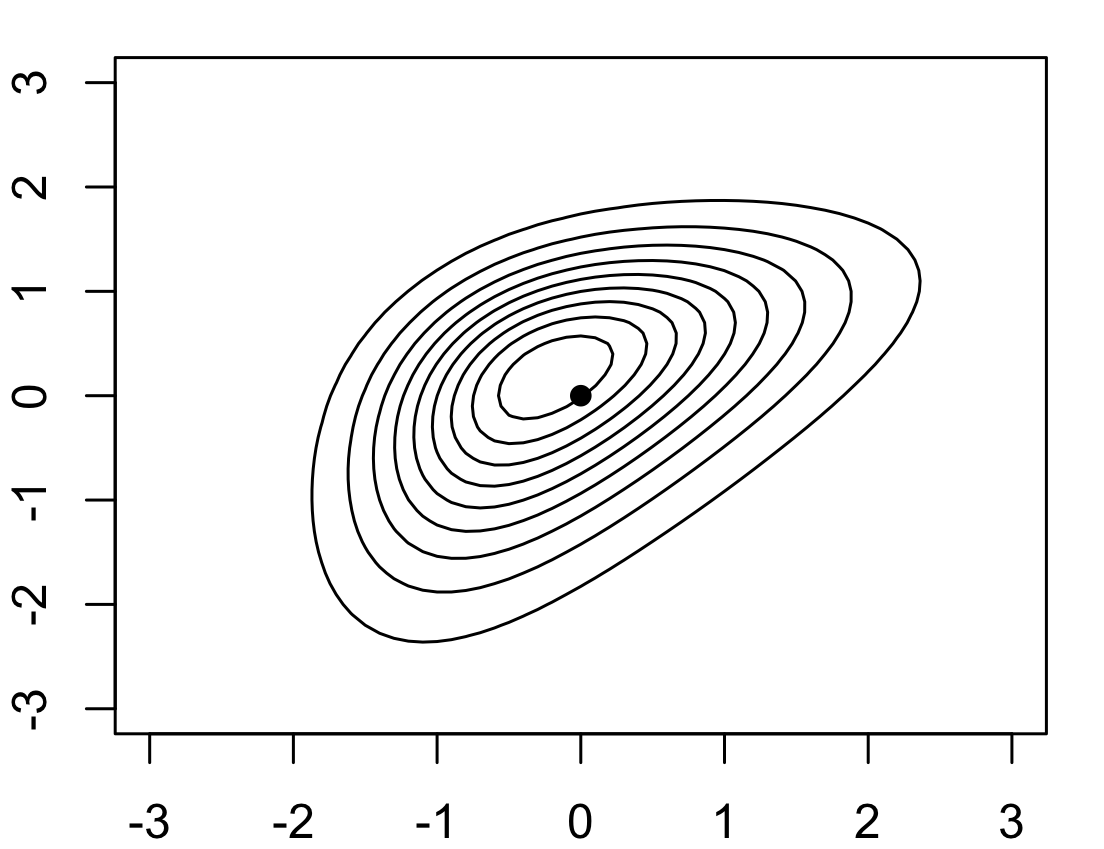} &
		\includegraphics[width = 2cm]{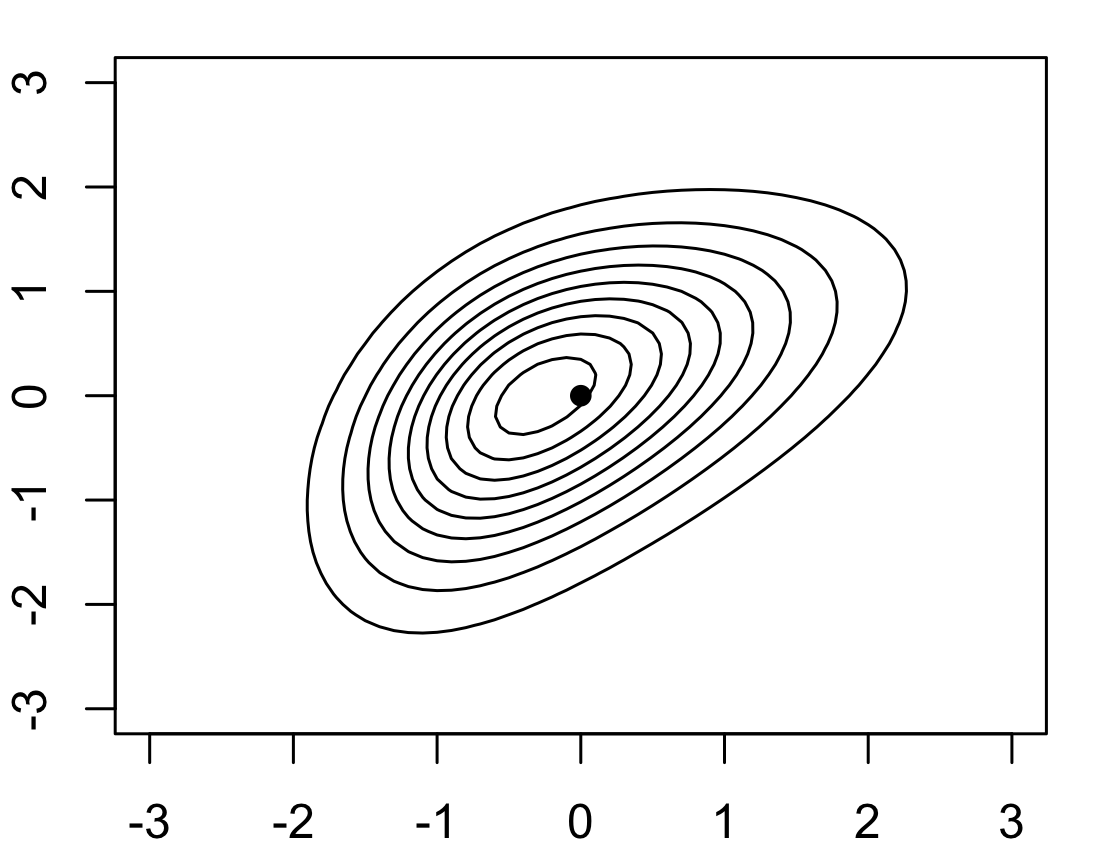} &
		\includegraphics[width = 2cm]{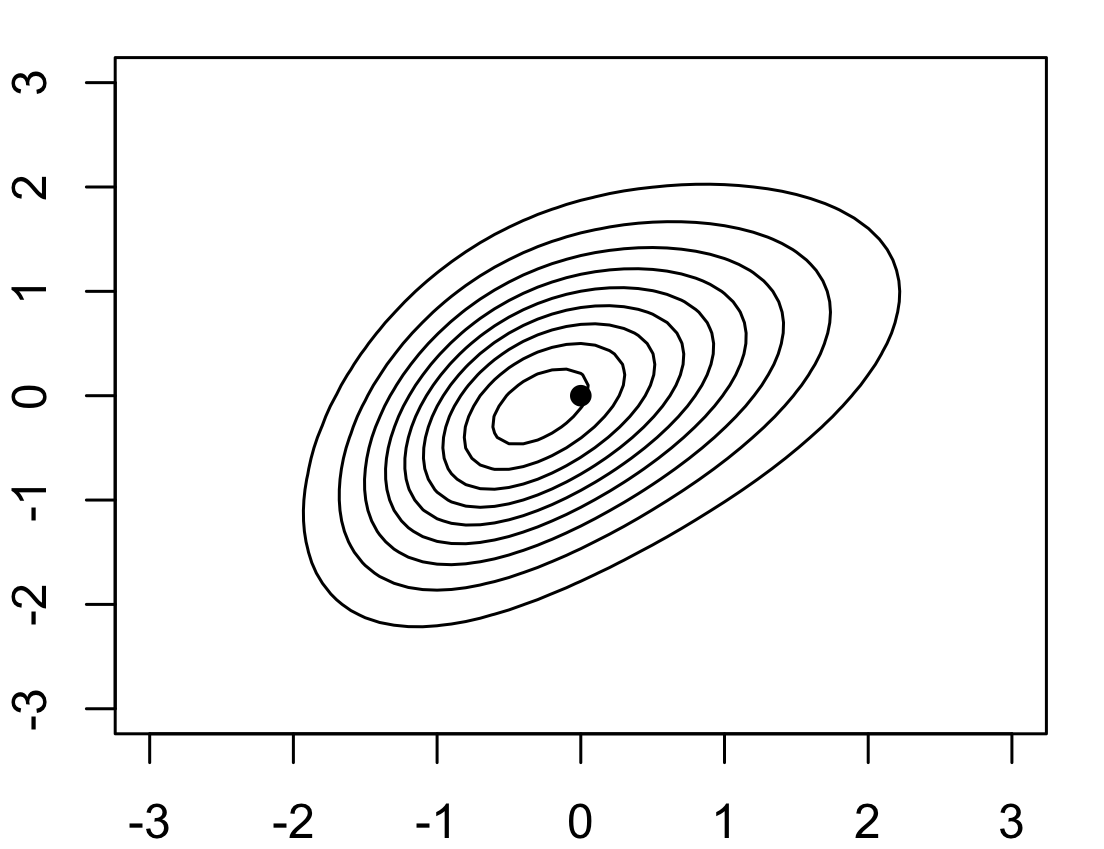} &
		\includegraphics[width = 2cm]{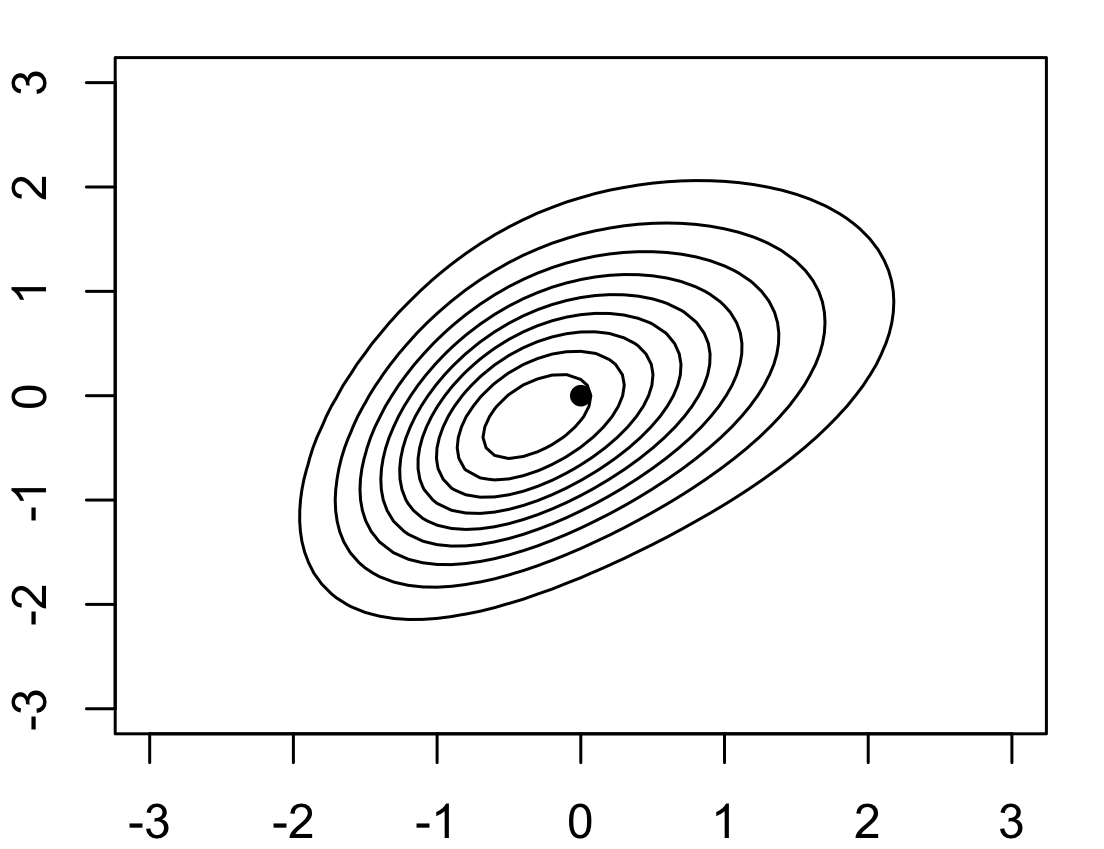} &
		\includegraphics[width = 2cm]{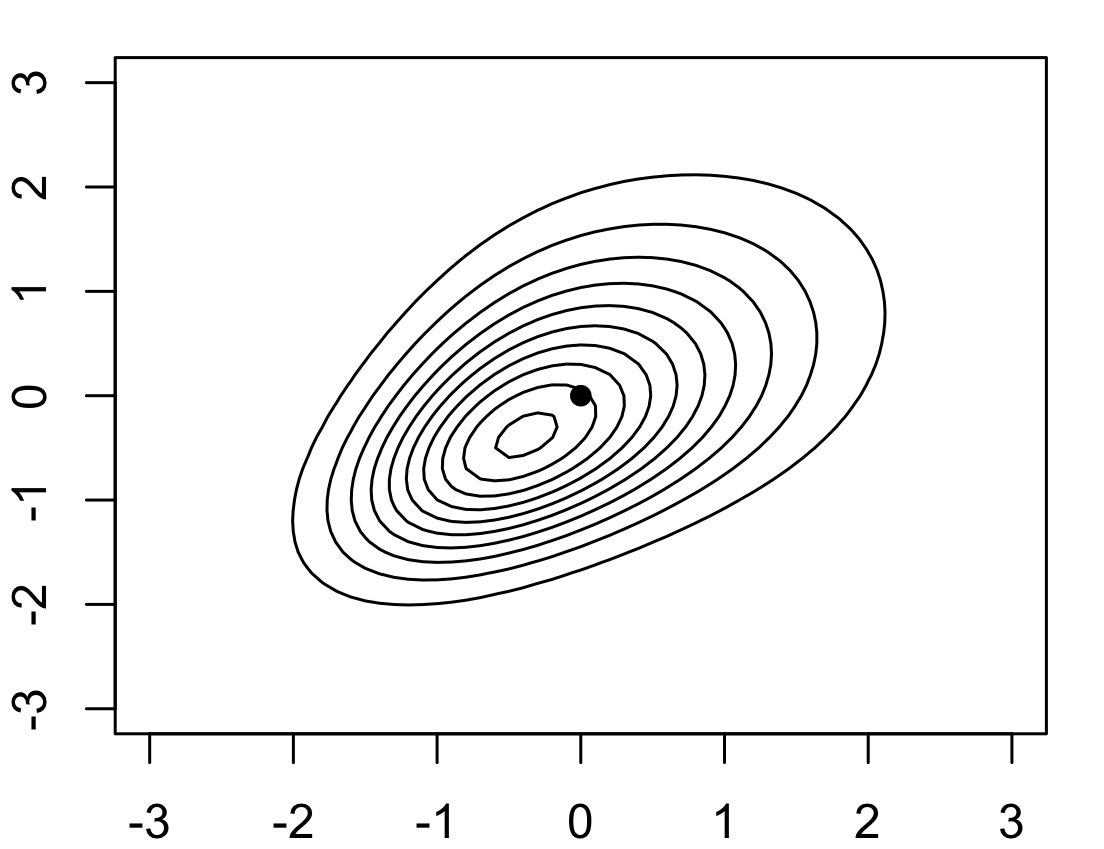} &
		\includegraphics[width = 2cm]{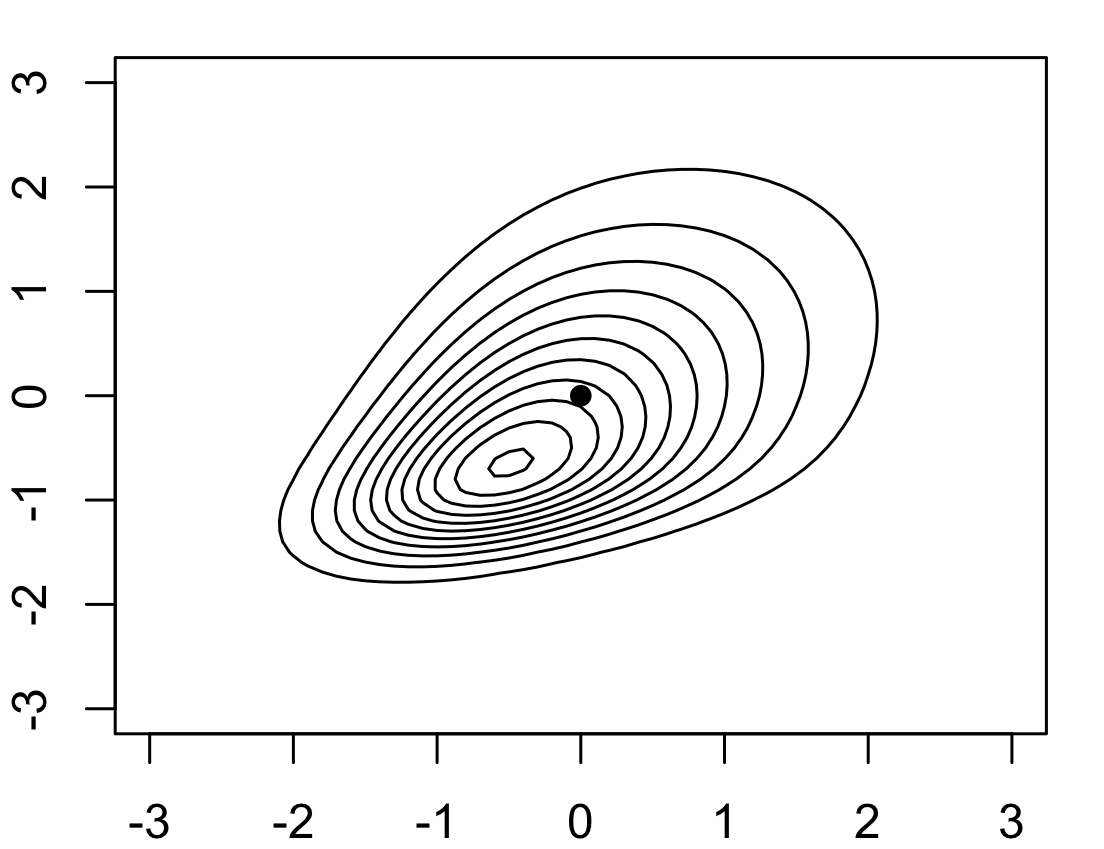}\\ \hline
		\rotatebox{90}{$s_1 = 0.8$} &\includegraphics[width = 2cm]{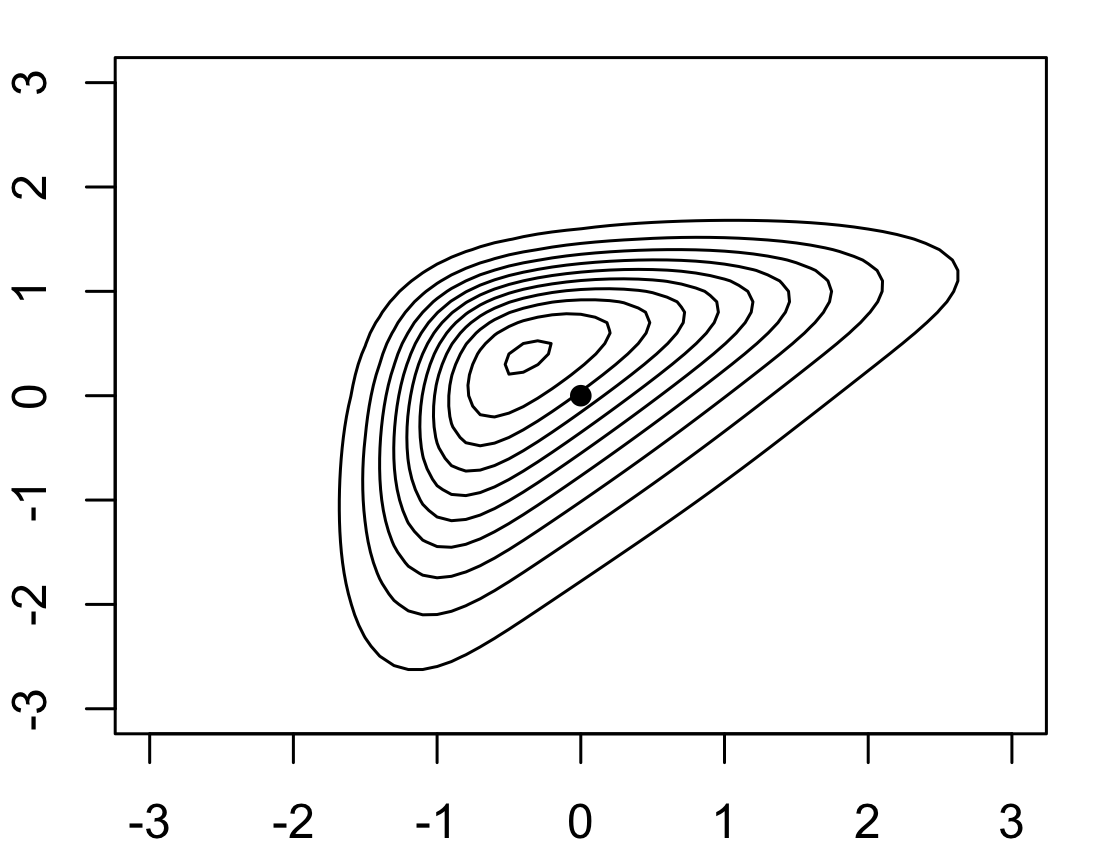} &
		\includegraphics[width = 2cm]{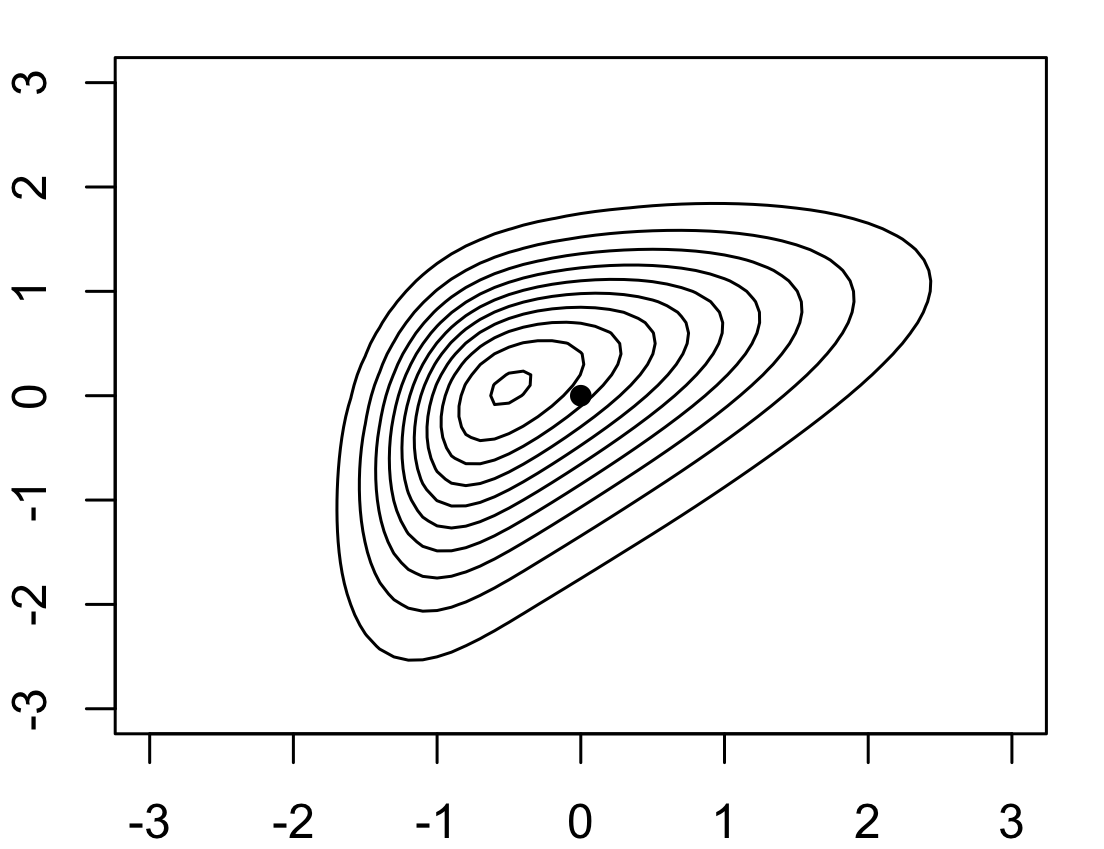} &
		\includegraphics[width = 2cm]{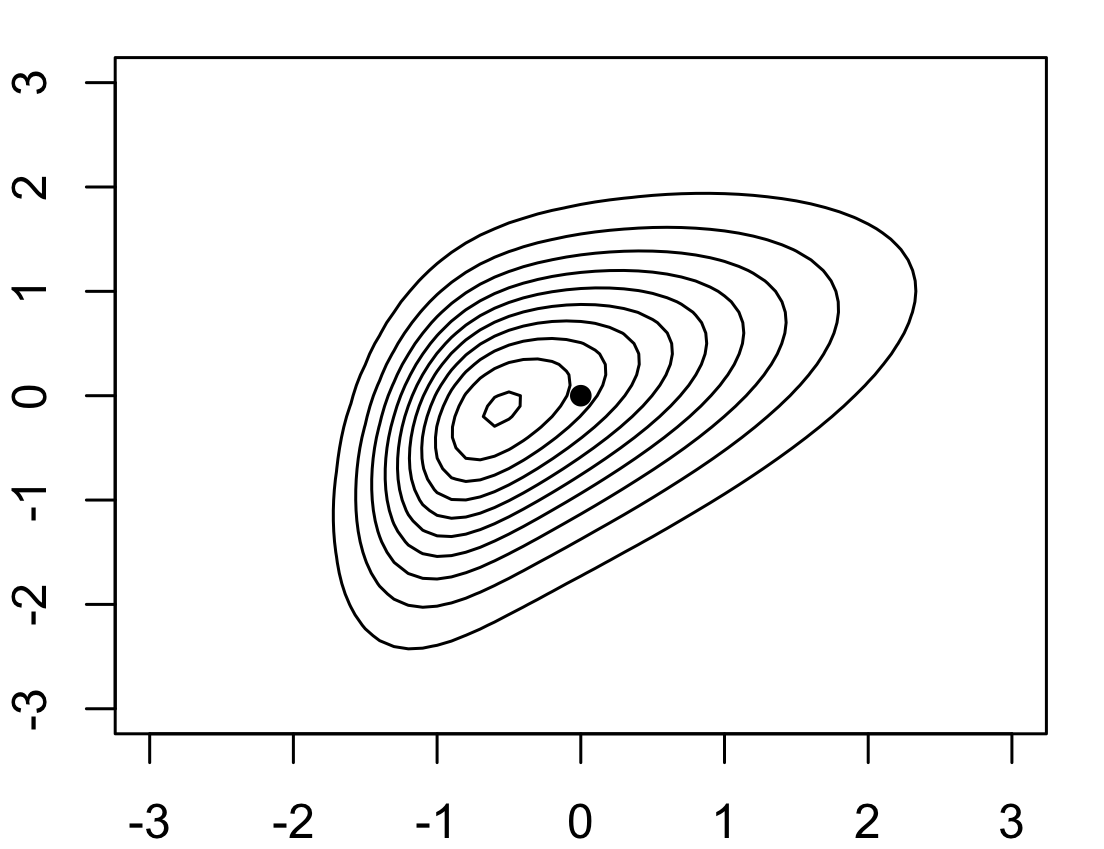} &
		\includegraphics[width = 2cm]{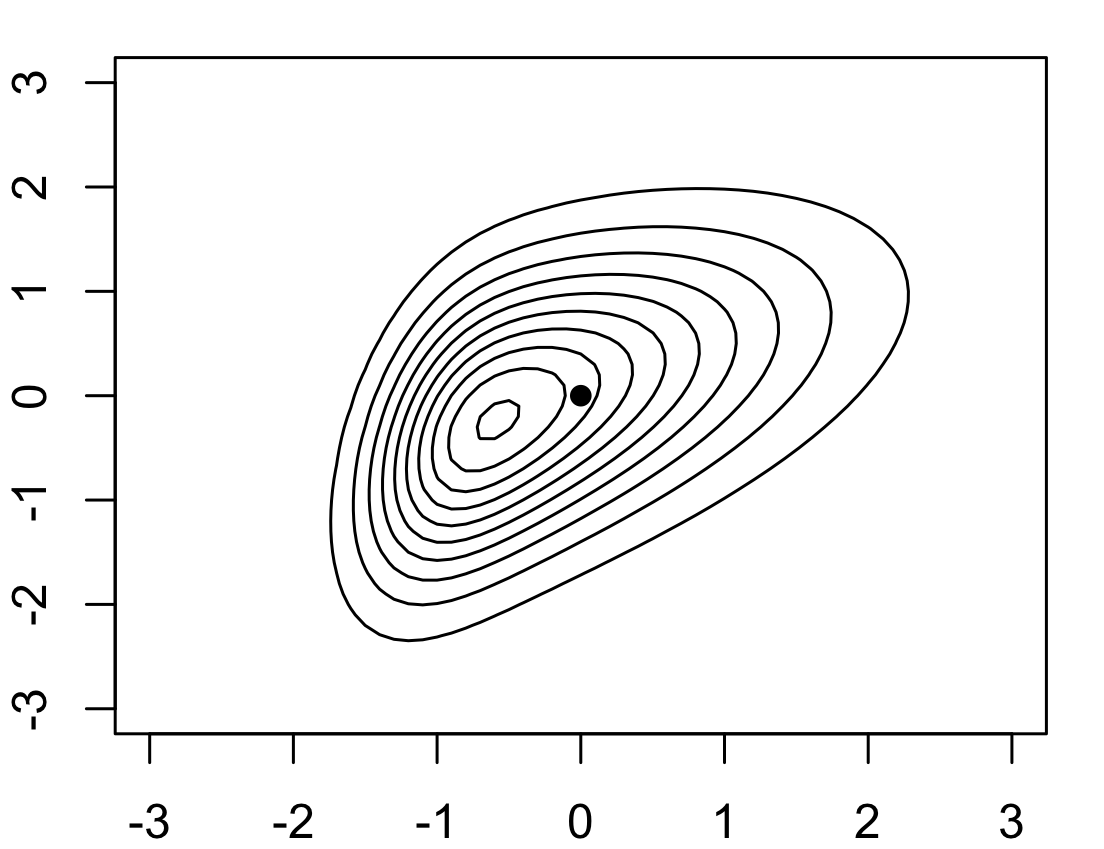} &
		\includegraphics[width = 2cm]{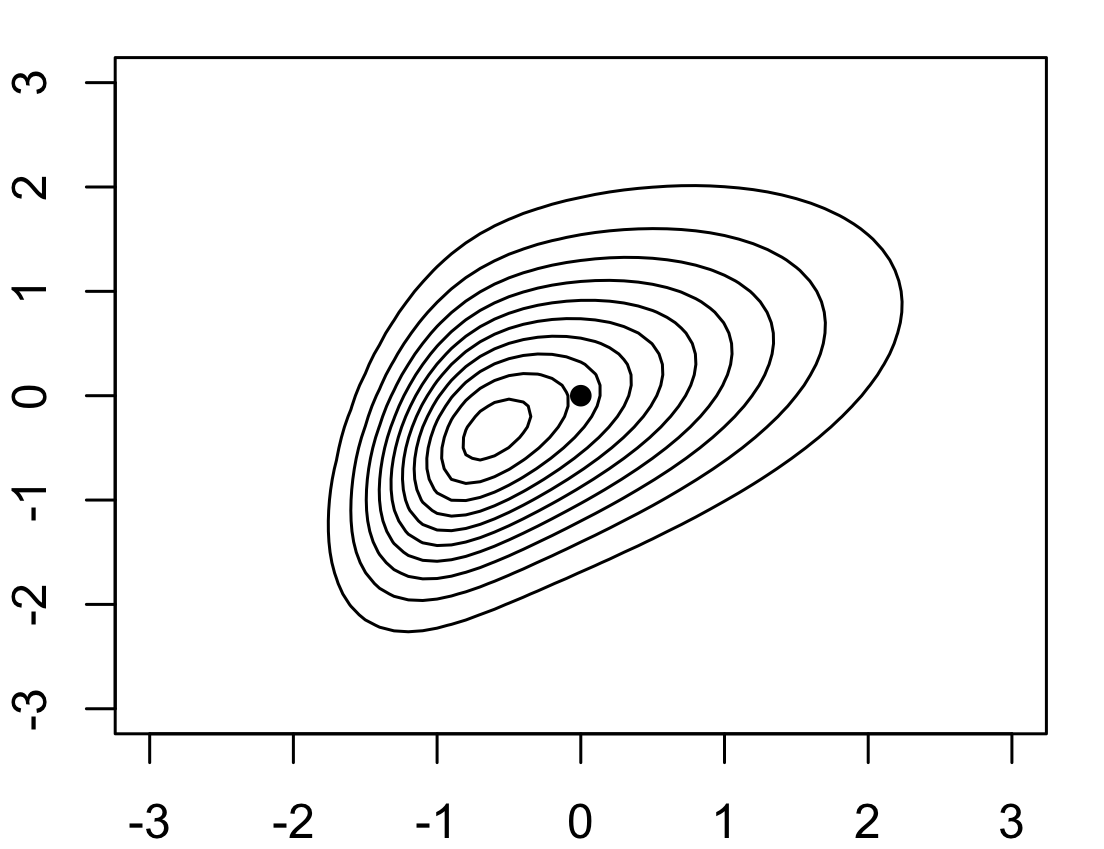} &
		\includegraphics[width = 2cm]{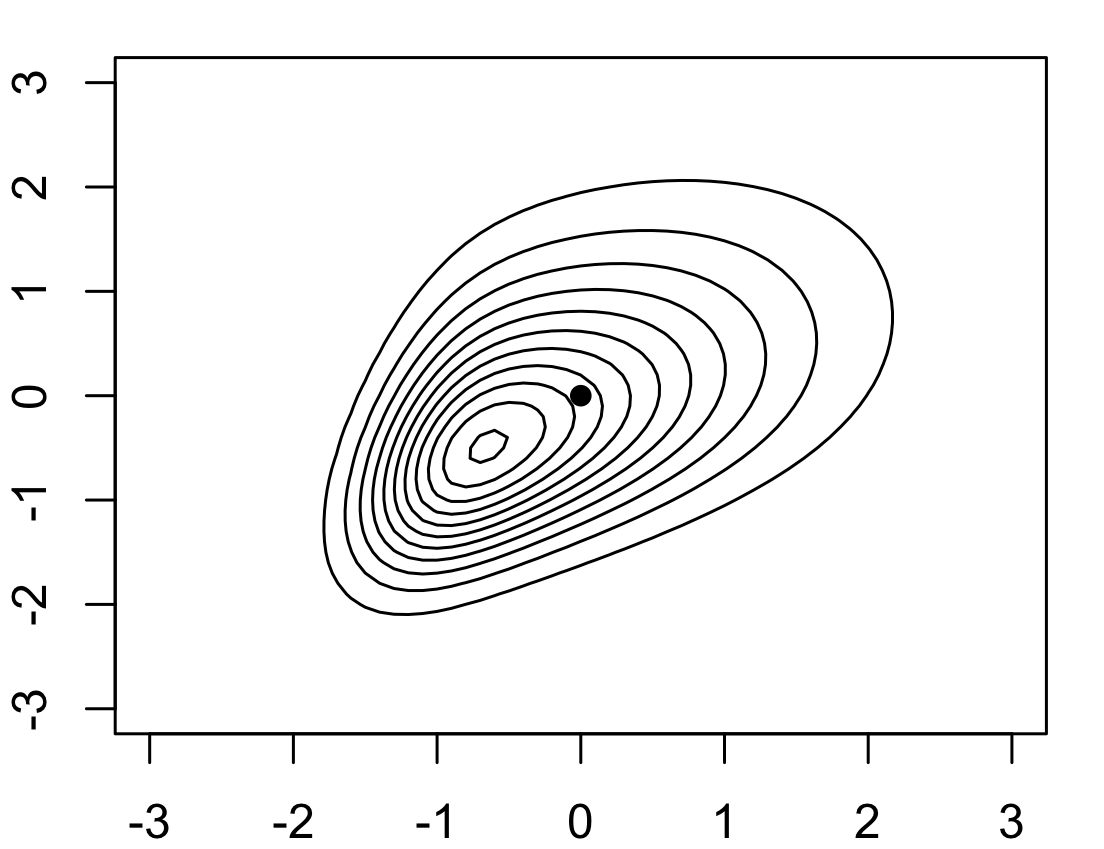} &
		\includegraphics[width = 2cm]{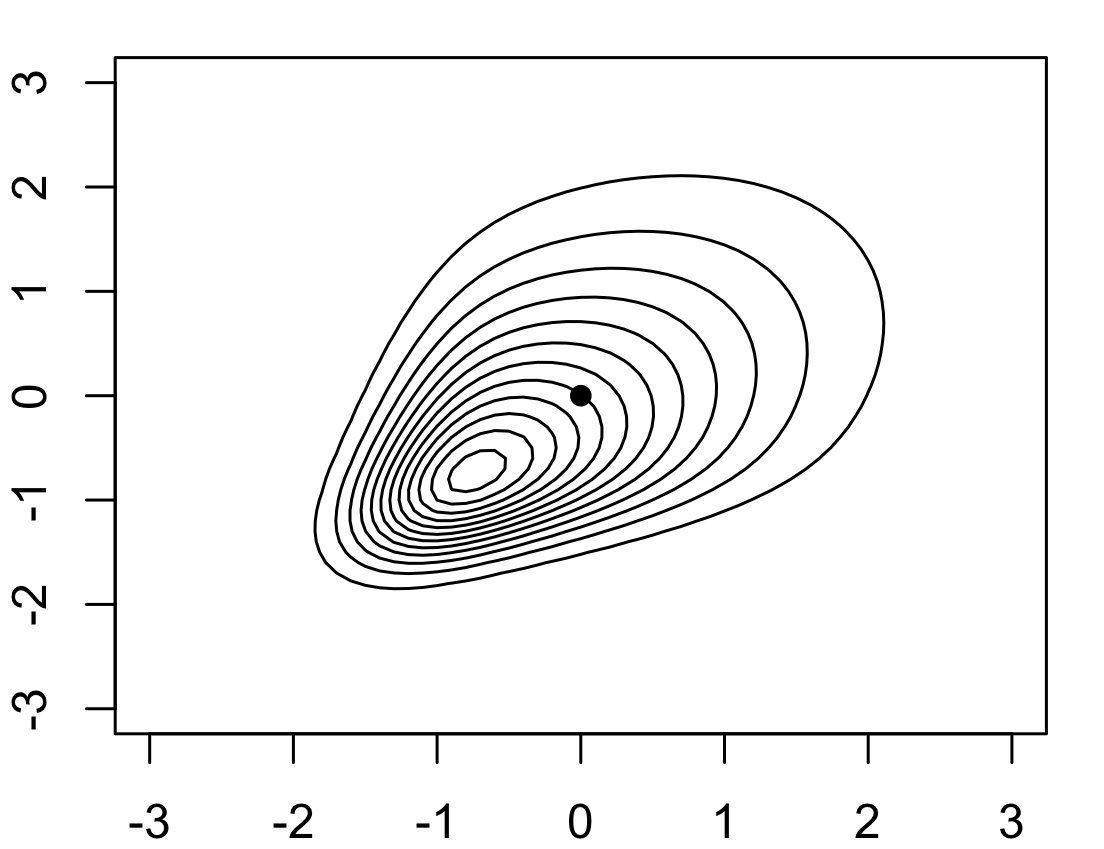} \\ \hline
	\end{tabular}
	
	\caption{Contour plots of the SGC density for a bivariate vector with mean vector $[0,0]$, Covariance matrix $0.5\pmb I + 0.5$ and marginal skewness $[s_1,s_2]$. The black dot indicates the mean at $[0,0]$.}
	\label{fig:sgc}
\end{figure}

In Figure \ref{fig:sgc} we present the contour plots of a density for a bivariate vector with the SGC distribution with mean vector $[0,0]$, Covariance matrix $0.5\pmb I + 0.5$ and marginal skewness vector $[s_1, s_2]$. From left to right illustrates $s_2 \in \{-0.8,-0.5,-0.2,0,0.2,0.5,0.8\}$ and from top to bottom illustrates $s_1 \in \{-0.8,-0.5,-0.2,0,0.2,0.5,0.8\}$. The black dot indicates the mean at $[0,0]$. Note that for $s_1 = s_2 = 0$, we recover the multivariate Gaussian, as expected. The flexibility of this distribution is clear from the many obtainable shapes by allowing skewed marginals.

\subsection{Inference of the marginal standardized skewness using Variational Bayes (SGC-VB)}
In this Section we use the results from Section \ref{sec:sgc} and apply them to the conditional posterior of the latent field of an LGM (see \eqref{eq:LGM} for details) based on the INLA methodology (see Section \ref{sec:INLA}).\\

\noindent Consider the corrected Gaussian approximation for the conditional posterior of $\pmb f$, $p_N(\pmb f;\pmb{\mu_f}, \pmb{ Q_f})$. The variational family is the SGC family, $p_{\text{SGC}}(\pmb f;\pmb\mu, \pmb Q, \pmb s)$ as defined in \eqref{eqn5.7}, with the same mean and copula correlation matrix as $p_N(\pmb f;\pmb{\mu_f}, \pmb {Q_f})$.\\ The marginal standardized skewness is then inferred as follows
\begin{equation}
	\hat{\pmb s} = \argmin_{\pmb s}\left\{E_{\pmb{\eta}} [- \log L(\pmb{y} | \pmb{\eta}, \pmb{\theta})]
	+	\text{KLD}(p_N(\pmb f;\pmb\mu, \pmb {Q_f})||p_{\text{SGC}}(\pmb f;\pmb\mu, \pmb Q, \pmb s))\right\},
\end{equation}
where the expectation is over the random vector $\pmb\eta$, where $\pmb \eta = \pmb A\pmb f$, $\pmb f|\pmb\theta,\pmb y \sim SGC(\pmb\mu, \pmb Q, \pmb s)$.

We require the Kullback-Leibler Divergence (KLD) between the hypothesized posterior and the Gaussian prior, as well as the expected negative log-likelihood of the linear predictor. These components will be addressed in the subsequent sections. The associated computational complexity from changing $m$ of the $p$ components, $m\ll p$ to non-zero skewness, does not add much to the cost of the Laplace method which is roughly $O(p^{3/2})$ for spatial models, for the factorization of a sparse precision matrix of the latent field of size $p$, of an LGM as defined in \eqref{eq:LGM}.

\subsubsection{Expected Negative Log likelihood Computation} \label{sec:loglik}

Contrary to Section \ref{sec:meanvarcorr}, the linear predictors $\pmb\eta$ does not follow a multivariate Gaussian anymore, since now it is a linear combination of dependent Gaussian and skew-normal random variables. This complicates the computation of the expected log-likelihood significantly compared to Section \ref{sec:meanvarcorr}. This complexity necessitates the use of variable transformations to achieve independence in the transformed latent field, allowing for a more manageable and efficient calculation. 

We want to calculate

\begin{equation} \label{eqn5.11}
	\sum_{i = 1}^{n} E_{\eta_i} \left[ - \log L(y_i \mid \eta_i, \pmb{\theta}) \right]
\end{equation}

where $\eta_i = \pmb{A}_i \pmb{f}$. Consider 
$\pmb{\gamma} = \pmb{L}^{-1} \pmb f$, where $\pmb L$ is the lower triangular Cholesky decomposition of $\pmb Q^{-1}$, i.e. $\pmb Q^{-1} = \pmb L\pmb L^\top$. Thus $\gamma_1 = l_{11}f_1, \quad \gamma_2 = l_{21}f_1+l_{21}f_2, \quad \gamma_3 = l_{31}f_1+l_{32}f_2 +l_{33}f_3$, and so on. Now, the orthogonal $\gamma$ variables are corrected for skewness. Note however, that the skewness vector of $\pmb f$, our target, only corresponds to the skewness vector for $\pmb \gamma$ in the first component since multiplying by a constant does not change the skewness except for the sign. Subsequently, the vector $\pmb f$ should be reordered $m$ times while we retain only the first element of the subsequent optimized skewness vector at each iteration. These $m$ iterations can be done in parallel.\\
Note that $\eta_i = \pmb A_i\pmb f = \pmb A_i\pmb L\pmb\gamma = \sum_j c_{ij}\gamma_j$, where $\pmb c_i = \pmb A_i\pmb L$. Thus, $\eta_i$ is a sum of orthogonal terms and thus the distribution of $\eta_i$ can be numerically computed from its characteristic function using the Discrete/Fast Fourier Transformation (FFT) \citep{brigham1988fast}.\\ \\ 
This method is only efficient for latent fields with dimensions up to about $1000$. This limitation arises because the process involves the full inversion of the sparse precision matrix, $\pmb{Q}$, which is computationally expensive and cannot be accommodated within the memory constraints of a standard computer for higher dimensions.

Therefore, we need an alternative method for calculating the negative log-likelihood in the case of a higher dimensional latent field. This alternative approach, using blocking, is detailed in the next section.

\subsubsection{Alternative way to compute Expected Negative Log likelihood for higher dimensional latent field} \label{sec:loglik_large}

Irrespective of the dimension of the latent field, we can perform a partial inversion of $\pmb{Q}$. We can decompose this partial inversion by using a rearrangement. This approach allows us to manage higher dimensions effectively by strategically breaking down the inversion process, thus making it computationally feasible and efficient. By reordering and restructuring the components of the partial variance covariance matrix, we ensure that the partial inversion is both accurate and manageable, even as the dimensionality of the latent field, $p$, increases.

To calculate the linear predictor, we need to multiply the latent field $f$ with the design matrix $\pmb{A}$.
\begin{align}
	\eta_i &= \pmb{A}_{r(i)}\pmb{f} \notag \\
	&= \begin{bmatrix}
		\pmb{A}_{r(i)}^{1} & \pmb{A}_{r(i)}^{2}
	\end{bmatrix}
	\begin{bmatrix}
		\pmb{f}^1 \\
		\pmb{f}^2
	\end{bmatrix} \label{eqn5.16} \\
	&= \pmb{A}_{r(i)}^{1}\pmb{f}^1 + \pmb{A}_{r(i)}^{2}\pmb{f}^2 \notag \\
	&= \pmb{z}_1 + \pmb{z}_2 \label{eqn5.17}
\end{align}

In \eqref{eqn5.16}, $\pmb{f} = \begin{bmatrix}
	\pmb{f}^1 & \pmb{f}^2
\end{bmatrix}^{T} $, being decomposed, where we are trying to optimise the marginal skewness of $\pmb{f}^2$ part and the $i-$th row of the design matrix being decomposed as $ \pmb{A}_{r(i)} = \begin{bmatrix}
	\pmb{A}_{r(i)}^{1} & \pmb{A}_{r(i)}^{2}
\end{bmatrix}$. $\pmb{A}_{r(i)}^{1}$ and $\pmb{A}_{r(i)}^{2}$ correspond to the coefficient of $\pmb{f}^1$ and $\pmb{f}^2$ respectively.

In \eqref{eqn5.17}, $\pmb{z}_1$ and $\pmb{z}_2$ corresponds to $\pmb{A}_{r(i)}^{1}\pmb{f}^1 $ and $\pmb{A}_{r(i)}^{2}\pmb{f}^2$ respectively.

Similar to Section \ref{sec:loglik}, we do a change of variable for the $\pmb{f}_2$ part,
\begin{equation} \label{eqn5.18}
	\pmb{\gamma}^2 = \pmb{L}_2^{-1}\pmb{f}^2
\end{equation}

In \eqref{eqn5.18}, \( \pmb{L}_2 \) represents the Cholesky factorization of the variance-covariance matrix corresponding to \( \pmb{f}^2 \), which is derived from partial inversion of \( \pmb{Q}_{\pmb f}^{corr} \). 
We deliberately designate the first component of \( \pmb{f}^2 \) as the marginal that we aim to correct (this should be rearranged in multiple iterations to calculate the marginal skewness for all the components as in Section \ref{sec:loglik}). 

Note that

\begin{equation} \label{eqn5.19}
	\eta_i= \pmb{A}_{r(i)}^{1}\pmb{f}^1 + \pmb{A}_{r(i)}^{2}\pmb{f}^2 
	= \pmb{A}_{r(i)}^{1}\pmb{f}^1 +  \pmb{A}_{r(i)}^{2}\pmb{L}_2\pmb{\gamma}^2
	= \pmb{z}_1 + \pmb{z}_2
\end{equation}

Next, we derive the distribution of \( \pmb{z}_2 \) by summing up the components of \( \pmb{\gamma}^2 \), again utilizing the FFT. Since these components are independent, FFT provides an efficient way to compute the distribution. This approach allows us to handle the summation of distributions quickly and accurately, leveraging the computational power of FFT to streamline the process.

Thereafter, we derive the distribution of $\eta_i$ by resorting to convolution. 

\begin{equation} \label{eqn5.201}
	p(\eta_i) = \int p_{\pmb{z}_1 \mid \pmb{z}_2^*} \left( \eta_i - \pmb{z}_1 \right) p_{\pmb{z}_2}(\pmb{z}_2) d\pmb{z}_2
\end{equation}

In \eqref{eqn5.201}, $\pmb{z}_2^*$ represents the Gaussian part of the sum of skew-normal distributions, found from the copula.

Once, we have the distribution of each $\eta_i$, we can calculate the expected negative log likelihood,\\ $\sum_{i = 1}^{n} E_{\eta_i} \left[ - \log L(y_i \mid \eta_i, \pmb{\theta}) \right]$ numerically for an arbitrarily large dimensional latent field. In this case, we need to compute the individual skew corrections by reordering the vector $\pmb f_2$ and repeating the procedure, as in Section \ref{sec:loglik}.

\subsubsection{KLD between SGC and Multivariate Gaussian} \label{sec:kld}

The KLD between the proposal $p_{\text{SGC}}(\pmb f;\pmb\mu, \pmb Q, \pmb s)$ and $p_N (\pmb{f};\pmb\mu, \pmb {Q_f})$ is non-trivial to compute in a scalable way. The KLD is 
\begin{eqnarray} \label{eqn5.9}
	\text{KLD} \left( p_{\text{SGC}}(\pmb{f}) \mid \mid p_N (\pmb{f})  \right) 
	&=& \int p_{\text{SGC}}(\pmb f)\log\left( \frac{p_{SGC}(\pmb f)}{p_N(\pmb f)}\right)d\pmb f\notag\\
	&=& \int p_N(g^{-1}(\pmb f))\mathcal{J}\log\left( \frac{p_N(g^{-1}(\pmb f))\mathcal{J}}{p_N(\pmb f)}\right)d\pmb f\notag\\
	&=&\int p_N(\pmb f^*)\mathcal{J}\log\left( \frac{p_N(\pmb f^*)\mathcal{J}}{p_N(g(\pmb f^*))}\right)\mathcal{J}^{-1}d\pmb f^*\notag\\
	&=&\int p_N(\pmb f^*)\log\mathcal{J}d\pmb f^* + \int p_N(\pmb f^*)\log\left( \frac{p_N(\pmb f^*)}{p_N(g(\pmb f^*))}\right)d\pmb f^*\notag \\ &=& I_1 + I_2
\end{eqnarray}
from \eqref{eqn5.7}. Now note that 
\begin{equation*}
	I_1 = \int p_N(\pmb f^*)\log\mathcal{J}d\pmb f^* = \sum_i\int p_N( f^*_i)\log\mathcal{J}_i d f^*_i,
\end{equation*}
which can be computed using numerical methods. For the second part,
\begin{eqnarray*}
	I_2 &=& \int p_N(\pmb f^*)\log\left( \frac{p_N(\pmb f^*)}{p_N(g(\pmb f^*))}\right)d\pmb f^*\notag\\
	&=& \int p_N(\pmb f^*)\left( -\frac{1}{2}\pmb f^{*\top}\pmb Q\pmb f^{*} + \frac{1}{2}g(\pmb f^*)^{\top}\pmb Q g(\pmb f^*) \right)d\pmb f^*\notag\\
	&=&\int p_N(\pmb f^*)\left( -\frac{1}{2}\pmb f^{*\top}\pmb Q\pmb f^{*} + \frac{1}{2}\left(\sum_j\frac{c_j}{j!}f^{*j}\right)^{\top}\pmb Q \left(\sum_{j'}\frac{c_{j'}}{j'!}f^{*j'}\right) \right)d\pmb f^*\notag\\
	&=&\sum_{jj'}C_{jj'}\sum_{kl}p_N(\pmb f^*)f^{*j}_kf^{*j'}_ld\pmb f^*
\end{eqnarray*}
if we express $f_i = g_i(f_i^*) = c_{0i} + c_{1i}f_i^* + \frac{1}{2}c_{2i}f_i^{*2} + \frac{1}{6}c_{3i}f_i^{*3}+...$. The integral in $I_2$ reduces to expectations of bivariate moments with respect to the multivariate Gaussian. Hence, some terms are 0 since the odd moments are 0 by definition. The even moments can be explicitly calculated using specific entries of $\pmb Q^{-1}$. In practice, the $C_{jj'}$ terms can be pre-computed for different values of skewness and recalled in the computation as necessary. 

\subsection{Illustrative example}
To evaluate the effectiveness of our method in tracking marginal skewness, we implemented a straightforward, yet rigorous example. We chose a model characterized by a binomial likelihood with a fixed hyperparameter. The linear predictor for this model includes four fixed effects and one random effect with an autoregressive structure. This setup allows us to thoroughly test and demonstrate the robustness of our approach in accurately capturing and correcting marginal skewness under varying conditions. By selecting a model that combines simplicity with extremity, we aim to showcase the precision and reliability of our methodology in practical applications (see \citet{durante2024skewed} and \citet{zhou2024tractable} for similar experiments).

The model is given as follows:

\begin{eqnarray}
	y_i|\eta_i &\sim& \text{Binomial}(2, \text{logit}(\eta_i) ), \quad i = 1, 2, \ldots, n \notag \\
	\eta_i &=& \beta_0 + \beta_1 X_{1i} + \beta_2 X_{2i} + \beta_3 X_{3i} + u_i, \quad i = 1, 2, \ldots, n \notag \\ 
	\beta_k &\sim& N\left(0, 1/100\right) \quad \text{and} \quad u_i|u_{i-1}\sim N(\rho u_{i-1}, 1) \label{eq:skew_sim_eq}
\end{eqnarray}
We simulate a dataset of size $n$ with $\rho = \sqrt{3}/{2}, (\beta_0, \beta_1, \beta_2, \beta_3) = (-2, -3, -3, 1)$.
For inference, we fix $\rho$ at the true value $\rho = \sqrt{3}/{2}$, which ensures that the difference in methods is due to the explicit difference in the computational approach of the fixed effects and not implicitly due to differences in the hyperparameter estimates. \\
We compare the results from our proposal, SGC-VB, to those obtained from a long MCMC run. 

\subsubsection{Results}

In Table \ref{table5.1} the excessive zeros in our simulation is evident. This design results in the likelihood having limited information, and thus poses a challenge for inference of the parameters. This extreme scenario was chosen specifically to rigorously test the proposal.

\begin{figure}[h!]
	\centering
	\includegraphics[width=1\textwidth]{ 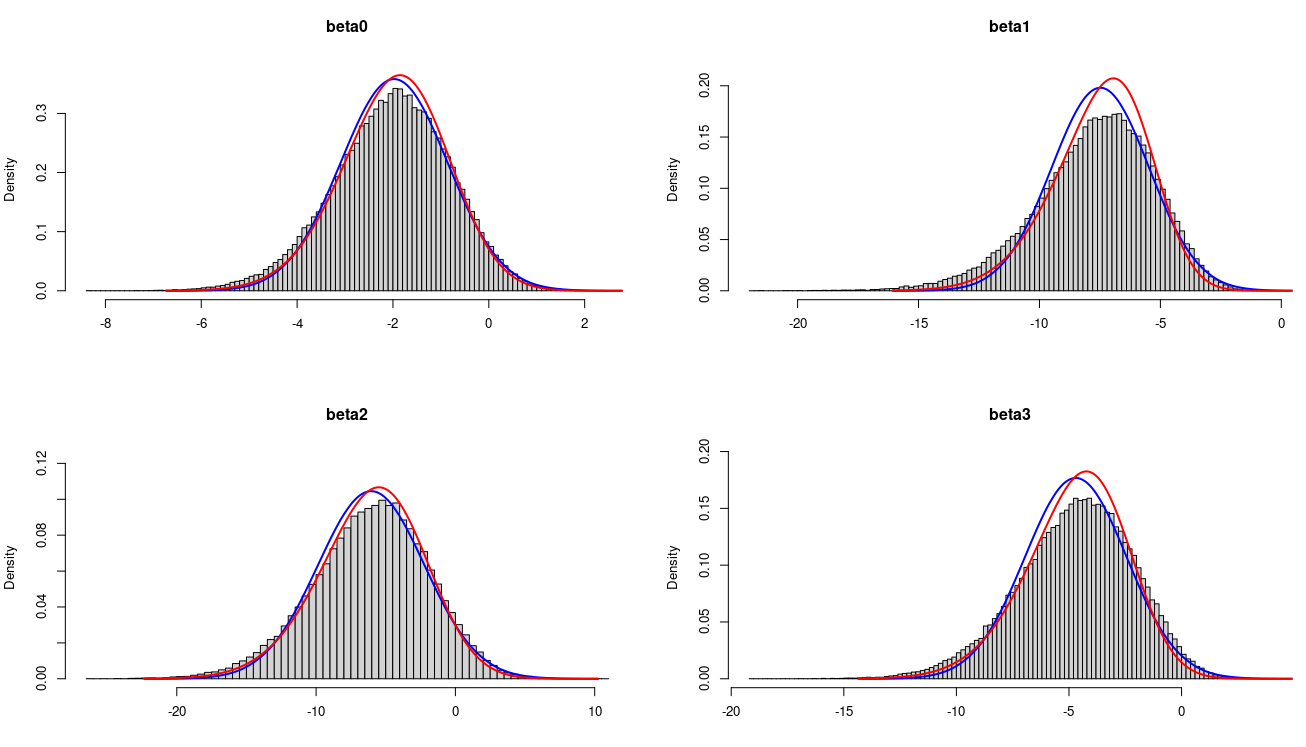}
	\caption{The marginal posterior distribution of the fixed effects for \eqref{eq:skew_sim_eq}, for $n=20$. The corrected Gaussian approximation (blue) and the proposed SGC-VB (red) posterior curves are superimposed on the histogram derived from an MCMC sample.} 
	\label{image5.120}
\end{figure}

\begin{figure}[h!]
	\centering
	
	\includegraphics[width=1\textwidth]{ 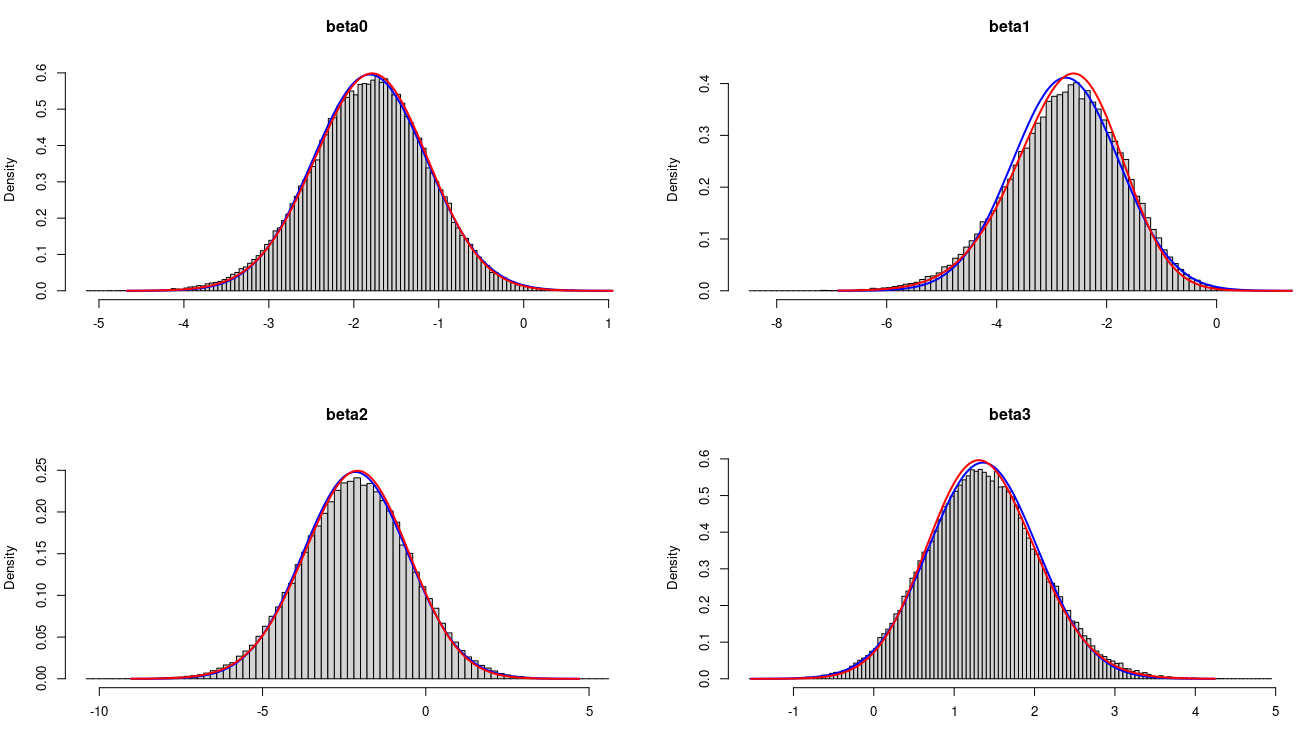}
	
	\caption{The marginal posterior distribution of the fixed effects for \eqref{eq:skew_sim_eq}, for $n=50$. The corrected Gaussian approximation (blue) and the proposed SGC-VB (red) posterior curves are superimposed on the histogram derived from an MCMC sample.} 
	\label{image5.150}
\end{figure}

\begin{figure}[h!]
	\centering
	
	\includegraphics[width=1\textwidth]{ 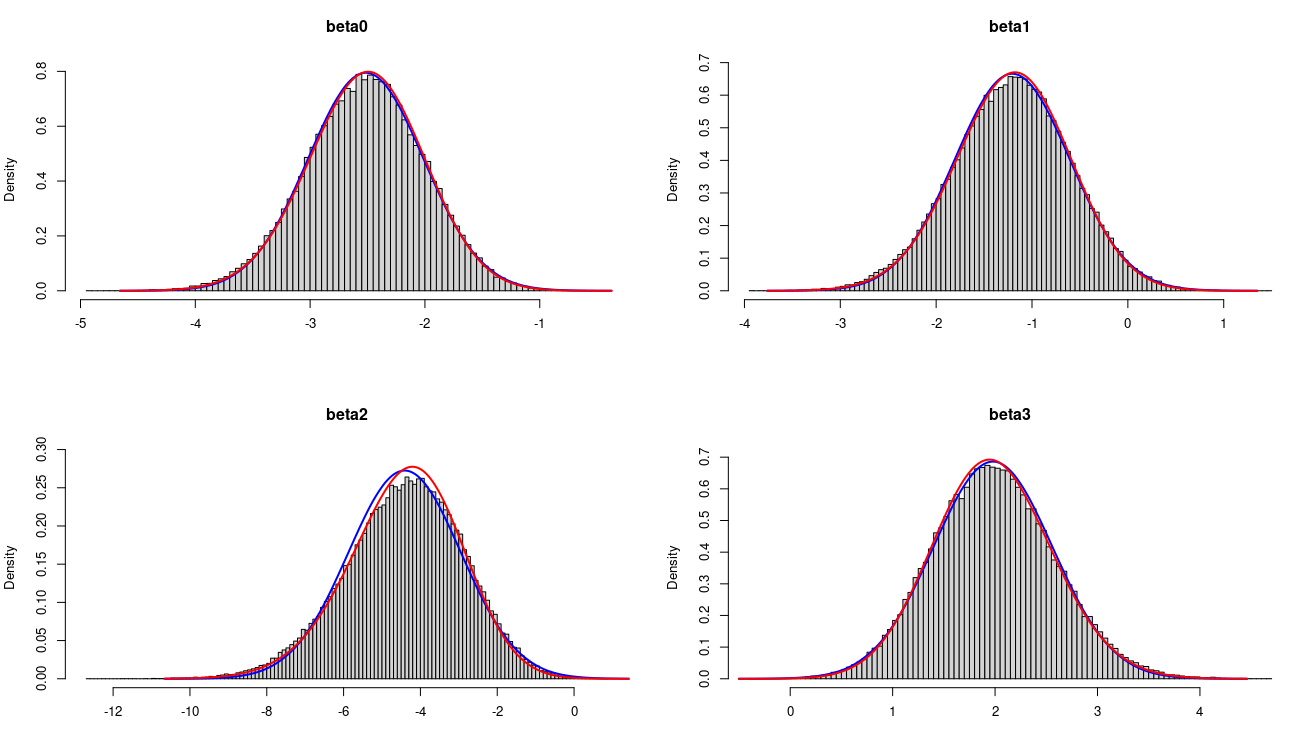}
	\caption{The marginal posterior distribution of the fixed effects for \eqref{eq:skew_sim_eq}, for $n=100$. The corrected Gaussian approximation (blue) and the proposed SGC-VB (red) posterior curves are superimposed on the histogram derived from an MCMC sample.} 
	\label{image5.1100}
\end{figure}

The results, as shown in Table \ref{table5.2}, reveal that our proposal is able to capture the  posterior marginal skewness comparably well to that from MCMC. This underscores the accuracy and ability of our proposal to correct for significant deviations from symmetry, in a computationally efficient way. The resulting marginal posterior distributions are shown in Figures \ref{image5.120}, \ref{image5.150} and \ref{image5.1100}.

\begin{table}[h]
	\centering
	\begin{tabular}{|c || c c c|} 
		\hline
		data &  0 &  1 & 2\\ 
		\hline
		frequency ($n=20$) &  16 & 1 & 3 \\
		\hline
		frequency ($n=50$) &  37 & 11 & 2 \\
		\hline
		frequency ($n=100$) &  80 & 16 & 4 \\
		\hline 
	\end{tabular}
	\caption{Description of the $n$ data points generated by \eqref{eq:skew_sim_eq}.}
	\label{table5.1}
\end{table}

\begin{table}[h]
	\centering
	\begin{tabular}{||c | c c | c c | c c ||} 
		\hline
		& \multicolumn{2}{c|}{$n=20$} & \multicolumn{2}{c|}{$n=50$} & \multicolumn{2}{c||}{$n=100$}  \\ 
		\hline\hline
		& SGC & MCMC & SGC & MCMC & SGC & MCMC \\ \hline
		$s_{\beta_0}$ &  -0.21 & -0.29  &  -0.08 & -0.13 & -0.08 & -0.11 \\ 
		\hline
		$s_{\beta_1}$ & -0.46 & -0.61  & -0.27 & -0.27 & -0.09 & -0.10   \\
		\hline
		$s_{\beta_2}$ & -0.27 & -0.31 & -0.09 & -0.07 & -0.26 & -0.27  \\
		\hline
		$s_{\beta_3}$ & -0.38 & -0.40 & 0.15 & 0.13 & 0.11 & 0.14 \\
		\hline 
	\end{tabular}
	\caption{The estimated marginal posterior skewness using the proposed approach (SGC-VB) and MCMC for different sample sizes.}
	\label{table5.2}
\end{table}


\section{Application to multiplex autoimmune assay data}\label{sec:app}
\begin{figure}[h!]
	\includegraphics[width = 6cm]{ 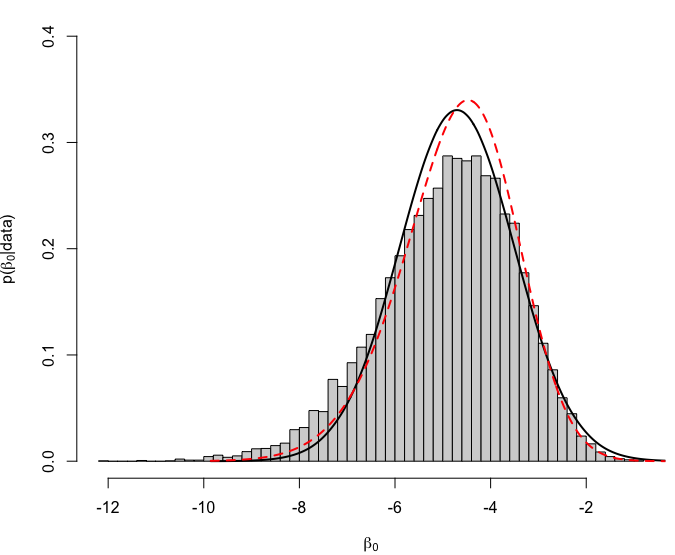}
	\includegraphics[width = 6cm]{ 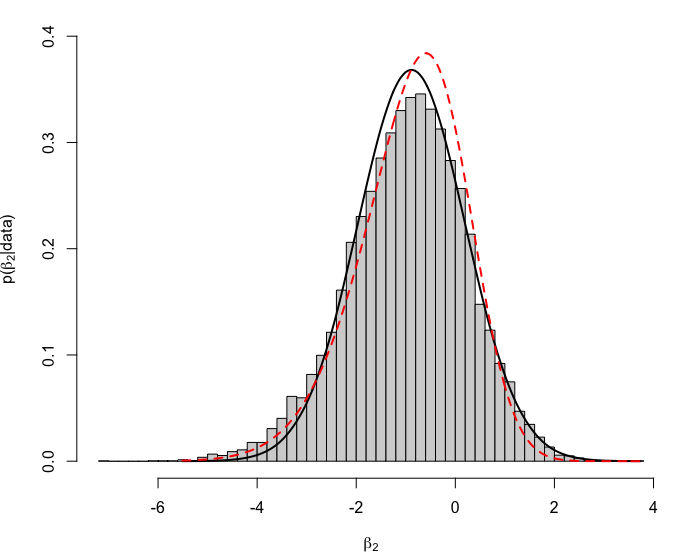}\\
	\includegraphics[width = 6cm]{ 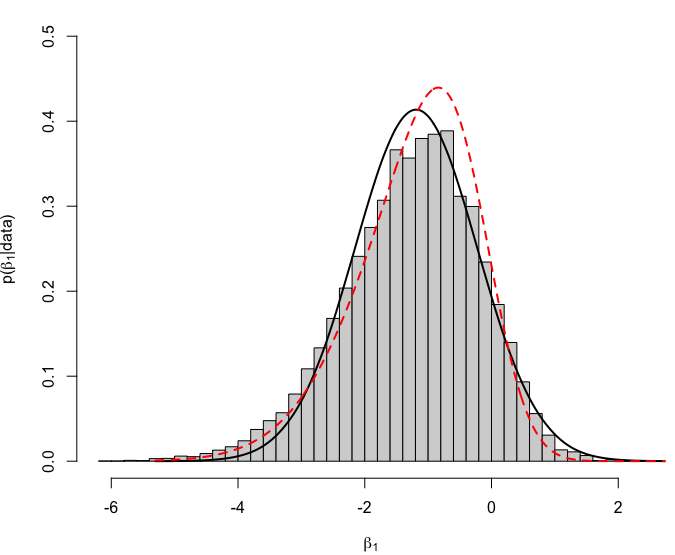}
	\includegraphics[width = 6cm]{ 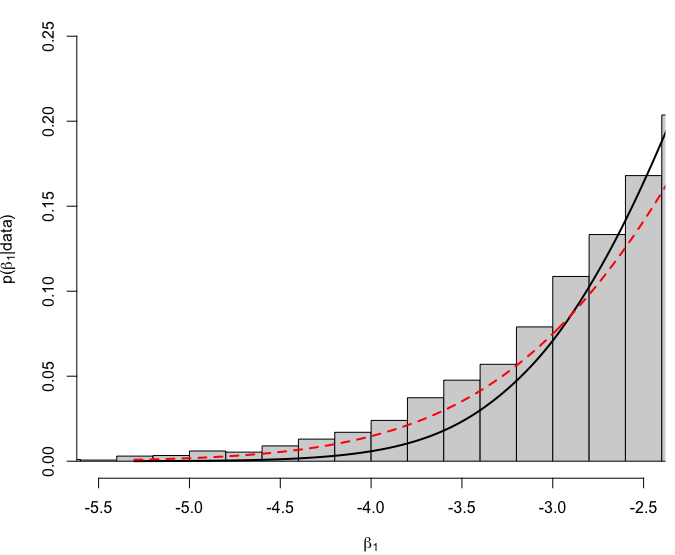}\\
	\includegraphics[width = 6cm]{ 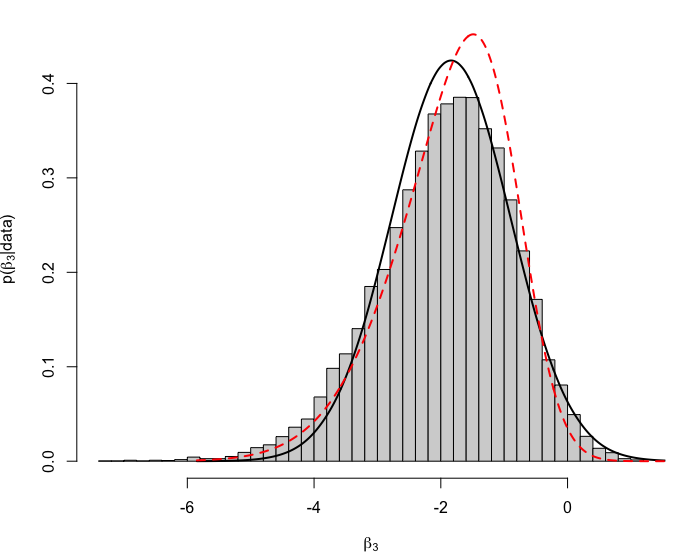}
	\includegraphics[width = 6cm]{ 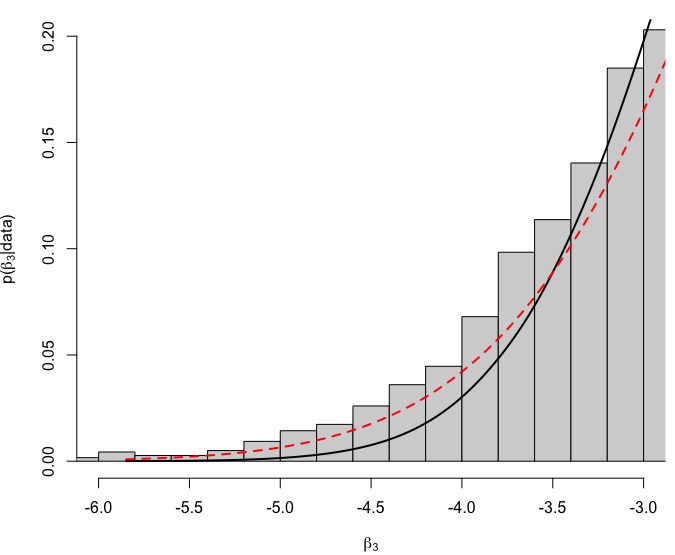}
	
	\caption{Posterior densities for the fixed effects using HMC (histogram), corrected Gaussian marginals (solid line) and skew-Gaussian marginals from SGC-VB (dashed line)}
	\label{fig:bimj}
\end{figure}

Class imbalanced data presents a challenge for classification frameworks, even though class imbalance occurs naturally in many situations, like rare diseases for example. The data relates to rheumatoid arthritis (RA) for which some biomarkers are available for a group of 10 patients with RA and 80 patients without (see \citet{schlieker2017multivariate} for more details). This class imbalance is often encountered in rare and neglected diseases. We consider a logistic regression model for the probability of having RA as follows

\begin{equation*}
	P(RA) = \text{logit}^{-1}(\eta),\quad \quad \eta = \beta_0 + \pmb\beta^\top\pmb X,
\end{equation*}
where $\pmb X$ is the design matrix constructed from three of the available biomarkers for the 90 patients. 
The priors assumed are weakly informative Gaussian with large variance. \\
We fit this model using three methods: Hamiltonian Monte Carlo (HMC) in STAN, INLA based on the mean and variance corrected Gaussian approximation (Section \ref{sec:meanvarcorr}), and INLA based on a SGC approximation using the SGC-VB proposal as presented in Section \ref{sec:skewcorr}. 
The respective results are presented in Figure \ref{fig:bimj}. The histogram is based on a sample from the posterior based on HMC, the black solid line is the corrected Gaussian approximation and the red broken line is the posterior approximated by the SGC-VB approach.

In Figure \ref{fig:bimj} we note that the SGC captures the posterior mode more accurately and can correct the symmetry of the Gaussian approximation sufficiently. The skewness correction allows for a better fit of the heavier tail of the posterior estimated from HMC. \\
The SGC-VB approach is more computationally efficient than HMC and it is clear that as the model complexity increases or the datasize grows, that the skewness-correcting approach provides an avenue for approximate Bayesian inference without the symmetry restriction, at a manageable computational cost.

\section{Discussion}\label{sec:disc}
Latent Gaussian models contain many well-known statistical models, and the implementation of the INLA methodology has increased their adoption in applied research and practice. Recent developments within the INLA framework enable even more efficient approximate Bayesian inference \citep{van2023new}. These developments are based on a low-rank Variational Bayes framework for one of the fundamental approximations in INLA. In this work, we adopt and extend the use of this Variational Bayes framework, to correct for departures from symmetry in the univariate conditional posteriors of fixed and random effects of LGMs (as described in Section \ref{sec:INLA}). \\ \\
Through simulation and a real data example, we show that the new approach captures significant departures from symmetry sufficiently well, when compared with "exact" sampling-based posteriors. It is important to note that the variance and skewness corrections are necessary and useful in specific cases such as heavy-tailed likelihoods, low-count Poisson models or binary regression models with large class imbalance. For most cases, the INLA framework provides sufficiently accurate posteriors based on only the mean correction approach (see Section \ref{sec:meancorr}), thus this approach is the default implementation of INLA in the current R-INLA library. The option for variance correction can be enabled with the command \begin{verbatim}
	control.inla = list(control.vb = list(strategy = "variance"))
\end{verbatim}and the skewness correction is not yet implemented.\\ \\
The premise of this paper is that we can use simple yet less accurate approximations, then correct and extend it in an efficient way, to infer our chosen target, contrary to inferring this in one step which can often be computationally expensive. Moreover, our proposal can be applied to other scenarios, such as modeling multivariate data and efficiently correcting posteriors after removing data in the case of cross-validation. The latter idea is currently being explored. In the case of complex models or large data, we might be left with only simple approximations as a viable efficient avenue and we envision that this work can be used 
to add more flexibility in such cases.\\
The code for reproducing the simulated and real examples is available at \url{https://github.com/JanetVN1201/Code_for_papers/tree/main/Skewing the Laplace method}.

\bibliographystyle{apalike}
\bibliography{References}

\end{document}